\documentclass[12pt,a4paper,,reqno]{amsart}
\usepackage{a4,}
\usepackage{amssymb,latexsym,,float}
\usepackage{enumerate}
\usepackage{graphicx,array}
\newcolumntype{?}{!{\vrule width 2pt}}

\makeatletter
\@namedef{subjclassname@2020}{2020 Mathematics Subject Classification}
\newcommand{\thickhline}{
\noalign{\ifnum 0=`}\fi \hrule height 2pt
\futurelet \reserved@a \@xhline}
\makeatother
\newcolumntype{?}{!{\vrule width 2pt}}
\usepackage{psfrag}
\newtheorem{theorem}{Theorem}[subsection]

\newtheorem{proposition}[theorem]{Proposition}
\newtheorem{corollary}[theorem]{Corollary}
\theoremstyle{definition}
\newtheorem{definition}[theorem]{Definition}
\newtheorem{definitiondis}[theorem]{Definitions and Discussion}

\newtheorem{example}[theorem]{Example}
\theoremstyle{remark}
\newtheorem{remark}[theorem]{Remark}
\numberwithin{equation}{section}
\def\FF{\mathbb{F}}
\def\H{\mathcal{H}}
\def\Z{\mathcal{Z}}
\def\K{\mathcal{K}}
\def\V{\mathcal{V}}
\def\E{\mathcal{E}}
\def\ZZ{\mathbb{Z}}
\allowdisplaybreaks
\begin{document}

\title[Optimal quantum tomography \ldots]
{Optimal quantum tomography with constrained \\ elementary measurements arising from unitary bases}
\author[S. Chaturvedi]{S. Chaturvedi}
\address{S. Chaturvedi, Department of Physics, Indian Institute of Science
Education and Research-Bhopal, Bhopal Bypass Road, Bhauri, Bhopal 462066, India}
\email{subhash@iiserb.ac.in}

\author[S. Ghosh]{Sibasish Ghosh}
\address{Sibasish Ghosh, Optics \& Quantum Information Group, The Institute of Mathematical Sciences, HBNI, C.I.T. Campus, Taramani, Chennai-600113, India}
\email{sibasish@imsc.res.in}

\author[K.R. Parthasarathy]{K.R. Parthasarathy}
\address{K.R. Parthasarathy, Theoretical Statistics and Mathematics Unit, Indian Statistical Institute, 7, S.J.S. Sansanwal Marg, New Delhi-110 016, India}
\email{krp@isid.ac.in}

\author[A. I. Singh]{Ajit Iqbal Singh}
\address{Ajit Iqbal Singh, INSA Emeritus Scientist, The Indian National Science Academy, New Delhi 110 002, India}
\email{ajitis@gmail.com}


\maketitle

\begin{abstract}
		The purpose of this paper is to introduce  techniques of obtaining optimal ways to determine a $d$-level quantum state or distinguish such states.
		It entails
		designing constrained elementary measurements extracted from maximal abelian subsets of  a unitary basis ${\bf U}$ for the operator algebra
		$\mathcal{B}(\mathcal{H})$ of a Hilbert space
		$\mathcal{H}$ of finite dimension $d>3$ or, after choosing an orthonormal basis for $\mathcal{H}$, for the $\star$-algebra $M_d$ of
		complex matrices of order $d>3$. Illustrations are given for the techniques. It is shown that the Schwinger basis ${\bf U}$ of
		unitary operators can give for $d$, a product of primes $p$ and $a$, the ideal number $d^2$ of rank one projectors that have a few quantum
		mechanical overlaps ( or, for that matter, a few angles between the corresponding unit vectors).
		Finally, we give a combination of the tensor product and constrained elementary measurement techniques to
		deal with all $d$, though with more overalaps or
		angles depending on the factorization of $d$ as a product of primes or their powers like
		$d=\prod_{j=1}^{k} d_j$ with $d_j=p_j^{s_j},~ p_1< p_2\cdots<p_k$, all primes,
		$s_j\geq 1$ for $1\leq j\leq k$, or other types. A comparison is drawn for different
		forms of unitary bases for the Hilbert space factors of the tensor product like
		$L^2(\mathbb{F}_t)$ or $L^2(\mathbb{Z}_u)$, where $\mathbb{F}_t $ is the Galois field of size $t=p^s$ and $\mathbb{Z}_u$ is the
		ring of integers modulo $u$. Even though as Hilbert spaces they are isomorphic, but quantum mechanical system -wise, these tensor products are different.

		In the process we also study the equivalence relation on unitary bases
		defined by R. F. Werner [{\it J. Phys. A: Math. Gen.} {\bf 34} (2001) 7081], connect it to local operations on maximally entangled vectors bases,
		find an invariant for equivalence classes in terms of certain commuting systems, called fan representations, and, relate it to mutually unbiased bases
		and Hadamard matrices.
		Illustrations are given in the context of latin squares and projective representations as well.
	\end{abstract}
	
	

	\tableofcontents
	\baselineskip13pt
	\section{Introduction}
	\def\P{\mathcal{P}}
	\def\FF{\mathbb{F}}
	\def\H{\mathcal{H}}
	\def\Z{\mathcal{Z}}
	\def\K{\mathcal{K}}
	\def\V{\mathcal{V}}
	\def\E{\mathcal{E}}
	\def\ZZ{\mathbb{Z}}
	
	Ivanovic \cite{5}, Wooters and Fields \cite{6}, Bandyopadhyay, Boykin, Roychowdhury and Vatan \cite{7}, Lawrence, Brukner
	and Zeilinger \cite{8}, Pittinger and Rubin \cite{9} and many other researchers constructed mutually unbiased bases (MUB's) for a
	$d$-level quantum system with $d$, $a$ prime and $d$, a prime power; they also pointed out obstructions to such a construction
	for certain composite numbers $d$. The starting point for them was quantum tomography. One aspect of quantum tomography is the study of
	identification of quantum states by means of
	a pre-assigned set of measurements. This set is usually taken to be a positive
	operator-valued measure (POVM) viz., a set $\mathbf{A} = \{A_j : 1 \leq j
	\leq v\}$ of positive operators on $\mathcal{H}$ with $\sum\limits_{j=1}^{\textit{v}}
	A_j = I_{\H}.$ The quantum state $\rho$ on $\mathcal{H}$ is then attempted to be
	determined via the tuple $\mathbf{\beta} = (\beta_j = (\rm{tr} \,\,(\rho
	A_j))^\textit{v}_{j=1}$ of
	measurement statistics. Because $\rm{tr} \,\,\rho = 1,$ we see that for any $j_0,$
	$\beta_{j_{0}} = 1- \sum\limits_{j_{0} \neq j=1}^{\textit{v}} \beta_j,$ and thus,
	only
	$v-1$ measurements are needed. If we can determine all states $\rho$ on
	$\mathcal{H}$ via $\mathbf{A}$, then $\mathbf{A}$ is said to be informationally complete. For
	that ${\bf A}$ spans $\mathcal{B}(\mathcal{H})$ and, therefore, $v$ has to be $d^2$ or more. For computational convenience, all $A_j$'s, except
	possibly one,
	are desired to have rank one. We call such a POVM a pure POVM. In a certain ideal sense, the problem shifts to finding a basis consisting of rank one projections of the ideal size $d^2$ for
	$\mathcal{B}(\mathcal{H})$. Pure states arising from a sought after complete system of MUB's work fine for this purpose and they are unbiased in the sense that any pair
	$P,P^\prime$ of distinct projections is either mutually orthogonal or has (quantum mechanical) overlap, i.e., $\text{tr}(PP^\prime)$ equal to $1/d$.
	But the problem is how to find them for a general $d$. For details, one can see
	for instance, the sources (\cite{5}, \cite{6}, \cite{7}, \cite{8}, \cite{9}) already referred to. In \S3 (Example \ref{e3.7.1}) we will display a basis consisting of rank one projections of the ideal size $d^2$ for
	$\mathcal{B}(\mathcal{H})$ with $\mathcal{H}=\otimes
	_{j=1}^{k} \mathcal{H}_j, \mathcal{H}_j= L^2(\mathbb{F}_{d_j})$, $d=\prod_{j=1}^{k}d_j,
	d_j=p_j^{s_j}$, $s_j\geq 1$ for $1\leq j\leq k$, $p_1<p_2<\cdots<p_k$, all primes,
	$\mathbb{F}_{d_j}$, the Galois field of size $d_j$  for $1\leq j\leq k$. But the number of quantum mechanical overlaps in this example is $2^k+1$.
	
	Equally relevant is the
	fundamental work on quantum designs by Zauner \cite{gz} or recent papers on symmetric informationally complete POVM's, in short, SIC--POVM's, like
	\cite{rksc}, \cite{sg}, \cite{mww}, \cite{gg} and \cite{aer},
	\cite{gs},\cite{fhs},\cite{abdf}.
	
	This problem is related to other equally
	subtle problem of equiangular lines, for instance, one can see Calderbank, Cameron, Kantor and Seidel \cite{10},
	Godsil and Roy \cite{11}, Kantor \cite{12}.
	
	Positive answers are known for all the problems above only for certain types of $d$ or special numerical values of $d$. The general case continues to
	remain at the conjecture level though the format has been changing from time to
	time. ( An elementary exposition
	of the concepts involved  may be found in \cite{14a}). So one has to look for alternative ways.
	
	The usual way is to begin with a unitary basis (UB),
	i.e., a collection $\mathbf{U}=\{U_x: x \in X\}$ of unitary operators
	$U_x$ on a $d$-dimensional Hilbert space $\H$
	such that ${\rm tr}(U_x^*U_y)= d \delta_{xy}$ for $x, y \in X$,
	where $A^*$ denotes the adjoint of a linear operator $A$ on $\H$.
	The reason for doing so is simply that they are systematically available for all $d$'s
	in a variety of ways, the most fundamental being the Schwinger basis in
	\cite{14b},
	also known as Weyl-Heisenberg operators.  This is an example of a unitary error basis in the theory of error correcting codes.
	There is a vast literature and we will refer to some of it as we go
	along in the paper.
	
	We may consider the case when $I_{\H}$, the identity operator on $\H$ is in $\mathbf{U}$ and then consider the unitary system
	$\mathbf{W}= \{ U_x:x \in X, U_x\neq I_{\H}\}$ instead. We can look for maximal abelian subsystems of $\mathbf{W}$, $\mathbf{W}$-MASS's so as to say.
	Then we can take a minimal set of $\mathbf{W}$-MASS's, say, $\V$, that covers $\mathbf{W}$. Any $\mathbf{W}$-MASS, say, $\mathbf{V}$
	has a common orthonormal basis of eigenvectors, say, $\E_{\mathbf{V}}$ and the corresponding system of one-dimensional projections, say,
	$\P_{\mathbf{V}}$. Then for each ${\bf V}$ , $\sum\{P: P\in \P_{{\bf V}}\}=I_{\H}$ . So except for one $ \mathbf{V}\in \V$, say ${\bf V}_0$ , we may
	ignore one $P$ in $\P_{\bf V}$ for other ${\bf V}$'s. We do so and continue to denote the truncated $\P_{{\bf V}}$ as $\P_{{\bf V}}$ only.
	Then $ \P_{\V}=\cup \{\P_{\bf V}:{\bf V}\in \V\}$ suffices in the sense
	that any operator  $\rho$ on $\H$ is determined by $\{{\rm tr}(\rho P): P\in \P_{\V}\}$.
	The trouble is that for a composite $d$, the size of $\P_{\V}$ may be more than the desired one, i.e., the ideal number $d^2$ for all ${\V}$'s.
	Such systems of smallest size are aimed at in the problem enunciated above.
	Theoretically speaking, any maximal linearly independent subset of $\P_{\V}$  will work fine. But we would like to have a method to obtain such sets and
	also desire some kind of mutual unbiasedness or only few angles or quantum mechanical overlaps like $\text{tr}(PP')$ for $P, P'\in \P_{\V}$.
	
	Parthasarathy \cite{3} gave a method to construct $\P_{\V}$ of size $1+(d -1)\prod\limits_{j=1}^k (d_j+1) $,
	where $d= \prod\limits_{j=1}^k d_j$ is the prime power factorization of $d$ with
	$d_j= p_j^{s_j}$, $p_1<p_2<\cdots<p_k$, all primes, $s_j\geq 1$ for$ 1\leq j\leq k$ with the help of tensor products of Weyl operators in the
	$L^2$-spaces over the finite fields $\mathbb{F}_{d_j}$
	of cardinality $d_j$, $1\leq j \leq k$. This motivated \S3.7, in fact.
	
	Chaturvedi, Mukunda and Simon \cite{2}
	termed their detailed study of the problem as Wigner distributions for finite-state systems without redundant phase-point operators,
	related it to isotropic lines in the lattice $\mathbf{Z}_d \times \mathbf{Z}_d$, well-studied by Albouy \cite{1}, for instance,
	and provided explicit methods to obtain certain $\P_{\V}$ in their set-up.
	Shalaby and Vourdas (\cite{shvo1,shvo2}) continued this further, gave a tomographically complete
	set of bases essentially for $d$, a product of two primes \cite{shvo1} and studied such so-called
	weak mutually unbiased bases \cite{shvo2}. The Hilbert space format here is $\mathcal{H}=L^2(\ZZ_d)$, with $\ZZ_d$ , the ring of integers modulo $d$,
	which is simpler and so are the unitary bases, such as Schwinger basis, than those in \cite{3} indicated above. The optimal constrained elementary
	measurements techniques introduced in \S3.1 help to obtain, in \S3.6, exactly $d^2$ projections of rank one with few overlaps
	viz. $0,1/p,1/d,1$ for $d=p^2$ and few overlaps including $0,1/p,1/a,1/d$ for $d=pa$ with $p$ and $a$ distinct primes.
	The tensor product technique in \S3.7 helps us to deal with all $d$'s albeit with the number of overlaps increasing to one more than the number of
	factors of $d$.
	
	The purpose of this paper is to introduce the techniques of constrained elementary measurements to obtain optimal ways to determine
	the state of a $d$-level quantum system or distinguish such states. Maximal abelian subsystems like $\mathbf{W}$-MASS's derived from unitary
	bases $\mathbf{U}$ are taken as the first step for these
	methods. The techniques are based on the observation that for any two such distinct ${\bf W}$-MASS's, say, ${\bf V}_1$ and ${\bf V}_2$ having
	nonempty intersection ${\bf V}_{12}$, common orthonormal systems in $\mathcal{H}$ and accordingly, the corresponding sets of rank one projections,
	say $\mathcal{P}_1$ and $\mathcal{P}_2$ for ${\bf V}_1$ and ${\bf V}_2$ respectively are constrained by the projections on eigenspaces of unitary
	operators in ${\bf V}_{12}$. This helps to reduce the number of rank one projections actually needed. Fan representation of ${\bf W}$-MASS's help
	in applying our techniques. It turns out that it can be well illustrated in
	the scenarios mentioned in the previous paragraph. For more situations, the
	construction and classification given by Werner \cite{14} is handy. Further details are developed in this paper with an eye on their merits
	or demerits vis-a-vis quantum tomography. To elaborate, we display different unitary bases, take one and first check if it has a small size minimal set
	$\V$ of ${\bf W}$-MASS's that cover ${\bf W}$, and, next, if we can have a method to systematically shorten $\mathcal{P}_{\V}$ to the ideal size of
	$d^2$, or, else, obtain an optimal size shortening. We also try to keep control on or an estimate of the number of quantum mechanical overlaps. Our
	techniques can be suitably modified for other bases like hermitian or normal bases for $\mathcal{B}(\mathcal{H})$ as well. We display that too
	for the so called Generalized Gell-Mann basis.
	
	This paper is organized as follows. The next section presents some basic material in a form that we need, with a tinge of
	novelty, mainly drawn from Werner \cite{14} and Vollbrecht and Werner \cite{13}, Parthasarathy \cite{3}, Albouy \cite{1},
	Chaturvedi, Mukunda and Simon \cite{2} and Shalaby and Vourdas \cite{shvo1,shvo2}. The third section gives the main techniques of optimal
	constrained measurements. It is followed by illustration of the
	methods to arrive at the ideal number $d^2$ of rank one projections with a few quantum mechanical overlaps. \S4 is devoted to other unitary bases. The last section contains our concluding remarks.
	
	\section{Basics of Unitary Bases and Elementary Measurements}
	It is convenient first to collect relevant material from known sources in a form that we need.
	\subsection{Unitary bases and unitary systems}
	We shall freely use \cite{14} and \cite{13} in this subsection. Let $\mathcal{H}$
	be a $d$-dimensional Hilbert space with $2\leq d< \infty$ and $X$ a set of $d^2$
	elements such as $\{1,2,\ldots,d^2 \}$ or $\{0,1,\ldots, d^2-1\}$ etc.\
	\subsubsection{} We begin with a unitary basis.
	\begin{itemize}
		\item[(i)] Let $\mathbf{U} = \{U_x : x \in X\}$ be a {\it unitary basis} in the context of
		$\mathcal{H}$, in
		short, UB, i.e. a collection $\mathbf{U}$ of unitary operators $U_x \in
		\mathcal{B}(\mathcal{H}),$ the $\ast$-algebra of (bounded) linear operators on
		$\mathcal{H}$ to itself, such that ${\rm tr} (U_x^{\ast}U_y)=d \delta_{xy}$ for $x,y
		\in X.$
		\item[(ii)] Rewording a part of the discussion after Proposition 9 \cite{14},
		we call two unitary bases $\mathbf{U}$ and $\mathbf{U}^{\prime}$ {\it equivalent}
		if there exist unitaries $V_1,$ $V_2$ in $\mathcal{B}(\mathcal{H})$ and a
		relabelling $x \rightarrow x^{\prime}$ of $X$ such that $U_{x^{\prime}}^{\prime}
		= V_1 U_x V_2$ for $x$ in $X.$
		\item[(iii)] We fix an orthonormal basis $\mathbf{e} = \{e_j: 1 \leq j \leq d\}$
		in $\mathcal{H}$ and identify $\mathcal{H}$ with $\mathbb{C}^d$ as such. This
		will permit us to identify $\mathcal{B}(\mathcal{H})$ with the $\ast$-algebra
		$M_d$ of complex $d \times d$ matrices. We set
		$\Omega = \frac{1}{\sqrt{d}} \sum\limits_j e_j \otimes e_j.$
		\item[(iv)] We recall a one-to-one linear correspondence $\mathcal{T}$ (in terms of
		this
		maximally entangled vector $\Omega$ in $\mathcal{H} \otimes \mathcal{H}$)
		between $\Psi \in \mathcal{H} \otimes \mathcal{H}$ and $A \in
		\mathcal{B}(\mathcal{H}),$ as set up in \cite{13}, Lemma 2 or \cite{14}, Proof
		of Theorem 1, for instance, via $\langle e_j| Ae_{k}\rangle = \sqrt{d} \langle
		e_j \otimes e_k, \Psi \rangle.$ At times we express this as $\Psi \rightarrow
		A_{\Psi}$ or, even $A \rightarrow \Psi_A$ and write $\mathcal{T} (\Psi)=A_{\Psi}.$

	\begin{itemize}
		\item[(a)] Note that $\mathcal{T}({\Omega}) = I$, the $d \times d$ identity matrix. The map $\mathcal{T}$ takes the set $\mathcal{M}$ of maximally entangled states in $\mathcal{H} \otimes \mathcal{H}$ to the set
		$\mathcal{U}(\mathcal{H})$ of unitaries on $\mathcal{H}.$
		\item[(b)] The rank of $A$ and the Schmidt rank of $\Psi_A$ are the same.
		\item[(c)] Entropy of $A^{\ast} A$ or certain variants are in vogue as measures
		of entanglement of $\Psi_A.$
	\end{itemize}
		\item[(v)] The {\it flip} (or {\it swap}) operation in $\mathcal{H} \otimes
		\mathcal{H}$ is the linear operator $\mathbb{F}$ determined by $\mathbb{F}
		(\varphi \otimes \psi) = \psi \otimes \varphi,$ for $\varphi,$ $\psi \in
		\mathcal{H},$ or equivalently, by $\mathbb{F}(e_j \otimes e_k) = e_k \otimes
		e_j$ for $1 \leq j,$ $k \leq d.$ Then $\mathbb{F}$ induces the {\it transpose
			operator } $A \rightarrow A^t$ on $\mathcal{B}(\mathcal{H})$ to itself defined
		in the basis $\mathbf{e}.$ We note the useful underlying fact: $\Psi = (A_{\Psi}
		\otimes I ) \Omega = (I \otimes A_{\Psi}^{t}) \Omega,$ where $I$ is the
		identity operator on $\mathcal{H}.$
		\item[(vi)] Members of the set $\mathcal{L} \mathcal{U} (\mathcal{H} \otimes
		\mathcal{H})$ of local unitaries, viz., $\{U_1 \otimes U_2 : U_1, U_2 \in
		\mathcal{U}(\mathcal{H})\}$, and $\mathbb{F}$ all take unit product vectors to
		unit product vectors and Proposition 4 in \cite{13} says that a unitary operator
		$U$ on $\mathcal{H} \otimes \mathcal{H}$ satisfies $U \mathcal{M} \subset
		\mathcal{M}$ if and only if $U$ is local up to a flip, i.e., there are unitaries
		$U_1$, $U_2$ on $\mathcal{H}$ such that either $U = U_1 \otimes U_2$ or
		$U = (U_1 \otimes U_2)$ $\mathbb{F}$.
	\end{itemize}
	\subsubsection{} We first define a unitary system.
	\setcounter{theorem}{1}
	\begin{definition}[Unitary systems]\label{def2.2}
		\begin{itemize}
			\item[]
			\item[(i)] A non-empty set $\mathbf{W} = \{ W_y : y \in Y\}$ of unitaries in
			$\mathcal{H}$ indexed by $Y$ will be called a {\it unitary system,} in short,
			US, if ${\rm tr} \,W_y =0$ and ${\rm tr} \, (W_x^{\ast} W_y)=d \delta_{xy}$ for $x,$ $y$
			in $Y.$ The number $s = \# Y$ will be called the {\it size} of $\mathbf{W}.$
			
			\item[(ii)] An {\it abelian unitary system,} in short, { AUS}, is a unitary
			system $\mathbf{W}$ with $W_xW_y=W_y W_x$ for $x,y$ in $Y.$
			
			\item[(iii)] {\it A maximal abelian subsystem of a unitary system} $\mathbf{W},$
			in short, $\mathbf{W}$-MASS, is a subset $\mathbf{V}$ of $\mathbf{W}$ which is
			an AUS and maximal with this property.
		\end{itemize}
	\end{definition}
	We note that ${\rm tr}\, A=0$ for each $A$
	in the linear span $\mathcal{L}$ of $\mathbf{W}.$ So $I \not\in \mathcal{L}.$
	Also ${\rm tr}\, (W_x^{\ast} W_y) = d \delta_{xy}$ for $x,y$ in $Y$ forces $\{W_x : x
	\in Y\}$ to be linearly independent. So we have $s=\# Y \leq d^2-1$ and, thus,
	we may consider $Y$ as a proper subset of $X,$ if we like.
	\subsubsection{}\label{r2.1.3} We now briefly describe the relationships of the above-mentioned notions amongst themselves together with MUBs and Hadamard matrices.
	\begin{itemize}
		\item[]
		\item[(i)] Given a US $\mathbf{W},$ $\widetilde{\mathbf{W}} = \mathbf{W} \cup
		\{I\}$ will be called the {\it unitization} of $\mathbf{W}.$ We note that
		$\widetilde{\mathbf{W}}$ is a system of mutually orthogonal unitaries and it is
		a UB if and only if the size of $\mathbf{W}$ is $d^2-1.$
		
		\item[(ii)] AUS of size $d-1$ have been very well utilized by Wootters and
		Fields \cite{6} and Bandyopadhyay et. al. \cite{7} to study {\it mutually
			unbiased bases} in $\mathcal{H},$ (see also \cite{8},
		[16], [20], [27], [28], \cite{Sz}, \cite{msp}, \cite{psmw}, \cite{PPGB}).
		Lemma 3.1 in \cite{7} records the basic fact that the size of an AUS can at most be $d-1.$
		
		\item[(iii)] In fact, if $\mathbf{W} = \{W_x: x \in Y\}$ is an AUS of size $s,$
		then there is a unitary $U$ on $\mathcal{H}$ and mutually orthogonal operators
		represented by mutually orthogonal diagonal matrices $D_x,$ $x \in Y$ with
		entries in the unit circle $\mathbf{S}^1$ such that ${\rm tr}\, D_x=0,$ $W_x=U D_x
		U^{\ast},$ $x \in Y.$ As a consequence, the $(s+1) \times d$ matrix $H$ formed
		by the diagonals of $I$ and $D_x,$ $x \in Y,$ as rows is a partial complex
		Hadamard matrix in the sense that $HH^{\ast} = d I_{s+1}$ and $|H_{jk}|=1$ for
		each entry $H_{jk}$ of $H.$ This forces $s \leq d-1.$ History and development
		of Hadamard matrices is very long and fascinating. We just mention a few sources like
		\cite{TZ}, \cite{Sz}, \cite{deLL}, \cite{Ha}, \cite{V1},
		\cite{V2}, \cite{Ho} which we can directly~use.
		
		\item[(iv)] (\cite{7}, Theorem 3.2. (See also [20], Section 2, which generalizes the notion of MUB's as well) says that if a  UB $\mathbf{U}$
		containing $I$ can be partitioned as a union of $(d+1)$ $\mathbf{U}\backslash
		\{I\}$-MASS's of size $(d-1)$ each together with $I,$ then one can construct a
		complete system of $(d+1)$ MUB's of $\mathbb{C}^d.$ The converse is given by
		them as (\cite[Theorem 3.4]{7}). Concrete illustration is given for $d=p^r,$
		with $p$ a prime.
		
		\item[(v)] Obstructions to construction of MUB have occupied many researchers,
		e.g., see \cite{6}, \cite{7}, \cite{PDB}, \cite{msp}, \cite{psmw},
		\cite{PPGB}. Even when, say for, $d$ a prime power, complete systems of MUB's
		exist, some subsystems of certain UB's may not be extendable to complete
		systems of MUB's, this is explained well by Mandayam, Bandyopadhyay, Grassl and
		Wootters \cite{psmw} for
		$d=2^2, 2^3, \ldots.$. A pictorial illustration with explanation for $d=2^2$ is
		provided in Example \ref{e3.2.6} in the next section.
	\end{itemize}
	\subsubsection{\it Example (Shift and multiply)}\label{e2.1.4}
	This is based on (\cite{13}, III.A) or (\cite{14}, Proposition 9). Let $Y_d =
	\{1,2,\ldots, d \}.$ For $m \in Y_d,$ let $H^{(m)} = [H_{jk}^{m}]$ be a complex
	$d \times d$ Hadamard matrix, i.e. $H^{(m)}$ $H^{(m)\ast} = dI$ and
	$|H_{jk}^{m}| =1$ for all $m,j,k.$ Let $\lambda$ be a latin square, i.e.
	$\lambda$ is a map on $Y_d \times Y_d$ to $Y_d$ satisfying $k \rightarrow
	\lambda (k,\ell)$ and $k \rightarrow \lambda (\ell, k)$ are injective for every
	$\ell.$ We shall write $e_k$ as $|k \rangle$ for $k \in Y_d.$
	
	For $m,n \in Y_d,$ let $U_{mn}$ (or, at times, also written as $U_{m,n}$ or
	$U_{(m,n)}$) be the operator which takes $|k\rangle$ to $H_{mk}^n$ $|\lambda
	(n,k)\rangle,$ $k \in Y_d.$. Then $\mathbf{U} = \{U_{m,n} : m, n \in Y_d \}$ is a UB. We note that
	its indexing set is $X=Y_d \times Y_d.$ Further, $U_{m,n} = I$ if and
	only if $\lambda(n,k)=k$ and $H_{mk}^n = 1$ for each $k \in Y_d.$ In this case
	$\mathbf{W} = \mathbf{U}\backslash \{I\}$ is a unitary system.
	
	At times, we will find it more convenient to replace the indexing set $Y_d$ by $\mathbb{Z}_d=\{0,1,\cdots,d-1\}$ and accordingly,
	take basis $\{|j\rangle\}:j=0,1,\cdots,d-1\}$ and index the tuples in $\mathbb{C}^d$ also by $0,1,\cdots,d-1$. This will impact the labelling of
	matrix entries as well as formats for $U_{m,n}$'s.
	
	This includes various known classes by careful choice of the latin square as a group and Hadamard matrices as coming from characters on the group.
	We shall confine our
	attention to examples detailed in \S2.2 and \S2.3 (essentially the Schwinger basis) in \S3 and look at more general ones in \S4.
	
	\subsection{Weyl operators based techniques}
	Ampliations of Pauli matrices (which constitute the first stage unitary bases for $d=2$) and their compositions are familiar techniques in
	Quantum Mechanics. Here, for tensor product $\H=\otimes_{j=1}^{k}\H_j$ of Hilbert spaces $\H_j$'s of respective dimensions $d_j\geq 2$, and
	operator $A$ on $\H_{j_0}$ to itself,
	the ampliation of $A$ is the operator $A^{(j_0)}$ on $\H$ to itself given by $A^{(j_0)}= \otimes_{j=1}^{k}A_j$ with $A_{j_0}=A$ and $A_j = I_{\H_{j}}$ for
	$j\neq j_0$.

	\subsubsection{}\label{2.2.1} Parthasarathy \cite{3} carried the technique further to advantage to give a unitary basis $\mathcal{F}$ for a
	composite $d=\prod\limits_{j=1}^k p_j^{s_j}$
	by identifying the $d$-dimensional Hilbert space $\H$ with $\bigotimes_{j=1}^k \H_j$, where $\H_j=L^2(\FF_{d_j})$, $\FF_{d_j}$ being the finite
	field of cardinality
	$d_j=p_j^{s_j}, ~p_1<p_2<\cdots <p_k,~\text{all primes and}~s_j\geq 1~\text{for}~1\leq j\leq k$. He used for $1\le j\le k$, Weyl operators $\{W(a_j,x_j),, a_j\in \FF_{d_j}\cup\{\infty_j\},x_j\in \FF_{d_j}\}$ on $\H_j$.
	To elaborate, for $q=p^s$,
	a prime power, fix any non-trivial character $\chi$ of the additive group $\FF_q$ and put $\langle x,y\rangle=\chi(xy)$, $x,y\in\FF_q$. Then
	$\langle \cdot,\cdot\rangle$ is a non-degenerate symmetric bicharacter on $\FF_q$. Using the counting measure on $\FF_q$ and writing the indicator
	function of
	$\{x\}$ as $|x\rangle$, $\{|x\rangle:x\in \FF_q\}$ is an orthonormal basis for $\K=L^2(\FF_q)$. For $a,b\in \FF_q$, let $U_a$ and $V_b$ be the
	unitary operators on $\K$
	determined by relation $U_a|x\rangle=|a+x\rangle$ and $V_b|x\rangle=\langle b,x\rangle|x\rangle$, $x\in \FF_q$.
	He manipulates phase factors $\alpha(a,x)$ in terms of $\chi$ to obtain for $a\in\FF_q\cup\{\infty\}=\tilde\FF_q$, (say), $x\in\FF_q$, unitary
	operators on $\K$
	given by
	\begin{align}
		W(a,x)=
		\begin{cases}
			\alpha(a,x)U_xV_{ax},&\text{if} \ a\in\FF_q, \ x\in \FF_q\\
			V_x,&\text{if} \ a=\infty,\ x\in \FF_q
		\end{cases}
		\label{2.1}
	\end{align}
	which in addition to satisfying Weyl commutation relations, have a neat\break property $W(a,x)W(a,y)=W(a,x+y)$ for $a\in\tilde\FF_q$, $x,y\in\FF_q$.
	
	\def\B{\mathcal{B}}
	\def\F{\mathcal{F}}
	The unitary basis so constructed for $B(\K)$ is simply $\F_{\K}=\{I_{\K},W(a,x): x\in\FF_q\setminus \{0\}, a\in\tilde\FF_q\}$. It can be partitioned
	into a complete system of MUB's given by $\F_{a}=\{W(a,x): x\in\FF_q\setminus \{0\}\}, a\in\tilde\FF_q,  $ together with $I_{\K}$.

	Finally, the announced unitary basis for $\B(\H)$ is
	\begin{align*}
		\F
		&=\{I_{\H},W^{(i_1)}(a_{i_1},x_{i_1})W^{(i_2)}(a_{i_2},x_{i_2})\ldots W^{(i_r)}(a_{i_r},x_{i_r}),a_{i_s}\in\tilde F_{d_{i_s}},x_{i_s}\!\in\!\FF_{d_{i_s}}
		\!\setminus \{0\},\\
		&\qquad s=1,2,\ldots,r , \ 1\le i_1<i_2<\cdots<i_r\le k, \ r=1,2,\ldots,k\}.
	\end{align*}
	Here, for $1\le j\le k$, $A\in\B(\H_j)$, $A^{(j)}$ is the ampliation of $A$ to $\H$.
	
	For further use, we may write the members in compact form:
	\begin{align*}
		&J=\mathbf{i}=(i_s)_{s = 1}^{r}, (1\le i_1<i_2<\ldots<i_r\le k),\quad \mathbf{a}=(a_{i_s})_{s=1}^r,\\
		&\mathbf{x}=(x_{i_s})^r_{s=1},\quad W(\mathbf{a},\mathbf{x})=\prod\limits_{s=1}^r W^{(i_s)}(a_{i_s},x_{i_s}),
	\end{align*}
	where the order is immaterial as $W^{(i_s)}(a_{i_s},x_{i_s})$'s commute with each other.

	\subsubsection{}\label{2.2.2}
	We continue with relevant formulation of excerpts from \cite{3}.\
	For $q$ and other entities as in \S \ref{2.2.1} above, (\cite{3}, Theorem~2.2) can be restated as:\\
	There exist orthogonal rank one projection operators $\{P(a,y):a\in\tilde \FF_q,y\in\FF_q\}$
	that satisfy for $a\in\tilde\FF_q$, $x,z\in\FF_q$,
	\begin{itemize}
		\item[(i)] $W(a,x)=\sum\limits_{y\in\FF_q}\langle x,y\rangle P(a,y)$,
		
		\item[(ii)] $P(a,z)=q^{-1}\sum\limits_{y\in\FF_q}\overline{\langle z,y\rangle} W(a,y)$,
		
		\item[(iii)] $\mathrm{tr}\,P(a,x)P(b,z)=q^{-1}$ for $a\neq b$,
		
		\item[(iv)] $P(a,x) P(a,z)=0$ for $x$ not equal to $z$, and
		
		\item[(v)] $\sum\limits_{y\in\FF_q} P(a,y)= I_{\K}$.
	\end{itemize}
	\subsubsection{}
	\S $(\ref{2.2.2})$ gives rise to projections on $\H$ of the type $P(\mathbf{a},\mathbf{x}) =$  $\prod\limits_{s=1}^r P^{(i_s)}(a_{i_s},x_{i_s})$
	on the lines of \S $(\ref{2.2.1})$ above. Its rank is $\prod\limits_{j\notin \{i_1,i_2,\ldots,i_r\}}d_j$. The important point is that a density
	$\rho$ on $\H$ can be recovered from the probabilities
	$\mathrm{tr}\, [\rho P(\mathbf{a},\mathbf{x})]$ and projections $P(\mathbf{a},\mathbf{x})$'s in the
	following sense.
	
	For $J=(\mathbf{i})$, $(1\le i_1< i_2<\ldots<i_r\le k)$, let $S_\rho(J)=\sum \mathrm{tr}(\rho P(\mathbf{a},\mathbf{x})) P(\mathbf{a},\mathbf{x})$ with $(\mathbf{a},\mathbf{x})$ varying in
	$\prod\limits_{s=1}^r\tilde\FF_{d_{i_s}}\times \prod\limits_{s=1}^r \FF_{d_{i_s}}$. Further, set $S_\rho(\emptyset)=I_\mathcal{H}$.
	
	Then (\cite{3}, Theorem~3.1) says that
	\begin{align*}
		\rho=\sum (-1)^{k-\#J}S_{\rho}(J),
	\end{align*}
	where summation is over all $J$ as specified above.
	\subsubsection{}
	Now the ranks of the projections $P(\mathbf{a},\mathbf{x})$ involved are $>1$ unless $r=k$.
	So measurements may not be easy.\ On the other hand, if we consider $J=(1<2<\ldots<k)$ alone, the number of these projections is larger than
	the ideal number $d^2$. In fact, it requires $\prod\limits_{j=1}^k (d_j+1)$ elementary measurements (see the remark after Theorem~3.1 in \cite{3}).
	We shall show  in \S3 that the number of projections can be systematically reduced to the ideal number $d^2$ by our techniques and it has
	$2^k+1$ quantum mechanical overlaps or angles.
	\subsection{Cyclic group case of example (shift and multiply)}
	We consider the special case of \S\ref{e2.1.4} above, when the latin square $\lambda$ is the group table of
	the cyclic group $\mathbb{Z}_d = \{0, 1, \ldots, d - 1\}$ with addition modulo $d$, and $H_{j,k}^{n} = (\eta_d)^{jk},$ where $\eta_d =e^{\left(\frac{2 \pi i}
		{d}\right)}$.
	It is essentially the Schwinger basis [19]. We note that $U_{0,0}=I_d$, in short, $I$ and $\tilde{\bf W}$ coincides with ${\bf U}$.
	Further, $U_{mn}$ commutes with $U_{m'n'}$ if and only if $mn'-m'n=0$;
	we write this also as $(m,n)\Delta(m',n')$.
	This subsection is based on \cite{1}. Consider now the discrete phase space $X=\ZZ_d^2 = \mathbb{Z}_d \times \mathbb{Z}_d$. We consider
	$\ZZ_d$ as a ring and $X$ as a module over it. Next, at times for the sake of convenience, we will not distinguish between $U_{(m,n)}$ and $(m,n)$.
	
	\subsubsection{}
	An \emph{isotropic line} is a set of $d$ points in the lattice $X$ such that the symplectic product $w(\sigma,\sigma')=mn'-m'n$ of any
	two points $\sigma=(m,n)$
	and $\sigma'=(m',n')$ is zero (mod $d$). The \emph{orthogonal} of a submodule $M$ of $X$ is denoted $M^w$, i.e.,
	$M^w=\{\sigma\in X:\forall \ \sigma'\in M,w(\sigma,\sigma')=0 \ \pmod{d}\}$.
	
	\emph{Isotropic submodules} are those that satisfy $M\subset M^w$. And \emph{Lagrangian submodules} are the maximal isotropic submodules for inclusion,
	which is equivalent to $M=M^w$.
	
	Albouy (\cite[\S2]{1}) identifies all Lagrangian submodules, first for $d$, a power of prime and then for the general case $d$ with
	$\prod\limits_{i\in I}p_i^{s_i}$, the
	prime factor decomposition of $d$. He proves that the Lagrangian submodules are the same as isotropic lines of $X$ and determines the number of isotropic
	lines of $X$
	as $\prod\limits_{i\in I}(p_i^{s_i+1}-1)/(p_i-1)$. We just note that by definition, a $\mathbf{W}$-MASS together with $(0,0)$ in this context
	is just a Lagrangian submodule (and vice-versa)
	and record the consequence that follows immediately.
	\subsubsection{}
	
	\def\prodl{\prod\limits}
	\begin{itemize}
		\item[(i)] All $\mathbf{W}$-MASS's are of full size $d-1$ and their number is $\prodl_{i\in I}(p_i^{s_i+1}-1)/(p_i-1)$, where
		$\prodl_{i\in I}p_i^{s_i}$ is
		the prime power factorization of $d$.
		\item[(ii)] Albouy \cite{1} goes on to realize enumeration of isotropic lines as orbits under the action of $SL(2,\ZZ_d)$, when $\mathbb{Z}_d$ is
		treated as a ring. One can see \cite{1} and \cite{2} for more details.
		\item[(iii)] Albouy \cite{1} determines the isotropic lines through a point $x$ in terms of $p_i$-valuations $v_{p_i}(x)$ of $x$, $i\in I$. Again, for details one
		can see \cite{1} and for applications, \cite{2}. We shall come back to that in our next section.
	\end{itemize}
	\subsection{Wigner distributions for finite-state systems}
	
	With $d$ as above in \S2.3, viz., $d=\prodl_{i\in I}p_i^{s_i}$, Chaturvedi, Mukunda and Simon \cite{2} take the set-up of the cyclic group $\ZZ_{d}$
	considered as a ring expressed as the product of
	$\ZZ_{p_i^{s_i}}$, $i\in I$, facilitated by the Chinese Remainder Theorem instead.\
	They study the Quantum tomography problem
	by obtaining suitable Weyl operators $D(\sigma)$ with $\sigma\in X$ and
	extract a basis of orthonormal Hermitian i.e., self adjoint operators
	$\hat{W}^\sigma$ with $\sigma\in X$.Then they try to construct projection operators, of rank one, from them and point out difficulties in doing
	so. For $d=2^k$, they define
	projection operators $P(\lambda,i)$,
	$i=0,1,\ldots,d-1$; $\lambda$, any isotropic line in $X$. Here $(\lambda,i)=\{\sigma+(0,i):\sigma\in\lambda\}$ or $\{\sigma+(i,0):\sigma\in\lambda\}$
	depending on nature of $\lambda$.
	
	\subsubsection{}
	Explicit formulae are obtained in \cite{2} utilizing Albouy's streamlining indicated in \S2.3 above.
	At times, the description here is more transparent. For instance, the facts like the following give us more information about isotropic lines and,
	as a consequence $\mathbf{W}$-MASS's in this context.
	\begin{itemize}
		\item[(i)] Each isotropic line $\lambda$ is a subgroup of $\ZZ_d^2$. So for any $\sigma=(m,n)\in \lambda$, the subgroup $\lambda_\sigma$ generated
		by $\sigma$ is  a subset of $\lambda$. We express this fact by saying that $\lambda_\sigma$ is a `move together' for $\sigma$. More generally,
		any subset $Z$ of $\tilde{\bf W}$  that satisfies the condition that any
		$\tilde{{\bf W}}$-MASS containing $\sigma$ must contain $Z$  will be called a {\it move together} for $\sigma$.
		
		\item[(ii)] For $\sigma=(m,n)\neq (0,0)$, let $h$=HCF of $m$, $n$, $d$, considering $0$ as $d$.
		\begin{itemize}
			\item[(a)] If $h=1$, then $\lambda_\sigma$ is the unique isotropic line containing $\sigma$.\ For example, we may take isotropic lines given
			by $\lambda_\sigma$
			with $\sigma=(m,1)$, $m\in\ZZ_d$ and
			$(1,n)$, $n$, zero or a factor of $d$ other than $1$ and $d$. Alternatively we may consider $\sigma=(m,1)$, $m$ zero or a factor of $d$ other
			than $1$ and $d$ and $(1,n)$, $n\in\ZZ_d$. Of course, there are more such instances for a composite $d$, the simplest being $\sigma=(d_1,d_2)$
			with $d_1$ and $d_2$ being coprime factors of $d$.
			
			\item[(b)] If $h>1$, then $|\lambda_\sigma|=d/h$ and $\lambda_\sigma$ is a move together for $\sigma$ in different isotropic lines that contain $\sigma$.
			This gives different $\mathbf{W}$-MASS's overlapping, of course, in $\{U(m,n),(0,0)\neq (m,n)\in\lambda_\sigma\}$, at least.
		\end{itemize}
	\end{itemize}
	\subsubsection{}
	For $d=2^k, k\geq 2$, in case of overlapping $\lambda$'s and accordingly overlapping $\mathbf{W}$-MASS's, the condition
	of unbiasedness assumes a modified form. The expression (109) of \cite{2} can be written as:
	
	$\mathrm{tr}(P_{(\lambda,i)}P_{(\lambda',j)})=\dfrac{1}{d}$ [number of points common to the lines $(\lambda,i)$ and $(\lambda',j)$].
	
	\subsubsection{}
	A rough estimate for the number of probabilities from the said projection operators is more than the ideal number $d^2$,
	of course, except when $d=2$ or $3$. This brings us to a search of techniques to reduce the rank or the number of projections.
	\subsection{Tomographically complete sets of orthonormal bases and weakly mutually unbiased bases}
	Shalaby and Vourdas \cite{shvo1,shvo2} continued with the approach in 2.3 and 2.4 above. They emphasized that for tomographical
	purposes all isotropic lines are not needed. In fact, only full-size cyclic ones are enough for this task. In \cite{shvo1}, they determined their number to be the Dedekind function $\psi(d)=d \prod_{i\in I}(1+1/p_i)$,
	and identified them for the case when $d$ is a product
	of two odd primes, to define and obtain tomographically complete sets of bases.\
	This was pursued further by them \cite{shvo2} to define and study
	weak mutually unbiased bases. Here again, the number of probabilities from the said projections is
	more than the ideal number $d^2$, in general. However, our new techniques will reduce the number, which we proceed to describe.
	\section{Optimal Quantum Tomography}
	We begin with the notion of constrained elementary measurements. It is pivotal for the development of the results and applications in this Section,
	the heart of the paper.
	\subsection{Constrained elementary measurements} The concepts and techniques here draw upon tactics in Linear Algebra.
	\subsubsection{} We start with a definition.
	\def\P{\mathcal{P}}
	\begin{definition}[Constrained elementary measurement]\mbox{}\\
		Let $\mathbf{Q}=\{Q_t:1\le t\le \tau\}$ be a family of mutually orthogonal projections on a
		Hilbert space $\K$ of finite dimension $q$ that add to $I_{\K}$ such that $q_t= \text{rank} ~Q_t\geq 2$ for some $t$. Such a $\mathbf{Q}$ will be called a ${\it constraint}$.
		A family $\P=\{P_j:0\le j\le q-1\}$ of mutually orthogonal rank one projections
		adding to $I_{\K}$ will be said to be \emph{$\mathbf{Q}$-constrained} if
		$\{j:0\le j\le q-1\}$ can be decomposed as (a disjoint union of)
		$\{I_t:1\le t\le \tau\}$ with $\sum\limits_{j\in I_t}P_j=Q_t$, $1\le t\le \tau$. $\P$ will be called a \emph{$\mathbf{Q}$-constrained elementary
			measurement}.
		
		Clearly, $1\leq \tau <q$. The case $\tau=1$ will be called the {\it
			trivial constraint}.
		
		Now, let $\tau>1$, unless stated otherwise. We note that $\# I_t=q_t$, so we may label members of $I_t$ by $\{i_u^t, 1\leq u\leq q_t\}$, $1\leq t\leq\tau$
		and express the condition as in Figure 1.
	\end{definition}
	\begin{figure}[H]\renewcommand{\arraystretch}{2}
		\centering
		\begin{tabular}{c|c|c|c?c}
			$P_{i_1^1}+P_{i_2^1}+\ldots+ P_{i^1_{q_1}}$&&&&$Q_1$\\\hline
			&$P_{i_1^2}+P_{i_2^2}+\ldots +P_{i^2_{q_2}}$&&&$Q_2$\\\hline
			&&$\ddots$&&$\vdots$\\\hline
			&&&$P_{i_1^\tau}+P_{i_2^\tau}+\ldots+P_{i^\tau_{q_\tau}}$&$Q_\tau$\\\thickhline
			$P_{i_1^1}+P_{i_2^1}+\ldots+P_{i^1_{q_1}}$&$P_{i_1^2}+P_{i_2^2}+\ldots+P_{i^2_{q_2}}$&$\cdots$
			&$P_{i_1^\tau}+P_{i_2^\tau}+\ldots+P_{i^\tau_{q_\tau}}$&$I_{\mathcal{K}}$\\
		\end{tabular}
		\caption{}
	\end{figure}
	\subsubsection{The Constraint Technique}
	Let $g\geq 2$. Let $\{\P_v:1\le v\le g\}$ be $\mathbf{Q}$-constrained elementary measurements with $\P_v=\{P_j^v:0\le j\le q-1\}$
	and decomposition $\{I_t^v:1\le t\le \tau\}$,
	$1\le v\le g$. Let  $T=\{t:\text{rank}\ Q_t>1\}$
	and   $T'=\{t:1\le t\le \tau,\,t\notin T\}$.\ For $2\le v\le g$, for $1\leq t\leq \tau$, fix any $j_t^v\in I_t^v$ and set $J_t^v =I_t^v\setminus \{j_t^v\}$.\
	Then $J_t^v$ is non-empty if and only if it $\in T$. We note that the family $\{\P_v:1\le v\le g\}$ can be replaced by the smaller family
	$\P'=\P_1\cup\bigcup\limits_{2\le v\le g}\bigcup\limits_{t=1}^\tau \{P_j^v:j\in J_t^v\}$ for estimation purposes.
	
	To see this, we only have to note that missing
	$P_{j_t^v}^v$ is $Q_t-\sum\limits_{j\in J_t^v}P_j^v=\sum\limits_{j\in I^1_t}P_j^1-\sum\limits_{j\in J_t^v}P_j^v$.
	It is as if $\mathcal{P}_1$ is acting as an anchor. Clearly, any $\mathcal{P}_v$ can act as an anchor.
	Figure 2 illustrates the case $g=2$.
	
	\begin{figure}[H]\renewcommand{\arraystretch}{2}
		\centering
		\begin{tabular}{c|c|c|c?c}
			$\begin{array}{l}P^1_{i_1^{1,1}}+P^1_{i_2^{1,1}}+\ldots +P^1_{i_{q_1}^{1,1}}\\
				P^2_{i_1^{2,1}}+P^2_{i_2^{2,1}}+\ldots+P^2_{i_{q_1}^{2,1}}\end{array}$ &&&&$Q_1$\\\hline
			&$\begin{array}{l} P^1_{i_1^{1,2}}+P^1_{i_2^{1,2}}+\ldots+ P^1_{i_{q_2}^{1,2}}\\
				P^2_{i_1^{2,2}}+P^2_{i_2^{2,2}}+\ldots+ P^2_{i_{q_2}^{2,2}}\end{array}$ &&&$Q_2$\\\hline
			&&$\ddots$& & $\vdots$\\\hline
			&&&$\begin{array}{l} P^1_{i_1^{1,\tau}}+P^1_{i_2^{1,\tau}}+\ldots +P^1_{i_{q_\tau}^{1,\tau}}\\ P^2_{i_1^{2,\tau}}+P^2_{i_2^{2,\tau}}+\ldots+P^2_{i_{q_\tau}^{2,\tau}}\end{array} $ &$Q_\tau$\\\thickhline
			$\begin{array}{l} P^1_{i_1^{1,1}}+P^1_{i_2^{1,1}}+\ldots+ P^1_{i_{q_1}^{1,1}}\\ P^2_{i_1^{2,1}}+P^2_{i_2^{2,1}}+\ldots+ P^2_{i_{q_1}^{2,1}}\end{array} $&
			$\begin{array}{l} P^1_{i_1^{1,2}}+P^1_{i_2^{1,2}}+\ldots+ P^1_{i_{q_2}^{1,2}}\\ P^2_{i_1^{2,2}}+P^2_{i_2^{2,2}}+\ldots+ P^2_{i_{q_2}^{2,2}}\end{array} $
			&$\cdots$&$\begin{array}{l} P^1_{i_1^{1,\tau}}+P^1_{i_2^{1,\tau}}+\ldots+ P^1_{i_{q_\tau}^{1,\tau}} \\ P^2_{i_1^{2,\tau}}+P^2_{i_2^{2,\tau}}+\ldots+P^2_{i_{q_\tau}^{2,\tau}}\end{array} $&$I_{\mathcal{K}}$\\
		\end{tabular}
		\caption{}
	\end{figure}

	\begin{remark}\label{r3.1.3}
		\begin{enumerate}[(i)]
			\item For $1\le t\le \tau$, $I_t=\{j:P_j Q_t=P_j\}=\{j:P_jQ_t\neq 0\}$.
			
			\item We have  $\#\mathcal{P}'\leq q+(g-1)(q-\tau)=g(q-\tau)+\tau$. Right hand side is the number of reduced common labelling without worrying
			about distinct entries. Equality can occur in different situations, for instance, when $\mathcal{P}_v$'s are mutually disjoint; for a different
			situation, see numerical case of Example \ref{e3.1.4}.
			
			\item If for each $t$ with $1\le t\le \tau$, $q_t=q^\prime\ge 2$, then $q=q^\prime\tau$. We will say in this case that $\mathbf{Q}$
			is a {\it uniform constraint} of size $q^\prime$ .
			In this scenario $\#\mathcal{P}^\prime\leq q+(g-1)(q-\tau)= g(q-\tau)+\tau=(g(q^\prime-1)+1)\tau$.
			\item The trivial constraint is a particular case of (iii) above with
			$q=q^\prime$ and, therefore, $\#\mathcal{P}^\prime \leq g(q-1)+1$. And for
			the case of the existence of a complete system of $(q+1)$ MUB's simply by substituting $g=q+1$, we can recover the well known fact
			that $\#\mathcal{P}^\prime$ is the ideal number $q^2$ as mentioned towards the end of the first paragraph in \S1.
		\end{enumerate}
	\end{remark}
	\begin{example}\label{e3.1.4}
		For a normal operator $T$ on $\mathcal{K}$ to itself, let $\sigma(T)$ be its spectrum i.e., the set of eigenvalues of $T$.
		For $\lambda\in \sigma(T)$, let $E_\lambda^T =\{x\in \mathcal{K}:Tx=\lambda x\}$ and
		$P_\lambda^T$ the projection on $E_\lambda^T$. Finally, let ${\bf Q}^T$
		= $\{Q_\lambda^T=P_\lambda^T:\lambda\in \sigma(T)\}$ to be called the spectral projection set or the set of eigenprojections for $T$.
		Then $\# {\bf Q}^T=\#\sigma(T)$.
		
		Let $A, B, C$ be normal operators on $\mathcal{K}$ to itself such that $AB\neq BA$,
		$AC=CA$, $BC=CB$. Then there exist  common orthonormal systems of eigenvectors $\{\xi_j:0\leq j\leq q-1\}$ and $\{\eta_j:0\leq j\leq q-1\}$
		for $A,C$ and for $B,C$ respectively. Let $\mathcal{P}_1=\{P_j=
		|\xi_j\rangle\langle\xi_j|:0\leq j\leq q-1\}$ and
		$\mathcal{P}_2=\{R_j=
		|\eta_j\rangle\langle\eta_j|:0\leq j\leq q-1\}$. Then ${\bf Q}^C$ is a constraint and $\mathcal{P}_1$ and $\mathcal{P}_2$ are ${\bf Q}^C$-constrained.
		To see this, we first note that if $C$ has only simple eigenvalues then $\mathcal{P}_1=\mathcal{P}_2$ and therefore $AB=BA$,
		which is not so. So $C$ has at least one multiple eigenvalue, and, therefore, $\text{dim} E_\lambda^C\geq 2$ for some $\lambda\in \sigma(C)$.
		So ${\bf Q}^C$ is a constraint. Now $C=\sum_{j=0}^{q-1}\lambda_jP_j$
		for $\lambda_j$'s in $\sigma(C)$. So, for $\lambda\in\sigma(C)$,
		$ Q_\lambda^C=\sum_{\lambda_j=\lambda}P_j$. Therefore $\mathcal{P}_1$ is ${\bf Q}^C$-constrained. Similarly for $\mathcal{P}_2$. To give an idea,
		let $q=4, \mathcal{K}=\mathbb{C}^4$, and
		\begin{align}
			A=\left(\begin{array}{cccc}1&0&0&0\\0&-1&0&0\\0&0&1&0\\0&0&0&1\end{array}\right),
			~B=\left(\begin{array}{cccc}1&0&0&0\\0&\frac{1}{\sqrt{2}}&- \frac{1}{\sqrt{2}}&0\\0&\frac{1}{\sqrt{2}}&\frac{1}{\sqrt{2}}&0\\0&0&0&1\end{array}\right),
			~C=\left(\begin{array}{cccc}1&0&0&0\\0&-1&0&0\\0&0&-1&0\\0&0&0&i\end{array}\right)
		\end{align}
	\end{example}
	\begin{theorem}\label{t3.1.5} Let $\left({\bf Q}^{(s)}\right)_{s=1}^{\nu}$ be a $\nu$-tuple of distinct non-trivial constraints
		${\bf Q}^{(s)}=\{ Q_t^s: 1\leq t\leq \tau_s\}$. For $1\leq s\leq \nu$, let $\mathcal{P}^{(s)} = \{ \mathcal{P}_v^{s}:
		1\leq v\leq g_s\}$ be a ${\bf Q}^{(s)}$-constrained family of elementary measurements $\mathcal{P}_v^s =\{ P_{s,v,j} :
		0\leq j\leq q-1\}$ with $g_s\geq 2$ for $1\leq s \leq \nu$, then for estimation purposes the family
		$\{\mathcal{P}^{(s)}:1\leq s\leq \nu\}$, or for that matter, the family $\Pi =\{P_{s,v,j} : 1\leq s\leq\nu,
		1\leq v\leq g_s, 0\leq j\leq q-1\}$ can be replaced by a  smaller family $\Pi^\prime$ with
		$\#$ $\Pi^\prime \leq \left(\sum_{s=1}^{\nu}g_s\right)q-\sum_{s=1}^{\nu}(g_s-1)\tau_s-(\nu-1)$.
	\end{theorem}
	
	\begin{proof} The case $\nu =1$ has already been given in 3.1.2, The Constraint Technique. Now we consider the case
		$\nu\geq 2$. Basically it is a repeated application of the Costrained Technique  to each $\mathcal{P}^{(s)}$ with a little care.
		
		Take any $s$ with $1\leq s \leq \nu$ and fix it temporarily. There is a $t_s$ with $1\leq t_s \leq \tau_s$ such that
		$q_{t_s}^s = \text{dim} Q_{t_s}^s\geq 2$. For $1\leq v\leq g_s$, by Remark \ref{r3.1.3} (i), $I_{s,v, t_s}=\{ j: P_{s,v,j}Q_{t_s}^s
		= P_{s,v,j}\}$ has at least two distinct members. Fix any such $j_s^\prime$ for $v=1$. Then, for $2\leq v\leq g_s$, there is a $j_{s,v}$ available
		in $I_{s,v,t_s}$ with $P_{s,v,j_{s,v}}\neq P_{s,1,j_s^\prime}$. We now apply the Constraint Technique to $\Pi_s
		=\{P_{s,v,j}: 1\leq v\leq g_s, 0\leq j\leq q-1\}$ to replace it, for estimation purposes, by $\Pi_s^\prime$ with extra care to use
		$j_{s,v}$ to be removed from $I_{s,v,t_s}$. Then $\# \Pi_s^\prime \leq  g_s q-(g_s-1)\tau_s$.
		Now $\{\mathcal{P}_1^s ; 1\leq s\leq \nu\}$ has trivial constraint $\{I_\mathcal{K}\}$. So we may replace it by $\mathcal{P}_1^1\cup_{s=2}^{\nu}
		(\mathcal{P}_1^s\smallsetminus\{P_{s,1,j_s^\prime}\})$ for estimation purposes. So pooling together, we may replace, for $2\leq s\leq \nu$,
		for estimation purposes, $\Pi_s^\prime$ ( and, therefore, $\Pi_s$) by $\Pi_s^{\prime\prime}= \Pi_s^\prime\smallsetminus\{P_{s,1,j_s^\prime}\}$.
		We take $\Pi_1^{\prime\prime}=\Pi_1^\prime$.  We now set $\Pi^\prime =\cup_{s=1}^{\nu} \Pi_s^{\prime\prime}$. Then $\Pi^\prime$
		works fine and $\# \Pi^\prime \leq \sum_{s=1}^{\nu} g_sq-\sum_{s=1}^{\nu}(g_s-1)\tau_s-(\nu-1)$.
	\end{proof}
	\subsubsection{The Simultaneous Constraints Technique}
	The method in 3.1.2. can be repeated if we have simultaneous constraints   for individual elementary measurements and if we can consolidate the reductions induced. Such consolidations are
	effective if and only if constraints are independent in a certain sense. It will be practical to fix anchors for all constraints and then
	deal with each individual elementary measurement for the remaining lot.
	We point out that, within anchors, the trivial constraint $I$ works as a constraint. This together with good choice of common anchors is illustrated in examples and theorems below and in later subsections after some preparation mainly in \S 3.2 and  \S 3.3.
	
	\begin{example}
		This is similar to Example \ref{e3.1.4} above and follows the notation and terminology there. We may take any unitary operator $U$ on $\mathcal{K}=\mathbb{C}^6$
		with distinct eigenvalues $\lambda_k=e^{\left(\frac{2\pi i k}{6}\right)}=e^{\left(\frac{\pi i k}{3}\right)}$ with a unit
		eigenvector $\xi_k$ for $k=0,1,2,3,4,5$  . Setting $P_k=|\xi_k\rangle\langle\xi_k|$ we have $U=\sum_{k=0}^{5}\lambda_kP_k$. We may take
		$U=\text{Diag}(1,e^{\left(\frac{\pi i }{3}\right)}, e^{\left(\frac{2\pi i }{3}\right)},-1, e^{\left(\frac{4\pi i }{3}\right)},
		e^{\left(\frac{5\pi i }{3}\right)})$, if we like. Next we take $V$ and $W$ as $U^2$ and $U^3$ ( or, for that matter, their multiples
		by scalars $\lambda$ and $\mu$ of modulus one). Then ${\bf Q}^V=\{P_0+P_3, P_1+P_4, P_2+P_5\}$ and
		${\bf Q}^W=\{P_0+P_2+P_4, P_1+P_3 +P_5\}$. So $\mathcal{P}_U=\{P_j, 0\leq j\leq 5\}$ has ${\bf Q}^V$ and ${\bf Q}^W$ as simultaneous
		constraints. We shall see more on this in \S 3.5 and \S 3.6.
	\end{example}
	\begin{remark}
		One can formulate Reduction Theorem like Theorem \ref{t3.1.5} for simultaneous constraints, but it will involve too much of notation. Instead we will
		illustrate special neat cases in \S 3.5 and \S 3.6.
	\end{remark}
	\subsection{Fan representation for $\mathbf{W}$-MASS's}
	We now turn to minimal covering systems of ${\bf W}$-MASS's that cover a given unitary system ${\bf W}$. They can be pictorially  represented as a fan,
	for certain examples. This helps us in implementing  the optimal constraint techniques discussed above.
	\begin{theorem}\label{thm3.2.1}
		Let $\mathbf{W}$ be a unitary system.
		\begin{itemize}
			\item[(i)] There is a unique maximal family
			$\mathcal{V}_{\mathbf{W}}=\{\mathbf{V}_{\alpha}: \alpha \in \Lambda \}$ of
			$\mathbf{W}$-MASS's, such that $\mathbf{W}=\underset{\alpha \in \Lambda}{\cup}
			\mathbf{V}_{\alpha}.$\\
			\item[(ii)] If a ${W}$ in  $\mathbf{W}$ has simple eigenvalues, then it belongs
			to unique ${\bf V}_\alpha$. In particular, if
			each $W$ in $\mathbf{W}$ has simple eigenvalues,
			then $\mathbf{V}_{\alpha}$'s are mutually disjoint.
		\end{itemize}
	\end{theorem}
	\begin{proof}
		\begin{itemize}
			\item[(i)] Let $\mathbf{W} = \{W_y: y \in Y \}.$ For
			any $y \in Y,$ $\{W_y\}$
			is an AUS $\subset \mathbf{W}.$ Let $\mathcal{W} = \{\mathbf{A} : \mathbf{A}
			\subset \mathbf{W} \,\mbox{and} \, \mathbf{A} \, \mbox{is an AUS} \} $ and order
			it by inclusion. Then $\mathcal{V}_{\mathbf{W}}$ is made up of maximal elements
			of $\mathcal{W}.$
			\item[(ii)] It follows from elementary Linear Algebra
			because a set of normal operators mutually commute if and only if they have a common
			orthonormal system of eigenvectors, and a normal operator with simple eigenvalues has a
			unique such system (to within multiplies by scalars of unit modulus).
			The rest follows immediately from this.
		\end{itemize}
	\end{proof}
	\begin{remark}\label{r3.2.2}
		\begin{itemize}
			\item [(i)] The role of Hadamard matrices has already been indicated in \S2. To elaborate a bit, for each
			$\alpha \in \Lambda,$ there is a unitary $U_{\alpha}$ and a (partial) Hadamard
			$s_{\alpha} \times d$ matrix $H_{\alpha}$ with $s_{\alpha} =
			\,\,\mbox{size}\,\,\mathbf{V}_{\alpha}$ such that $\mathbf{V}_{\alpha}$ consists
			of operators of the type $U_{\alpha} D U_{\alpha}^{\ast},$ $D$ is a diagonal
			matrix whose diagonal forms a row of $H_{\alpha}.$ The ordering of rows
			corresponds to that of operators in $\mathbf{V}_{\alpha}.$ To within that
			$H_{\alpha}$ is unique up to a permutation of columns, and the corresponding
			$U_{\alpha}$'s will undergo changes accordingly. For each $\alpha \in \Lambda,$
			the augmented matrix $\widetilde{H}_{\alpha}$ formed by adding a top row of all
			$1$'s is also a Hadamard matrix and it arises from the
			$\widetilde{\mathbf{W}}$-MASS $\widetilde{\mathbf{V}}_{\alpha} =
			\mathbf{V}_{\alpha} \cup \{I\}$ with same $U_{\alpha}$ in force.
			\item[(ii)] If each $W_y$ in $\mathbf{W}$
			has simple eigenvalues, then $\mathbf{V}_{\alpha}$'s are
			mutually disjoint. This happens for the case $d=2$ and $d=3$ but may not be so
			for larger $d$ because the requirement is that eigenvalues of $W_y$ lie on the
			unit circle and (counted with multiplicities) add to zero.
			Towards the end of this section, different Remarks and Theorems for $d>3$
			give concrete situations. It is as if there is a fan
			of these subsets $\mathbf{V}_{\alpha}$'s (possibly overlapping) hinged at $I$
			and, accordingly, {\it a
				fan of abelian subspaces of} $\mathcal{B}(\mathcal{H})$ (possibly overlapping),
			hinged at the linear span of $I.$
			\item[(iii)] A $\mathbf{W}$-MASS of size $d-1$ together with $I$ generates a
			maximal abelian subalgebra of $M_d,$ in short, a MASA in $M_d.$ Theory of
			orthogonal MASA is well developed by \cite{Ha}, \cite{Wei}, \cite{Ch}. In fact,
			\cite{Ch} even defines an entropy $h(A/B)$ between a pair $(A,B)$ of MASA's and
			proves that $A,B$ are orthogonal if and only if $h(A/B)$ takes the maximum
			value, and then the value is $\log \,d.$
		\end{itemize}
	\end{remark}
	\begin{definition}[Fan representation]
		In view of above, we call $\mathcal{V}_{\mathbf{W}}$
		in Theorem \ref{thm3.2.1}, the {\it fan representation of} $\mathbf{W}.$
	\end{definition}
	\begin{remark}
		\begin{itemize}
			\item[(i)] One can figure out the $\mathbf{W}$-MASS fan representation through a common eigenvector system approach. It is neat when eigenvalues of each $W_x$ in $\mathbf{W}$ are simple and becomes quite involved when some of them are multiple.
			\item[(ii)] In fact, our interest is in a suitable, possibly smallest set, say ${\Lambda^\prime}$ such
			that ${\bf W}=\cup \{ {\bf V}_\alpha : \alpha\in {\Lambda^\prime}\}$.
		\end{itemize}
	\end{remark}
	\begin{example}\label{e3.2.5}
		In the context of \S 2.3 and \S 2.4, interesting pictorial representation for different $\lambda$'s, and thus of $\tilde{\mathbf{W}}$-MASS's have been
		given in \cite{2}, \S 4. We give the same information a bit differently, looking like fans.
		To facilitate a neat picture, the lattice for $d=4$ has been drawn, ordering the elements of $\ZZ_4$ as $0,2,1,3$ in
		Figure~3 below. The figure has overlapping subgroups closer to $(0,0)$ like a hinge and generating elements for full size cyclic ones fanning out.
		\vspace{-0.5cm}
		\begin{figure}[H]
			\centering
			\includegraphics[width=2.5in]{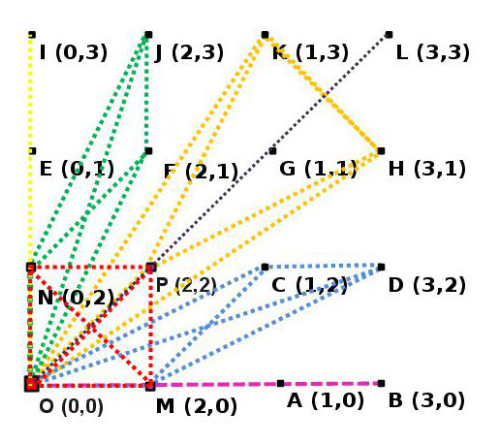}
			\caption{}
		\end{figure}
	\vspace{-0.5cm}
		And for similar reasons of clarity for $d=6$, move-together pairs like
		$\{(1,1),(5,5)\}$, $\{(1,4),(5,2)\}$, $\ldots$ have been drawn as single points assigned on a part of a circle, so as to say, in
		Figure~4. The figure has three overlapping subgroups of order 2 viz., those generated by $(3,0),(0,3)$ and $(3,3)$, and, four overlapping subgroups
		of order 3 generated by even number pairs like $\{(2,0),(4,0)\}$ closer to $(0,0)$ like a hinge and generating element pairs
		for the full size cyclic ones fanning out. We enlist the two collections for proper illustration.
		\begin{figure}[H]
			\centering
			\includegraphics[clip, trim=1.5cm 0.0cm 0.0cm 13.5cm, width=0.6\textwidth]{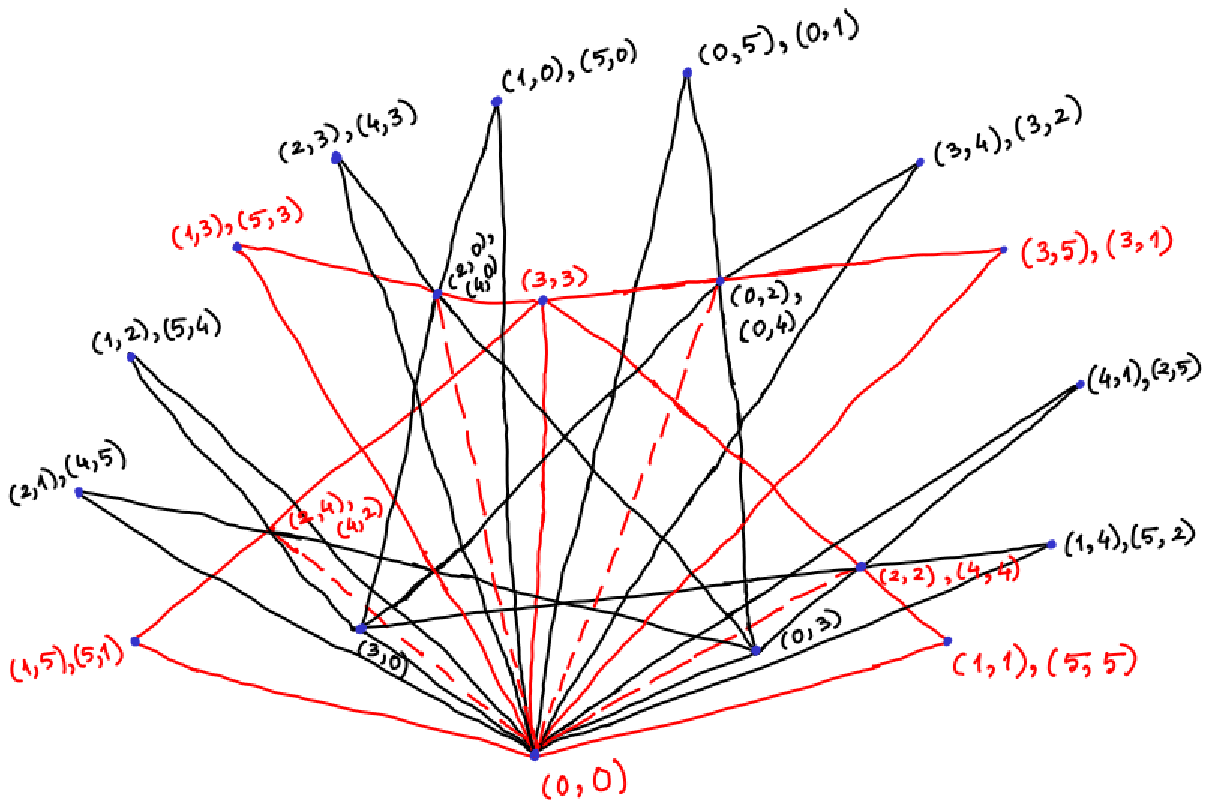}
			\caption{}
		\end{figure}
		\begin{itemize}
			\item[(a)] $\mathbf{d=4}.$ The seven $\mathbf{W}$-Mass's are
			\begin{align}
				{\bf V}_1&=  \{(1,0), (2,0), (3,0)\}, \nonumber \\
				{\bf V}_2&= \{(0,1), (0.2), (0,3) \}, \nonumber\\
				{\bf V}_3&= \{(1,1), (2,2), (3,3)\}, \nonumber\\
				{\bf V}_4&= \{(2,1), (0,2),(2,3)\}, \nonumber\\
				{\bf V}_5&=  \{(1,2),(2,0),(3,2) \},\nonumber \\
				{\bf V}_6&= \{(3,1), (2,2), (1,3)\}, \nonumber\\
				\text{and finally},~{\bf V}_7&= \{(2,0),(0,2),(2,2)\}.
			\end{align}
			\item[(b)] $\mathbf{d=6}.\,\,$ The
			twelve $\mathbf{W}$-MASS's are:
			\begin{align}
				{\bf V}_1&=\{(1,0),(2,0),(3,0),(4,0),(5,0)\},\nonumber\\
				{\bf V}_2&=\{(0,1),(0,2),(0,3),(0,4),(0,5)\},\nonumber\\
				{\bf V}_3&=\{(1,1),(2,2),(3,3),(4,4),(5,5)\},\nonumber\\
				{\bf V}_{4}&=\{(1,2),(2,4),(3,0),(4,2),(5,4)\},\nonumber\\
				{\bf V}_5&=\{(2,1),(4,2),(0,3),(2,4),(4,5)\},\nonumber\\
				{\bf V}_6&=  \{(1,3),(2,0),(3,3),(4,0),(5,3)\},\nonumber\\
				{\bf V}_7&=\{(3,1),(0,2),(3,3),(0,4),(3,5)\},\nonumber\\
				{\bf V}_{8}&=\{(2,3),(4,0),(0,3),(2,0),(4,3)\},\nonumber\\
				{\bf V}_{9}&=\{(3,2),(0,4),(3,0),(0,2),(3,4)\},\nonumber\\
				{\bf V}_{10}&=\{(5,2),(4,4),(3,0),(2,2),(1,4)\},\nonumber\\
				{\bf V}_{11}&=\{(4,1),(2,2),(0,3),(4,4),(2,5)\},\nonumber\\
				{\bf V}_{12}&=\{(5,1),(4,2),(3,3),(2,4),(1,5)\}.
			\end{align}
		\end{itemize}
	\end{example}
	\begin{example}\label{e3.2.6}[Pauli matrices technique] This works for $d,$ a
		power of $2$ and is based on \cite{5}, \cite{6}, \cite{7}, \cite{8}, Sych
		and Leuchs \cite{dsgl} and \cite{psmw}.\ It could come also
		under ``Shift and Multiply''-type by considering the group
		$\mathbb{Z}_2 \times \mathbb{Z}_2$ and real Hadamard matrix. But it is
		interesting to display the use of Pauli matrices as done in papers
		cited above, particularly, \cite{dsgl} and \cite{psmw}.
		\begin{itemize}
			\item[(i)] It follows immediately that there are 15 $\mathbf{W}$-MASS's
			of seven types as below:
			\begin{center}
				\begin{align}
					\mathcal{C}_X &= \left \{X \otimes I, I \otimes X, X \otimes X \right \}, \nonumber\\
					\mathcal{C}_Y &= \left \{Y \otimes I, I \otimes Y, Y \otimes Y \right \}, \nonumber\\
					\mathcal{C}_Z &= \left \{Z \otimes I, I \otimes Z, Z \otimes Z \right \}; \nonumber\\
					\mathcal{D}_X &= \left \{X \otimes X, Y \otimes Z, Z \otimes Y \right \}, \nonumber\\
					\mathcal{D}_Y &= \left \{Y \otimes Y, Z \otimes X, X \otimes Z \right \}, \nonumber\\
					\mathcal{D}_Z &= \left \{Z \otimes Z, X \otimes Y, Y \otimes X \right \}; \nonumber\\
					\mathcal{C}_{YZ} &= \left \{Y \otimes I, I \otimes Z, Y \otimes Z \right\}, \nonumber\\
					\mathcal{C}_{ZX} &= \left \{Z \otimes I, I \otimes X, Z \otimes X \right\}, \nonumber\\
					\mathcal{C}_{XY} &= \left \{X \otimes I, I \otimes Y, X \otimes Y \right\}; \nonumber\\
					\mathcal{D}_{YZ} &= \left \{I \otimes Y, Z \otimes I, Z \otimes
					Y \right\}, \nonumber\\
					\mathcal{D}_{ZX} &= \left \{I \otimes Z, X\otimes I, X \otimes
					Z \right\}, \nonumber\\
					\mathcal{D}_{XY} &= \left \{I \otimes X, Y\otimes I, Y \otimes
					X \right\}; \nonumber\\
					\text{together with}\quad\mathcal{E} &= \left \{X \otimes X, Y \otimes Y,
					Z \otimes Z \right\}, \nonumber\\
					\mathcal{F} &= \left \{X \otimes Y, Y \otimes Z, Z \otimes X \right\},\nonumber\\
					\mbox{and,}\quad\mathcal{G} &= \left \{Y \otimes X, Z \otimes Y, X \otimes
					Z \right\}.
				\end{align}
			\end{center}
			\item[(ii)] Each unitary in $\mathbf{W}$ occurs in exactly three of them. Figure 5
			gives an idea.
			\begin{figure}[H]
				\centering
				\includegraphics[width=3.75in]{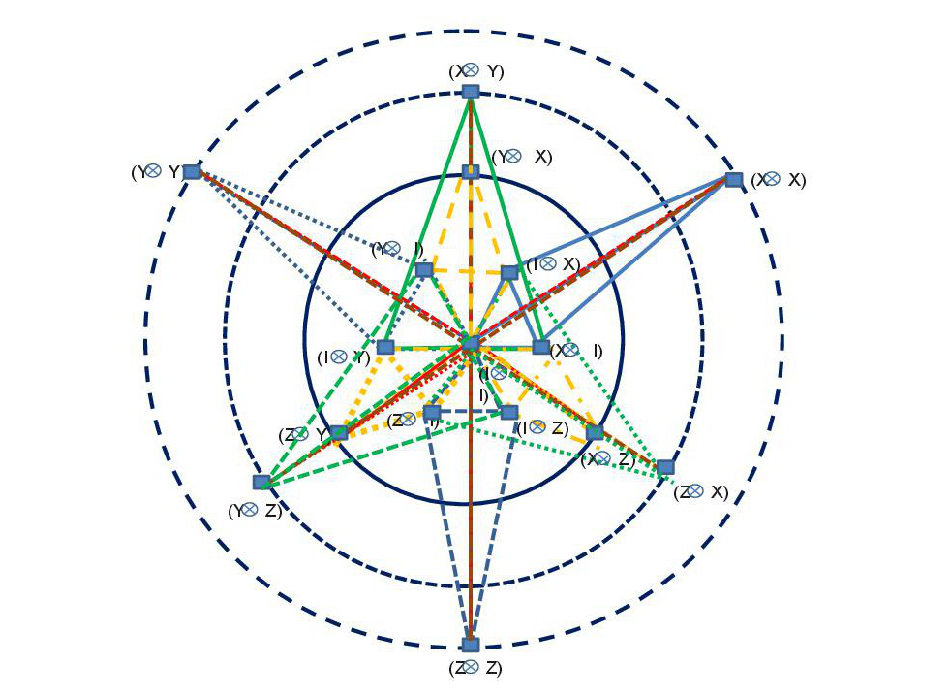}
				\caption{}
			\end{figure}
			\item[(iii)] The figure facilitates
			a pictorial understanding of the contents of \cite{psmw} as well. We elaborate in words
			as follows.
			\begin{itemize}
				\item[(a)] Three navy blue circles put together overlap with the remaining subsystems.
				\item[(b)] Three maroon lines put together overlap with the remaining subsystems.
				\item[(c)] Three yellow quadrilaterals put together can be combined with the middle and the outer circle.
				But, they overlap with green and sky-blue on the hexagon and also with maroon and navy-blue on the inner circle.
				\item[(d)] Three sky-blue quadrilaterals put together can be combined with the middle and the inner circle.
				But, they overlap with green and yellow on the hexagon and also with maroon and navy-blue on the outer circle.
				\item[(e)] Three green triangles put together can be combined with the inner and the outer circle.
				But, they overlap with yellow and sky-blue on the hexagon and also with maroon and navy-blue on the middle circle.
			\end{itemize}
		\end{itemize}
		The first sentence in each part of (c), (d), (e) gives an MUB.
	\end{example}
	
	\subsection{Positive operator-valued measures}
	
	Theoretically speaking, if $\mathbf{U}$ is a unitary basis then the tuple
	$({\rm tr} \,\, (\rho U_x))_{x \in X}$ determines the state $\rho.$ The same is true
	if we take any unitary system $\mathbf{W} = \{W_y : y \in Y\}$ of size $(d^2-1)$ and consider the
	tuple $({\rm tr} \,\,(\rho W_y))_{y \in Y}$ as a representative of $\rho.$
	But usually such measurements are performed with respect to a set of projections or positive
	operators, the so-called, positive-operator-valued measure (POVM) viz., a set $\mathbf{A} = \{A_j : 1 \leq j
	\leq v\}$ of positive operators on $\mathcal{H}$ with $\sum\limits_{j=1}^{\textit{v}}
	A_j = I$.  For computational convenience, these $A_j$'s are taken to be suitable positive multiples of rank one projections except possibly for one, because we may
	take the remaining one to make the sum equal to $I$.  Such POVM's are called pure POVM's.
	Alternatively, we may consider an informationally complete system of rank one projections,
	say $\{P_j, 1\leq j\leq \nu\}$, such that
	$\left(\text{tr}\rho P_j)\right)_{j=1}^\nu$ determine the state $\rho$. And the aim will remain to have $\nu$ the ideal number $d^2$,
	or, an optimal number by a specific method. Whenever possible, the method should also ensure that the number of
	quantum mechanical overlaps, $\text{tr}(P_jP_k), 1\leq j\neq k\leq \nu$ is small as well.
	\subsubsection{}
	We draw upon \cite{7}, particularly excerpts given in Remark \ref{r3.1.3}(iv).
	\begin{itemize}
		\item[(i)] If $\mathbf{W}$ can be partitioned as union of $(d+1)$
		$\mathbf{W}$-MASS's, say $\{\mathbf{V}_s : 0 \leq s \leq d\}$ of size
		$(d-1),$ then the complete system of $(d+1)$ orthonormal bases, say, $\left \{
		\{\mathbf{b}_t^s : 0 \leq t \leq d-1\} , 0 \leq s \leq d \right \},$ of
		$\mathbb{C}^d$ obtained via [3], Theorem 3.2 (viz., $\{\mathbf{b}_t^s : 0 \leq
		t \leq d-1 \}$ is a common orthonormal eigenbasis for $\mathbf{\mathbf{V}_s},$
		$0 \leq s \leq d$) gives rise to $(d^2-1)$ pure states
		$\{\rho_j : 1 \leq j \leq d^2-1\}$ with $\rho_{s(d-1) +t}$ determined by the
		unit vector $\mathbf{b}_t^s,$ $1 \leq t \leq d-1,$ $0 \leq s \leq d.$ Thus we
		obtain a pure POVM $\{ A_j = \frac{1}{(d+1)} \rho_j, \,\,0 \leq j \leq d^2-1
		\},$ where $\rho_0 = (d+1) I - \sum\limits_{j=1}^{d{^2}-1} \rho_j.$ See also Remark \ref{r3.1.3}(iv).
		\item[(ii)] As noted in Theorem \ref{thm3.2.1} (ii) above, if each $W_y$ in $\mathbf{W}$ has
		simple eigenvalues, then $\mathbf{V}_{\alpha}$'s constituting the fan
		representation of $\mathbf{W}$ are mutually disjoint. This is a situation
		similar to (i) above except that there is no guarantee that the sizes of
		$\mathbf{V}_{\alpha}$'s are all $(d-1)$ or, equivalently, that of $\Lambda$ is
		$(d+1).$ We do not yet have an example for that. In any case, the technique
		indicated in (i) above does give a pure POVM of size $(d-1) \,\#\Lambda + 1.$ Rewording, the Constrained Technique 3.1.2
		for the trivial constraint gives an informationally  complete system of $(d-1)\#\Lambda +1$ rank one projections.
		\item[(iii)] In general, for a composite $d$ which is not a prime power, or, even when
		$d$ is a prime power, a given $\mathbf{W}$ may not be decomposable as in (i)
		above. This can be seen in
		Figures 3 and 4 for Example \ref{e3.2.5} and Figure 5 for Example \ref{e3.2.6}.
		Figure 4 makes it clear that the whole fan is needed to cover $\mathbf{W}$
		and in Figure 3, only the red part can be ignored to obtain a smaller subset of the
		fan to cover $\mathbf{W}$. Once again, Theorem \ref{thm3.2.1}, Remark \ref{r3.2.2}(ii) and Example \ref{e3.1.4} tell us that
		overlapping $W_y$'s have to possess some multiple eigenvalues and, therefore,
		their spectral projection sets are all constraints. In the rest of this
		section we make an attempt to obtain pure POVM's or informationally complete systems of rank one projections of optimal size for such
		situations.
	\end{itemize}
	\subsubsection{}
	What comes in handy for our purpose is a minimal subset, say,
	$\mathcal{M}_{\mathbf{W}},$ of $\mathcal{V}_{\mathbf{W}}$ satisfying
	$\mathbf{W}= \cup \{ \mathbf{V}
	: \mathbf{V} \in \mathcal{M}_{\mathbf{W}}\}.$ We may write
	$\mathcal{M}_{\mathbf{W}} =
	\{\mathbf{V}_{\alpha} : \alpha \in \Lambda_1\}$ with $\Lambda_1 \subset
	\Lambda,$ if we like.
	\begin{itemize}
		\item[(i)]
	\S3.3.1 above, particularly, part (iii) throws some light on it together with its size and
		various subsections of \S2 tell the general story
		for some special $\mathbf{W}$. The smallest general one is, of course, as given
		in \S2.2.4. In the context of Example taken up in \S2.3 to \S2.5, the next one is as in \S2.5.
		We emphasize that the $\mathbf{W}$-MASS's coming from cyclic subgroups only are needed for the corresponding $\mathcal{M}_{\mathbf{W}}$.
		\item[(ii)]  A crude way to obtain a pure POVM would be to consider a common
		orthonormal eigenbasis $\{{\bf b}_t^\alpha: 0\leq t\leq d-1\}$ for ${\bf V}_\alpha$ with $\alpha\in \Lambda_1$  and construct
		a pure POVM as in (i) above, say ${\bf A}=\{A_j:0\leq j\leq (d-1)\#\Lambda_1\}$.  We can
		refine this construction to obtain a pure POVM of smaller size.
	\end{itemize}
	\subsection{Utilization of the techniques in \S 3.1 to obtain an optimal measurement set}
	As noted in 3.3.1 (iii), overlapping $W_y$'s
	have to possess some multiple eigenvalues.\ The projections on the corresponding
	eigenspaces can be taken to be $Q_t$'s
	giving rise to a constraint $\mathbf{Q}$.\
	In the rest of this section
	we make an attempt to obtain pure POVM's of optimal size for such situations
	using our technique outlined in \S 3.1.
	
	\subsubsection{ Reorganization of elementary measurements for ${\bf W}$- MASS's}
	Let $\mathcal{M}_{\bf W}=\{{\bf V}_\alpha :\alpha\in \Lambda_1\}$ be a minimal system as in 3.3.2 above.
	\begin{itemize}
		\item[(i)] Let $Y_1=\{y\in Y : {\bf W}_y\in~\text{a unique}~{\bf V}_\alpha~\text{with}~\alpha\in\Lambda_1\}$ and $Y_2=Y\setminus Y_1$=
		$\{ y\in Y: {\bf W}_y\in {\bf V}_\alpha \bigcap {\bf V}_{\alpha^\prime}~\text{ for
			some}~\alpha\neq \alpha^\prime~\text{in}~\Lambda_1\}$. The case $Y_2 = \emptyset$ has been considered in 3.3.1 (ii).
		We consider the case $Y_2\neq \emptyset$.
		Then each $y\in Y_2$ determines a costraint, say, ${\bf Q}^{{\bf W}_y}$, viz., the set of eigenprojections of ${\bf W}_y$ with distinct eigenvalues.
		Different $y$'s in $Y_2$ may determine the same constraint as we will see in \S3.5 and \S3.6; for all such $y$'s, the corresponding $W_y$'s commute.
		Let $\Lambda_2$=
		$\{ \alpha\in \Lambda_1: {\bf V}_\alpha~\text{contains a}~{\bf W}_y~\text{for some}~y\in Y_2\}$,
		$\Lambda_3$= $\{ \alpha\in \Lambda_1: {\bf V}_\alpha~\text{consists of}~{\bf W}_y\text{'s}~\text{with}~y\text{'s}\in Y_1\}
		=\Lambda_1\setminus \Lambda_2$.
		
		\item[(ii)] We note that for Example \ref{e3.2.5} (a) above $\#\Lambda_1=6$ by taking the first six ${\bf W}$-MASS's, $\#Y=15$, $\#Y_1=12$, $\#Y_2=3, \Lambda_3=
		\emptyset$ (see Figure 3). But, for Example \ref{e3.2.5}(b), $\Lambda= \Lambda_1$
		with $\#\Lambda =12, \#Y=35$, $\#Y_1=24$, $\#Y_2=11, \Lambda_3=\emptyset$ (see Figure 4). On the other hand, for Example 4.4 (v)(e), $\#Y=35$,
		$\#\Lambda_1=22$, $\#Y_1=34$, $\#Y_2=1$, $\#\Lambda_2=4$,
		$\#\Lambda_3=18$ (see Figure 17 in \S4).
		
		\item[(iii)] For $\alpha\in \Lambda_1$, there is a complete common unit eigenvector system
		for $W_y$'s in ${\bf V}_\alpha$ and thus the set $\mathcal{P}_\alpha$ of corresponding rank one projections, say, $P_{\alpha,j}, 0\leq j\leq q-1$.
		
		\item[(iv)] We now collect distinct constraints ${\bf Q}^{{\bf W}_y}$ with $y$ in  $Y_2$ , say
		$\{{\bf Q}^{(s)} : 1\leq s\leq \nu\}$.
		Suppose $\alpha\in \Lambda_2$. Then there is a $y\in Y_2$ with ${\bf W}_y$ belonging to ${\bf V}_\alpha$. Then $\mathcal{P}_\alpha$ is
		${\bf Q}^{{\bf W}_y}$-constrained.
		Also ${\bf Q}^{{\bf W}_y} = Q^{(s)}$ for some $1\leq s\leq \nu$. If ${\bf Q}^{ {\bf W}_z}= {\bf Q}^{(s)}$ for some $z\in Y_2$, then, in view of (iii) above, $W_z$
		commutes with each $W_t$ in ${\bf V}_\alpha$. So by maximality of ${\bf V}_\alpha$, we have that ${\bf W}_z\in {\bf V}_\alpha$. Therefore, we may talk in
		terms of constraints $Q^{(s)}, 1\leq s\leq \nu$ only.
		
		\item[(v)] For $1\leq s\leq \nu$, let $\mathcal{P}^{(s)}=\{\mathcal{P}_\beta: \mathcal{P}_\beta~\text{is}~Q^{(s)}~
		\text{constrained}\}$ = $\{\mathcal{P}_v^s:1\leq v \leq g_s\}$, say. Then
		$\cup_{1\leq s\leq \nu}\mathcal{P}^{(s)}=\{\mathcal{P}_\alpha: \alpha \in \Lambda_2\}$. Hence $\{\mathcal{P}_\alpha:\alpha \in \Lambda_1\}$=
		$\{\mathcal{P}_v^s: 1\leq s\leq \nu, 1\leq v\leq g_s\}\cup \{\mathcal{P}_\alpha :
		\alpha\in\Lambda_3\}$ and for $1\leq s\leq \nu$, $\mathcal{P}_v^s$ is
		${\bf Q}^{(s)}$-constrained for $1\leq v\leq g_s$.
		
		\item[(vi)] We note that the above reorganization leaves the scope of
		\begin{itemize}
			\item[(a)] overlaps amongst $\{\mathcal{P}^{(s)}, 1\leq s\leq \nu\}$ and $\{\mathcal{P}_\alpha: \alpha\in \Lambda_3\}$,
			\item[(b)] overlaps amongst
			$\{P_{s,v,j}: 0\leq j\leq q-1\}, 1\leq s\leq \nu, 1\leq v \leq g_s$, and also
			$\{P_{\alpha,j}:0\leq j\leq q-1\}, \alpha\in \Lambda_3$.
		\end{itemize}
	\end{itemize}
	But, all the same, we note below an application of this and Theorem \ref{t3.1.5} for an attempt to
	reduce the number of rank one projections as opposed to the crude limit presented in 3.3.2 (ii).
	\begin{theorem} For estimation purposes the collection
		$\Pi =\{P_{\alpha,j}: \alpha\in \Lambda_1, 0\leq j\leq q-1\}$ can be replaced by a smaller collection $\Pi^\prime$ with
		$\#\Pi^\prime \leq (\sum_{s=1}^\nu g_s +\#
		\Lambda_3 )q-\#\Lambda_3-\sum_{s=1}^{\nu}(g_s-1)\tau_s-(\nu-1)$.
	\end{theorem}
	\begin{proof} If $\Lambda_3=\emptyset$, then the result coincides with Theorem \ref{t3.1.5}.
		The only extra point needed to be noted is that if $\#\Lambda_3\geq 1$,  then we
		may take anchor as any $\mathcal{P}_\alpha$ with $\alpha\in \Lambda_3$. So $\{\mathcal{P}_\alpha, \alpha\in \Lambda_3\}$ require
		$q+(\#\Lambda_3-1)(q-1)$ rank one projections and we may ignore one more
		projection while applying Theorem \ref{t3.1.5}.
	\end{proof}
	\subsubsection{\it Simultaneous Constraints} We can apply the Simultaneous Constraints Technique 3.1.3 in case certain $\mathcal{P}_\alpha$'s have more than one constraint and
	reduce the number further. But the general case will involve two much of mathematical jargon. We will only illusrate this in \S3.6.
	\subsection{Some useful properties for the cyclic group case of shift and multiply (Schwinger Basis) as in \S2.3}
	\subsubsection{} We begin with the relationships between the unitary operators in the unitary basis. It will help to keep in view related examples
	above as well.
	\begin{proposition}\label{p3.5.1} In the context of \S 2.3 and Example \ref{e3.2.5}, for $m,n,j\in \ZZ_d$, we have the following facts.
		\begin{itemize}
			\item[(i)] $U_{jm,jn}= (\eta_d)^{-\frac{j(j-1)}{2}mn} (U_{m,n})^j$.
			\item[(ii)] $U_{m,n}^d =\pm I_d$, it is $I_d$ if and only if $mn(d-1)$ is even.
			\item[(iii)] If $\lambda_{(m,n)}$ is of full size, then $U_{(m,n)}$ has simple eigenvalues.
		\end{itemize}
	\end{proposition}
	\begin{proof} (i) It is trivially true for $j=0$ or $(m,n) = (0,0)$ because $U_{0,0}=I_d$. Now suppose $j\neq 0$. Consider any $(m,n)\neq (0,0)$ in
		$\ZZ_d\times \ZZ_d$. Let $k\in \ZZ_d$. Then $U_{jm,jn}|k\rangle = (\eta_d)^{jmk}|(jn+k)\rangle$. Also,
		\begin{align}
			U_{m,n}|k\rangle &= (\eta_d)^{mk}|(n+k)\rangle, ~\text{and, therefore,}\nonumber\\
			U_{m,n}^2|k\rangle &= (\eta_d)^{mk}(\eta_d)^{m(n+k)}|(2n+k)\rangle =  (\eta_d)^{2mk}(\eta_d)^{nm}|(2n+k)\rangle, \nonumber\\
			U_{m,n}^3|k\rangle & =   (\eta_d)^{mk}(\eta_d)^{m(n+k)}(\eta_d)^{m(2n+k)} |(3n+k)\rangle \nonumber\\
			&=(\eta_d)^{3mk}(\eta_d)^{nm(1+2)}|(3n+k)\rangle .
		\end{align}
		Proceeding like this
		\begin{align}
			U_{m,n}^j|k\rangle & =(\eta_d)^{jmk}(\eta_d)^{nm(1+2+\cdots+(j-1))}|(jn+k)\rangle \nonumber\\
			& =(\eta_d)^{jmk}(\eta_d)^{nm(\frac{j(j-1)}{2}}|(jn+k)\rangle.
		\end{align}
		Hence (i) is true.
		
		As to (ii), from the proof of (i) for $k\in\ZZ_d$,
		\begin{align}
			U_{m,n}^{(d-1)}|k\rangle = \eta_d^{(d-1)mn}\eta_d^{mn(d-2)(d-1)/2}|(d-1)n+k\rangle,
		\end{align}
		and, therefore
		\begin{align}
			U_{m,n}^{(d)}|k\rangle & = \eta_d^{(d-1)mk}\eta_d^{mn(d-2)(d-1)/2} \eta_d^{m((d-1)n+k)}|k\rangle =\eta_d^{mnd(d-1)/2}|k\rangle,\nonumber\\
			&=e^{\pi i mn(d-1)}|k\rangle.
		\end{align}
		Hence (ii) follows.
		
		Turning to (iii), note that $\{U_{(m,n)}, m,n \in \ZZ_d\}$ are linearly independent. But by (i), $U_{(jm,jn)} = (\eta_d)^{-\frac{j(j-1)}{2}mm}
		U_{(m,n)}^j$ for $0\leq j,m,n\leq d-1$. So, for $(m,n)$ with $\lambda_{(m,n)}$ of full size, $\{U_{(jm,jn)}; 0\leq j\leq d-1\}$ are linearly
		independent and, therefore $\{U_{(m,n)}^j; 0\leq j\leq d-1\}$ are linearly independent. Consider any such $(m,n)$. If $U_{(m,n)}$ has a multiple
		eigenvalue, say $\lambda_1$, with eigenprojection $S_1$ of dimension $\geq 2$, then $U_{(m,n)}=\sum_{k=1}^{s} \lambda_kS_k$ with $2\leq s < d$ with
		$\lambda_k$'s all distinct and $S_k$'s mutually orthogonal projections with $0\neq S_k\neq I_d$. So
		$\{U_{(m,n)}^j=\sum_{k=1}^{s}\lambda_k^j S_k : 0\leq j\leq d-1\}$ is contained in the span of $s$ linearly independent projections $S_k$'s, which
		has dimension $s<d$, a contradiction.
	\end{proof}
	
	\begin{corollary}\label{c3.5.2} Let $m, n, j\in \ZZ_d$.
		\begin{itemize}
			\item[(i)] Eigenvalues of $U_{m,n}$ are in the set $\{\eta_d^s:s\in \ZZ_d\}$ or $e^{\left(\frac{\pi i}{d}\right)}\{\eta_d^s:s\in \ZZ_d\}$ according
			as $mn(d-1)$ is even or odd.
			\item[(ii)] Eigenspaces of $U_{jm,jn}$ and $U_{m,n}^j$ are the same.
		\end{itemize}
	\end{corollary}
	\begin{proof} It is immediate from Proposition \ref{p3.5.1} or its proof above.
	\end{proof}
	\begin{theorem}\label{t3.5.3} Let $d$ be a composite number and $j$ a factor of $d$ other than $1$ and $d$. Set $u=d/j$. Let $m,n \in \ZZ_d$.
		\begin{itemize}
			\item[(i)] The set ${\bf Q}$ of the eigenprojections of $U_{jm,jn}$ is a constraint.
			
			\item[(ii)] For any orthonormal system $\{\xi_k: 0\leq k\leq d-1\}$ of eigenvectors of $U_{m,n}$, $\mathcal{P}=\{P_k=|\xi_k\rangle\langle\xi_k|:
			0\leq k\leq d-1\}$ is ${\bf Q}$-constrained.
			
			\item[(iii)] For $m^\prime, n^\prime \in \ZZ_d$ with $m^\prime\equiv m~(\text{mod}~u)$ and $n^\prime\equiv n~(\text{mod}~u)$, any orthonormal
			system $\{\xi_k^\prime : 0\leq k\leq d-1\}$ of eigenvectors of $U_{m^\prime,n^\prime}$,$\mathcal{P}^\prime=
			\{P_k^\prime=|\xi_k^\prime \rangle\langle\xi_k^\prime|,
			0\leq k\leq d-1\}$ is ${\bf Q}$-constrained.
			
			\item[(iv)] Theorem 3.3 can be applied to reduce the number of rank one projections obtained from a minimal $\mathcal{M}_{\bf W}$ for
			estimation purposes.
		\end{itemize}
	\end{theorem}
	\begin{proof} If $U_{m,n}$ has multiple eigenvalues, let $\lambda$ be such an eigenvlue with multiplicity $v$. Then $\lambda^j$ is an eigenvalue
		of $U_{m,n}^j$ of multiplicty at least $v$. So using Corollary \ref{c3.5.2} (ii) ${\bf Q}$ is a constraint. It is clearly, a constraint for $\mathcal{P}$
		by Spectral Mapping Theorem applied to powers of $U_{m,n}$.
		
		Now suppose $U_{m,n}$ has only simple eigenvalues. We have $d=ju$ with $u\geq 2$.
		For $0\leq v\leq j-1$, $0\leq s\leq u-1$, $(\eta_d^{s+vu})^j=\eta_d^{sj}$. So $U_{m,n}^j$ has eigenvalues $\{\eta_d^{sj} : 0\leq s\leq u-1\}$
		or $\{e^{\left(\frac{j\pi i}{d}\right)}\eta_d^{sj} : 0\leq s\leq u-1\}$ each of multiplicity $j$ according as $mn(d-1)$ is even or odd.
		
		So using Corollary \ref{c3.5.2} above, ${\bf Q}$ is a uniform constraint and $\mathcal{P}$ is ${\bf Q}$-constrained. Hence both (i) and (ii) hold.
		
		As to (iii) We have only to note that in this case $jm^\prime\equiv jm~(\text{mod}~d)$ and $jn^\prime\equiv jn~(\text{mod}~d)$.
		
		Turning to (iv), consider any $\sigma=(m,n)$  with $\lambda_\sigma$ of full size giving rise to a ${\bf W}$-MASS ${\bf V}_\sigma$. Then
		for  $\sigma^\prime=(m^\prime,n^\prime)$ with $m^\prime\equiv m~(\text{mod}~u)$ and $n^\prime\equiv n~(\text{mod}~u)$,
		${\bf V}_\sigma$ and ${\bf V}_{\sigma^\prime}$ overlap in ${\bf V}_{(jm,jn)}$ for sure. We now appeal to the discussion in the beginning
		of \S3.4.
	\end{proof}
	\begin{remark} Exact ideal number $d^2$ can be obtained for the case $d=p^2$, $d=pa$ with $p$ and $a$ both primes, $p<a$. This will
		be taken up in \S3.6.
	\end{remark}
	\subsubsection ~{\it A practical approach to obtain elementary measurements for the case of composite $d$ and $U_{m,n}$ having
		simple eigenvalues}. Let $j$ be a factor of $d$ other than $1$ or $d$ and $u=d/j$ and other notation and terminology
	also as in Theorem \ref{t3.5.3} above and its proof.
	\begin{remark}\label{r3.5.5} We explain the approach in different steps as follows.
		\begin{itemize}
			\item[Step (i)] It consists of first obtaining for $1\leq k\leq u$, eigenspaces $E_k$'s and any mutually orthogonal system of unit
			eigenvectors in $E_k$, say $\{ \xi_{k,v} : 0\leq v\leq j-1\}$ for $U_{jm,jn}$. We then set $Q_k=P_{E_k}$, the projection on $E_k$ for
			$1\leq k\leq u$ and ${\bf Q}=\{Q_k: 1\leq k\leq u\}$.
			\item[Step (ii)] It consists of considering the restrictions $U_{m,n}$ to $E_k$ and obtaining a mutually orthogonal system of unit
			eigenvectors for $U_{m,n}/E_k$, say $\{ \eta_{k,\beta}: 0\leq \beta\leq j-1\}$ in terms of
			$\{ \xi_{k,v}: 0\leq v \leq j-1\}$ and then pool them together varying $k$ from $1$ to $u$ to form a complete
			system of mutually orthogonal unit eigenvectors $\{\eta_{k,\beta}: 1\leq k\leq u, 0\leq \beta \leq j-1\}$ for
			$U_{m,n}$ and accordingly the set $\mathcal{P}$ of rank one mutually orthogonal projections
			$\{P_{k,\beta}=|\eta_{k,\beta}\rangle\langle\eta_{k,\beta}| : 1\leq k\leq u, 0\leq \beta \leq j-1\}$ for the
			spectral representation of $U_{m,n}$. Clearly, $\sum_{\beta=0}^{j-1} P_{k,\beta}=Q_k$ for $1\leq k\leq u$ and
			therefore, $\mathcal{P}$ is ${\bf Q}$-constrained.
			\item[Step (iii)] is to note that for another $(0,0)\neq (m^\prime,n^\prime)\in \ZZ_d\times \ZZ_d$ with $(jm^\prime,jn^\prime)=(jm,jn)$ we
			can repeat step (ii) with $(m,n)$ replaced by $(m^\prime,n^\prime)$ using ${\bf Q}$ etc. as in (i), above. Thus we can
			obtain various $\eta_{k,\beta}^\prime$, $P_{k,\beta}^\prime$ accordingly, $\mathcal{P}^\prime
			=\{ P_{k,\beta}^\prime=|\eta_{k,\beta}^\prime\rangle\langle\eta_{k,\beta}^\prime| : 1\leq k\leq u, 0\leq \beta \leq j-1\}$ which is
			${\bf Q}$-constrained because $\sum_{\beta=0}^{j-1} P_{k,\beta}^\prime=Q_k$ for $1\leq k\leq u$.
			\item[Step (iv)] Finally, we note that convenience lies in the fact that quantum mechanical overlaps for $\mathcal{P}$ and $\mathcal{P}^\prime$ are easy to obtain
			in the sense that for $1\leq k\neq k^\prime \leq u, 0\leq \beta, \beta^\prime \leq j-1$ we have $P_{k,\beta}P_{k^\prime,\beta^\prime}^\prime=0$
			and, therefore, $\text{tr}(P_{k,\beta}P_{k^\prime,\beta^\prime}^\prime)=0$. On the other hand, for $1\leq k\leq u, 0\leq \beta,\beta^\prime\leq j-1$,
			we have $\text{tr}(P_{k,\beta}P_{k,\beta^\prime}^\prime)= |\langle \eta_{k,\beta}|\eta_{k,\beta^\prime}^\prime\rangle|^2
			= |\sum_{v=0}^{j-1} \overline{\eta_{k,\beta,v}}{\eta_{k,\beta^\prime,v}^\prime}|^2$ where $\eta_{k,\beta}=\sum_{v=0}^{j-1}\eta_{k,\beta, v}\xi_{k,v}$ and
			$\eta_{k,\beta^\prime}^\prime =\sum_{v=0}^{j-1}\eta_{k,\beta^\prime, v}^\prime\xi_{k,v}$.
		\end{itemize}
	\end{remark}
	We shall demonstrate all this numerically for the case $d=p^2$ or $d=pa$  with $p$ and $a$ distinct primes in \S3.6 after a little
	more preparation.
	\subsubsection {\it Basic modular arithmetic}
	We consider the case $d=pa$ with $p$, $a$ prime, $p < a$.
	There exist $j_0,k_0,r_0 \in \mathbb{N}$ with $r_0<p$ such that $a = j_0 p+ r_0$ and $p^{k_0}< a <p^{k_0+1}$.
	\begin{remark}\label{r3.5.6}
		\begin{itemize}
			\item[(i)] The following observations from Modular Arithmetic will be helpful. We
			write equalities modulo $d$ unless otherwise specified.
			\begin{itemize}
				\item[(a)] For $n \in Z_d$ with $\text{HCF}(n,d)=1$, there exists a unique $\hat{n}$ with $n \hat{n}\equiv 1 ~(\text{mod}\,d)$.
				As a consequence, for $m \in Z_d$, $(m,n)=n(m\hat{n},1)$.
				\item[(b)] For $1 \leq r \leq p-1$, $\{{\rm tr} :1 \leq t \leq p-1\} (\mathrm{mod}\, p)= \{j :1 \leq j \leq p-1 \}$ and similarly, for $a$.
			\end{itemize}
			\item[(ii)] By (i)(a) there are $1\leq s_0\leq p-1$ and $u\geq 0$ that satisfy $r_0s_0=up+1$, i.e., $r_0s_0=1~(\text{mod}~p)$.
			\begin{itemize}
				\item[(a)] If $p$ and $a$ are both odd, then $p+2\leq a$. Therefore, $j_0$ and $r_0$ in $a=j_0p+r_0$ satisfy atleast one of $j_0$ and $r_0$ is $\geq 2$. Also
				$s_0=1$ if and only if $r_0=1$ and therefore, at least one of $j_0$ and $s_0\geq 2$.
				\item[(b)] $s_0a=(s_0j_0+u)p+1$, and therefore, $s_0a\equiv 1~(\text{mod}~p)$. This gives $s_0a^2=a$ in $\ZZ_d = \ZZ_{pa}$.
				\item[(c)] Let $f= a-\frac{s_0 a-1}{p}= \frac{(p-s_0)a+1}{p}$. Then $ \frac{a+1}{p}\leq f\leq \frac{(p-1)a+1}{p}=\frac{pa-(a-1)}{p}$. So
				$\frac{p+2}{p}\leq f\leq a-1$. As $f$ is an integer, we have $2\leq f\leq a-1$ . Next, $fp=(p-s_0)a+1\equiv 1~(\text{mod}~a)$. This gives
				$fp^2=p$ in $\ZZ_d$.
				\item[(d)] Let $g_1 = \frac{s_0a+p-1}{p} s_0 ~(\text{mod}~p)$. Set $g=p-1$ if $g_1=0$ and $g_1-1$, otherwise. Then $0\leq g\leq p-1$. So $2\leq ga+f\leq
				(p-1)a+f \leq pa-1$.
				\item[(e)] Also $\text{HCF} (s_0a+p,  pa) =1$, we can check that $(s_0a+p)(ga+f)\equiv 1~(\text{mod}~pa)$ so that $ga+f=(s_0a+p)^\wedge$.
				Indeed, $ga=\left(\frac{s_0a+p-1}{p}
				s_0a-a\right)$ $(\text{mod}~pa)$, and therefore, modulo $pa$,
				$ga+f=\frac{s_0a+p-1}{p}s_0a
				-\frac{s_0a-1}{p} = \frac{s_0a+p-1}{p}(s_0a-1)+1$. Using (b) $(ga+f)(s_0a+p)= \frac{s_0a+p-1}{p}(-p)+s_0a+p=1$.
				\item[(f)] To give an idea, for $p=2, r_0=s_0=1, s_0a+p=a+2, f=\frac{a+1}{2}, ga+f= \frac{a+1}{2}$ if it is odd and $a+\frac{a+1}{2}$, otherwise.
			\end{itemize}
		\end{itemize}
	\end{remark}
	\subsection{Ideal number $d^2$ for $d=p^2$ and $d=pa$ with $p$ and $a$ primes by optimal constraint technique for the shift and multiply, cyclic group case as in \S2.3 and \S3.5}
	We freely use the preparation in \S3.5 above to apply the results in \S3.1 to \S3.4 effectively.
	\subsubsection{}
	We begin with the case $d=p^2$ with $p$ any prime.
	\begin{theorem}\label{t3.6.1} (Minimal ${\bf W}$-MASS and constraints) Let $d=p^2$, $p$ any prime.
		\begin{itemize}
			\item[(i)] A convenient ${\bf W}$-MASS,  $\mathcal{M}_{{\bf W}}$, is given by
			$\{\lambda _{(ap+r,1)}, 0\leq a$, $r\leq p-1\}$ $\bigcup \{\lambda_{(1,bp)}:0\leq b\le p-1\}$.
			Its cardinality is $p^{2} +p=p(p+1)$.
			\item[(ii)]  For $0\leq r\leq p-1 $,
			\begin{align*}
				\lambda (rp,p)\subset \lambda _{(ap+r,1)}\quad\text{for} \ 0\leq a\leq p-1.
			\end{align*}
			Also $\lambda_{(p,0)}\subset \lambda_{(1,bp)}$ for $0\leq b\leq p-1$.
			
			Further, there are no more overlaps.
			
			\item[(iii)] Each overlap in (ii) above gives rise to a uniform constraint of size $p$.
		\end{itemize}
	\end{theorem}
	\begin{proof} (i) and (ii) follow from easy computations in view of basic modular arithmetic.
		
		For (iii), we divide the proof in a few parts below.
		\begin{itemize}
			\item[(a)] Consider $0\le r\le p-1$ and the corresponding unitary operator $U_{(rp,p)}$.
			For $0\leq k\leq d-1=p^{2} -1 $,
			\begin{align}
				U_{(rp,p)}|k\rangle =\exp \left(\frac{2\pi irk}{p} \right)|p+k\rangle.
			\end{align}
			So, for $0\leq t\leq p-1$, $0\leq v\leq p-1$
			\begin{align}
				U_{(rp,p)} |vp+t\rangle =\exp \left( \frac{2\pi irt}{p} \right) |(v+1)p+t\rangle.
			\end{align}
			Consider $0\leq u\leq p-1 $ and put $\beta =\exp \left( \dfrac{2\pi iu}{p} \right) $.
			Set $\xi_{u,t}=\sum\limits_{v=0}^{p-1}\bar{\beta } ^{v} |vp+t\rangle $.
			Then
			\begin{align}
				U_{(rp,p)}\xi _{u,t}=\exp \left( \dfrac{2\pi i(rt+u)}{p} \right) \xi _{u,t}.
			\end{align}
			For $0\leq c\leq p-1 $, let
			\begin{align}
				S_{r,c}=\{(u,t):rt+u=c\,(\textrm{mod}\,\, p)\}.
			\end{align}
			Then $\#S_{r,c}=p$ and, as $c$ varies, $S_{r,c}$'s are mutually disjoint with union equal to
			$\{(u,t):0\leq u$, $t\leq p-1\}$. Further, $U_{(rp,p)} $ has
			$e^{\left( \frac{2\pi ic}{p} \right)} $ as an eigenvalue
			of multiplicity $p $ with eigenspace $E_{r,c}$, the span
			of $\xi _{u,t}$ as $(u,t) $ varies in $S_{r,c} $.\ We set $Q_{r,c}=$
			the orthogonal projection on $E_{r,c} $.
			We note that $Q_{r,c}$ has rank $p$.
			\item[(b)] Let $0\leq t\leq p-1 $.\ Then for $0\leq v\leq p-1 $,
			$U_{(p,0)} |vp+t\rangle =e^{\left( \frac{2\pi it}{p} \right)} |vp+t\rangle $.
			So $e^{\left( \frac{2\pi it}{p} \right)}$ is an eigenvalue of
			multiplicity $p $ for $U_{(p,0)}$ with eigenspace
			$E_{t} $ as the span of $\{|vp+t\rangle :0\leq v\leq p-1\}$.
			Set $Q_{t} $, the orthogonal projection on $E_t$.
			We note that $Q_t$ has rank $p$.
			\def\Q{\mathbf{Q}}
			\item[(c)] Let $0\leq r\leq p-1 $. Set $\Q^{(r)}=\{Q_{r,c} :0\leq c\leq p-1\} $
			and for $0\leq a\leq p-1 $, let $\mathcal{P}^{(r)}_a$
			be a common orthonormal eigenvector system for $\lambda _{(ap+r,1)}$.
			Then each $\mathcal{P}_{a}^{(r)}$ is a $\Q^{(r)}$-constrained
			elementary measurement, $0\leq a\leq p-1 $.
			\item[(d)] Let $\Q^{(p)} =\{Q_{t}:0\leq t\leq p-1\} $
			and for $0\leq b\leq p-1$, $\mathcal{P}_b^{(p)}$ a common orthonormal eigenvector system
			for $\lambda _{(1,bp)}$.
			Then $\mathcal{P}_{b}^{(p)}$ is a $\Q^{(p)}$-constrained elementary measurement for $0\leq b \leq p-1$.
		\end{itemize}
	\end{proof}
	\begin{theorem}\label{t3.6.2}  The size of pure measurement obtained by our Optimal Constraint Technique for the case $d=p^2$ with $p$ prime is the ideal number
		i.e., the minimum possible , $d^2
		=p^4$.
	\end{theorem}
	\begin{proof} Theorem \ref{t3.6.1} above facilitates the use of Theorem \ref{t3.1.5} or, for that matter, repeated use of Remark \ref{r3.1.3}
		(iii) after some relabelling. So the size of the reduced set of pure measurements is $ (p+1)p\cdot p^2-(p+1)(p-1)p-p = p^4=d^2$.
		
		We now proceed to obtain one such system explicitly with the help of Theorem \ref{t3.6.1} and its proof above in a few steps.
		\begin{itemize}
			\item[(i)] We begin with part (iii)(a) and (c) of the proof of Theorem \ref{t3.6.1}.
			\begin{itemize}
				\item[(a)] Fix $r$ and $c$. For $0\leq t\leq p-1$, let $\xi_t=\frac{1}{\sqrt{p}}\xi_{(c-rt){~(\text{mod}~p),t}}$, i.e.,
				\begin{align}
					\xi_t&= \frac{1}{\sqrt{p}}\sum_{v=0}^{p-1}e^{-\left(\frac{2\pi iv}{p}(c-rt)~(\text{mod}~p)\right)}|vp+t\rangle\nonumber\\
					&=\frac{1}{\sqrt{p}}\sum_{v=0}^{p-1}e^{-\left(\frac{2\pi iv}{p}(c-rt)\right)}|vp+t\rangle.
				\end{align}
				Then $\xi_t$'s span $E_{r,c}$.
				\item[(b)] Fix $0\leq a\leq p-1 $ temporarily. For $0\leq t\leq p-1$,
				\begin{align}
					U_{ap+r,1}\xi_t& =\frac{1}{\sqrt{p}}\sum_{v=0}^{p-1}e^{-\left(\frac{2\pi iv}{p}(c-rt)\right)}
					e^{\left(\frac{2\pi i}{p^2}(ap+r)(vp+t)\right)}|vp+t+1\rangle\nonumber\\
					&=\frac{1}{\sqrt{p}} \sum_{v=0}^{p-1}e^{-\left(\frac{2\pi iv(c-rt-r)}{p}\right)}
					e^{\left(\frac{2\pi i(ap+r)t}{p^2}\right)}|vp+(t+1)\rangle\nonumber\\
					&=\frac{1}{\sqrt{p}}e^{\left(\frac{2\pi i}{p^2}(ap+r)t\right)}
					\sum_{v=0}^{p-1}e^{-\left(\frac{2\pi iv(c-r(t+1))}{p}\right)}|vp+(t+1)\rangle\nonumber\\
					&= e^{\left(\frac{2\pi i(ap+r)t}{p^2}\right)}\left\{
					\begin{array}{l}\xi_{t+1}~~~~~~~~~\text{if}~0\leq t\leq p-2\\
						e^{\left(\frac{2\pi ic}{p}\right)}\xi_0~\text{if}~t=p-1\end{array}\right.\nonumber\\
					&\text{Consequently}, ~ U_{ap+r,1} E_{r,c} = E_{r,c}.
				\end{align}
				\item[(c)] For a $p$-tuple ${\bf b}= (b_t)_{t=0}^{p-1}$ in $\mathbb{C}^p$, $\eta_{{\bf b}}
				=\sum_{t=0}^{p-1}b_t\xi_t$, we have
				\begin{align}
					~~~~~~~~U_{ap+r,1}\eta_{{\bf b}} = \sum_{t=0}^{p-2} b_t
					e^{\left(\frac{2\pi i(ap+r)t}{p^2}\right)}\xi_{t+1} +
					b_{p-1} e^{\left(\frac{2\pi i(ap+r)(p-1)}{p^2}\right)}e^{\left(\frac{2\pi ic}{p}\right)}\xi_0.
				\end{align}
				Now by (b) above, $U_{ap+r,1}E_{r,c}=E_{r,c}$ and, therefore,
				\begin{align}
					\left(U_{ap+r,1}/E_{r,c}\right)^{-1}= \left(U_{ap+r,1}\right)^{-1}/E_{r,c}.
				\end{align}
				So
				\begin{align}
					\left(U_{ap+r,1}/E_{r,c}\right)^{\ast}= \left(U_{ap+r,1}\right)^{\ast}/E_{r,c}.
				\end{align}
				So $U_{ap+r,1}/E_{r,c}$ has $p$ eigenvalues with eigenvectors in $E_{r,c}$. Let $\gamma$ be
				any such eigenvalue. Then by Proposition \ref{p3.5.1} (i),  $e^{-\left(\frac{2\pi i}{p^2}
					\frac{p(p-1)}{2}(ap+r)\right)}\gamma^p$ is an eigenvalue of $U_{p(ap+r),p}=U_{rp,p}$
				restricted to $E_{r,c}$ and, therefore,
				\begin{align}
					\gamma^p=  e^{\left(\frac{2\pi i}{p^2}\frac{p(p-1)}{2}(ap+r)\right)}e^{\left(\frac{2\pi i c}{p}\right)}.
				\end{align}
				Let $\gamma$ be any such eigenvalue with $\eta_{{\bf b}}$ as an eigenvector. Then
				\begin{align}
					\gamma b_{t+1}= b_t e^{\left(\frac{2\pi i}{p^2}(ap+r)t\right)}~\text{for}~0\leq t\leq p-2
				\end{align}
				and
				\begin{align}
					\gamma b_{0}= b_{p-1} e^{\left(\frac{2\pi i}{p^2}(ap+r)(p-1)\right)}e^{\left(\frac{2\pi ic}{p}\right)}.
				\end{align}
				The converse is also true. In particular, $\eta_{{\bf b}}\neq 0$ if and only if ${\bf b}\neq 0$ if and only if $b_t\neq 0$ for $0\leq t\leq p-1$.
				
				Now the system of equations becomes
				\begin{align}
					b_{t}&= \overline{\gamma}^{t} e^{\left(\frac{2\pi i}{p^2}(ap+r)\frac{(t-1)t}{2}\right)}b_0 ~\text{for}~1\leq t\leq p-1,\nonumber\\
					\gamma b_{0}&= b_{p-1}.
				\end{align}
				Considering $t=p-1$,
				\begin{align}
					b_{p-1}= \overline{\gamma}^{p-1} e^{\left(\frac{2\pi i}{p^2}(ap+r)\frac{(p-1)(p-2)}{2}\right)}b_0.
				\end{align}
				So
				\begin{align}
					b_{p-1} e^{\left(\frac{2\pi i}{p^2}(ap+r)(p-1)\right)}e^{\left(\frac{2\pi ic}{p}\right)}&= \overline{\gamma}^{p-1} e^{\left(\frac{2\pi i}{p^2}(ap+r)\frac{p(p-1)}{2}\right)}e^{\left(\frac{2\pi ic}{p}\right)}b_0.\nonumber\\
					&=\overline{\gamma}^{p-1}\gamma^pb_0\nonumber\\
					&=\gamma b_0
				\end{align}
				Thus the last equation is automatically satisfied.
				
				Now
				\begin{align}
					&\gamma=\gamma_s = e^{\left(\frac{\pi i}{p^2}(ap+r)(p-1)\right)}
					e^{\left(\frac{2\pi ic}{p^2}\right)}e^{\left(\frac{2\pi is}{p}\right)}\nonumber\\&~\text{for some}~s~\text{with}~0\leq s\leq p-1.
				\end{align}
				So
				\begin{align}
					\overline{\gamma}_s e^{\left(\frac{\pi i}{p^2}(ap+r)(t-1)\right)}
					=e^{\left(\frac{\pi i}{p^2}(ap+r)(t-p)\right)}e^{\left(-\frac{2\pi ic}{p^2}\right)}e^{\left(-\frac{2\pi is}{p}\right)}
				\end{align}
				Hence for a unit eigenvector $\eta_{{\bf b}_s}$ for $\gamma_s$, we may take
				\begin{align}
					{\bf b}_s= \frac{1}{\sqrt{p}}\left(\left( e^{\left(\frac{\pi i}{p^2}(ap+r)(t-p)\right)}
					e^{\left(-\frac{2\pi ic}{p^2}\right)}e^{\left(-\frac{2\pi is}{p}\right)}
					\right)^t\right)_{t=0}^{p-1}
				\end{align}
				\item[(d)] We consolidate the conclusions in (a), (b), (c) above in a notation that can be used for the whole system later.
				Let $0\leq r,a\leq p-1$. For $0\leq c,t\leq p-1$, we write
				$\xi(r,c,t)$ for $\xi_t$ as in (a) above, i.e.,
				\begin{align}
					\xi(r,c,t) = \frac{1}{\sqrt{p}} \sum_{v=0}^{p-1}
					e^{-\left(\frac{2\pi iv(c-rt)}{p}\right)}|vp+t\rangle.
				\end{align}
				We note that for $0\leq r^\prime,c^\prime, t^\prime\leq p-1$,
				\begin{align}
					\langle\xi(r,c,t)\xi(r^\prime,c^\prime,t^\prime)\rangle =0~
				\end{align}
				for $t^\prime \neq t$ and also for $r=r^\prime$ but $c\neq c^\prime$.
				
				Next, for $0\leq c,s\leq p-1$, we write $\gamma(r,a;c,s)$ for $\gamma_s$ as in (c)
				above, i.e.,
				\begin{align}
					\gamma(r,a;c,s) =  e^{\left(\frac{2\pi i}{p^2}(ap+r)(p-1)\right)}
					e^{\left(\frac{2\pi ic}{p^2}\right)}e^{\left(\frac{2\pi is}{p}\right)},\end{align}
				$\beta(r,a;c,s) =(\beta(r,a;c,s,t)_{t=0}^{p-1}$ with
				\begin{align}
					\beta(r,a;c,s,t)= \frac{1}{\sqrt{p}} e^{\left(\frac{\pi i}{p^2}(ap+r)((t-p)t)\right)}
					e^{\left(-\frac{2\pi ict}{p^2}\right)}e^{\left(-\frac{2\pi ist}{p}\right)}\end{align}
				for $0\leq t\leq p-1$; and , finally
				$ \eta(r,a;c,s)$ for $\eta_{{\bf b}_s}$ as in (c) above i.e.,
				\begin{align}
					\eta(r,a;c,s) =\sum_{t=0}^{p-1}\beta(r,a;c,s,t)\xi(r;c,t).
				\end{align}
				We note that $\mathcal{E}_a^{(r)}=\{\eta(r,a;c,s):0\leq c,s\leq p-1\}$ is a complete system of orthonormal unit eigenvectors corresponding to eigenvalues
				$\{\gamma(r,a;c,s): 0\leq c,s\leq p-1\}$ for
				$U_{ap+r,1}$.
				
				This gives us a family \\$\{\mathcal{P}_a^{(r)}=\{P(r,a:c,s) =|\eta(r,a;c,s)\rangle
				\langle \eta(r,a;c,s)| : 0\leq c,s, \leq p-1\}: 0\leq r,a\leq p-1\}$ \\of elementary
				measurements $\mathcal{P}_a^{(r)}$'s constrained by ${\bf Q}^{(r)}$ for $0\leq r, a\leq p-1$.
			\end{itemize}
			\item[(ii)] We now utilize (ii)(b) and (iii)(d) of the proof of Theorem \ref{t3.6.1} above.
		\begin{itemize}
				\item[(a)] $U_{1,0}$ has a complete set $\mathcal{E}_0^{(p)}$ of orthonormal eigenvectors $\{|j\rangle : 0\leq j\leq p^2-1\}$
				with corresponding eigenvalues $\{e^{\left(\frac{2\pi i j}{p^2}\right)} : 0\leq j\leq p^2-1\}$ . We may rewrite the parameter set $\{j: 0\leq j\leq p^2-1\}$
				as $\{vp+t:0\leq v,t\leq p-1\}$ as well.
				\item[(b)] Let $0\leq t\leq p-1$. Then for $0\leq v\leq p-1$,
				\begin{align}
					U_{1,p}|vp+t\rangle &= e^{\left(\frac{2\pi i}{p^2}(vp+t)\right)}|vp+t+p\rangle\nonumber
					\\ &= e^{\left(\frac{2\pi i v}{p}\right)}e^{\left(\frac{2\pi i t}{p^2}\right)}
					|(v+1)p+t\rangle
				\end{align}
				
				So as argued in (i)(c) above, $U_{1,p}/E_t$ has $p$ eigenvalues in $E_t$ and further, for any such eigenvalue $\gamma$,
				$e^{\left(-\frac{2\pi i }{p^2}\frac{p(p-1)}{2}p\right)}\gamma^p$ is an eigenvalue of $U_{p(1,p)}=U_{(p,0)}$
				restricted to $E_t$.
				Therefore, $e^{\left(-\pi i (p-1)\right)} \gamma^p=
				e^{\left(\frac{2\pi it}{p}\right)}$. Consequently $\gamma^p=\mp
				e^{\left(\frac{2\pi i t}{p}\right)} $ according as $p=2$ or $p>2$.
				
				Now for $p=2$, for $t=0, \gamma=\mp i$ and $\frac{1}{\sqrt{2}}(|0\rangle \mp i|2\rangle)$
				work as unit eigenvectors, whereas for $t=1$, $\gamma=\pm1$ and
				$\frac{1}{\sqrt{2}}(|1\rangle\pm i|3\rangle)$ work as eigenvectors.
				
				For $p>2$, $\gamma=\gamma_{t,s} = e^{\left(\frac{2\pi i t}{p^2}\right)}
				e^{\left(\frac{2\pi i s}{p}\right)}$ for some $s$ with $0\leq s\leq p-1$. Simple
				computations
				give that for $0\leq s\leq p-1$,
				\begin{align}
					\eta_{1;t,s}=\frac{1}{\sqrt{2}}\sum_{v=0}^{p-1}
					e^{\left(\frac{\pi i v(v-1)}{p}\right)}e^{\left(\frac{2\pi i sv}{p}\right)}|vp+t\rangle\end{align}
				works as a unit eigenvector corresponding to the eigenvalue $\gamma_{t,s}$ for
				$U_{1,p}/E_t$.
				
				\item[(c)] Let $p>2$ and $2\leq b\leq p-1$. Let $0\leq t\leq p-1$. Since $p$ is prime ,
			
			$\{kb~\text{mod}~p): 1\leq k\leq p-1\}$ =$\{v: 1\leq v\leq p-1\}$. So
				$\{v+pt: 0\leq v\leq p-1\} = \{kb~(\text{mod}~p) p+t : 0\leq k\leq p-1\}$=
				$\{(kbp+t)~(\text{mod}~p^2: 0\leq k\leq p-1\}$).
				Therefore, we may say that $E_t$, the span of $\{|vp+t\rangle: 0\leq v\leq p-1\}$, is same as the
				span of $\{|kbp+t\rangle : 0\leq k\leq p-1\}$ in the context of $\mathbb{C}^{p^2}$.
				
				Now
				\begin{align}
					U_{1,bp}|kbp+t\rangle &= e^{\left(\frac{2\pi i}{p^2}(kbp+t)\right)}|kbp+t+bp\rangle\nonumber
					\\ &= e^{\left(\frac{2\pi i t}{p^2}\right)} e^{\left(\frac{2\pi ikb}{p}\right)}
					|(k+1)bp+t\rangle
				\end{align}
				for $0\leq k\leq p-1$.
				So
				\begin{align}
					\eta_{b;t,s}=\sum_{k=0}^{p-1}
					e^{\left(\frac{2\pi i}{p^2}(\frac{k(k-1)}{2}bp)\right)}e^{\left(-\frac{2\pi ik s}{p}\right)}|kbp+t\rangle
				\end{align}
				is a unit eigenvector corresponding to the eigenvalue
				$\gamma_{t,s}= e^{\left(\frac{2\pi i t}{p^2}\right)}
				e^{\left(\frac{2\pi i s}{p}\right)}$
				of $U_{1,bp}/E_t$.
				\item[(d)] We note that $\eta_{1;t,s}$ in (3.33) in (b) above and the expression on RHS of $\eta_{b;t,s}$ in (3.35) in (c) above for $b=1$ are the same.
				So we may say that for $p>2$, $1\leq b\leq p-1$,
				$\{\eta_{b;t,s}:0\leq t,s\leq p-1\}$ is a complete system of unit eigenvectors of $U_{1,bp}$.
				\item[(e)] We set $\mathcal{P}_0^{(p)}=\{P_j=|j\rangle\langle j| : ~0\leq j\leq p^2-1\}$ and for
				$1\leq b\leq p-1$, $\mathcal{P}_b^{(p)}=\{P_{b,t,s}=|\eta_{b;t,s}\rangle\langle
				\eta_{b;t,s}| : ~0\leq t,s\leq p-1\}$ and obtain elementary measurements, constrained by ${\bf Q}^{(p)}$. For notational convenience, we can even write $|j\rangle$ as $\eta_{0;t,s}$ for $j=sp+t,~ 0\leq t,s \leq p-1$.
			\end{itemize}
			\item[(iii)] There are various ways in view of Theorem \ref{t3.6.1} above to obtain informationally complete families of rank one projections of the ideal size $d^2=p^4$ from
			$\{\mathcal{P}_b^{(p)} : 0\leq b\leq p-1\}\cup\{\mathcal{P}_a^{(r)} :
			0\leq r,a\leq p-1\}$. We just display one of them, say $\mathcal{P}^\prime$, below.
			
			For $1\leq b\leq p-1$ let $\{\mathcal{P}_b^{(p)\prime}=\{P_{b,t,s}: 0\leq t\leq p-1,
			1\leq s\leq p-1\}$.

			For $0\leq r\leq p-1$ let $\{\mathcal{P}_0^{(r)\prime}=\{P(r,0;c,s): 0\leq c,s\leq p-1,
			(c,s)\neq (0,0)\}$ and let for $1\leq a\leq p-1$ , $\mathcal{P}_a^{(r)\prime}
			=\{P(r,a;c,s): 0\leq c\leq p-1, 1\leq s\leq p-1\}$. Now set
			$$ \mathcal{P}^\prime = \mathcal{P}_0^{(p)}\cup \cup_{b=1}^{p-1}\mathcal{P}_{b}^{(p)\prime}\cup\cup_{0\leq r,a\leq p-1}\mathcal{P}_a^{(r)\prime}.$$
			We verify that size of
			$\mathcal{P}^\prime$ is $p^2+(p-1)(p^2-p)+(p(p^2-1)+p(p-1)(p^2-p))=p^4$.
		\end{itemize}
	\end{proof}
	We now come to quantum mechanical overlaps $\text{tr}(P_\eta P_{\eta^\prime})$, or,
	for that matter $\langle\eta,\eta^\prime\rangle|^2$ as indicated in Remark \ref{r3.5.5} (iv).
	We first give it for $d=2^2=4$ and then $d=p^2$ with $p$ an odd prime.

	\begin{example}  Let $d=4=2^2$. Then from the proof of Theorem \ref{t3.6.2}, parts (ii) and (i)(d) give the following which can be computed directly as well.
		\begin{align}
			\mathcal{E}_0^{(2)}&=\{|j\rangle: j=0,1,2,3\}~\text{coming from}~ U_{1,0}, \nonumber\\
			\mathcal{E}_1^{(2)}&=\{\frac{1}{\sqrt{2}}(|0\rangle-i|2\rangle), \frac{1}{\sqrt{2}}(|0\rangle+i|2\rangle), \frac{1}{\sqrt{2}}(|1\rangle+i|3\rangle),
			\frac{1}{\sqrt{2}}(|1\rangle-i|3\rangle)\}
			\nonumber\\
			&~\text{coming from}~ U_{1,2}, \\
			\mathcal{E}_0^{(0)}&=\{\frac{1}{2}((|0\rangle+|2\rangle) +(|1\rangle+|3\rangle)),
			\frac{1}{2}((|0\rangle+|2\rangle) -(|1\rangle+|3\rangle)),\nonumber\\
			&\frac{1}{2}((|0\rangle-|2\rangle) -i(|1\rangle-|3\rangle)),
			\frac{1}{2}((|0\rangle-|2\rangle) +i(|1\rangle-|3\rangle))\}\nonumber\\
			&~\text{coming from}~ U_{0,1}, \\
			\mathcal{E}_1^{(0)}&=\{\frac{1}{2}((|0\rangle+|2\rangle) -i(|1\rangle+|3\rangle)),
			\frac{1}{2}((|0\rangle+|2\rangle) +i(|1\rangle+|3\rangle)),\nonumber\\
			&\frac{1}{2}((|0\rangle-|2\rangle) - (|1\rangle-|3\rangle)),
			\frac{1}{2}((|0\rangle-|2\rangle) +(|1\rangle-|3\rangle))\}\nonumber\\
			&~\text{coming from}~ U_{2,1},\\
			\mathcal{E}_0^{(1)}&=\{\frac{1}{2}((|0\rangle+|2\rangle) +\left(\frac{1-i}{\sqrt{2}}\right)(|1\rangle-|3\rangle)),\nonumber\\
			&\frac{1}{2}((|0\rangle+|2\rangle) -\left(\frac{1-i}{\sqrt{2}}\right)(|1\rangle-|3\rangle)),\nonumber\\
			&\frac{1}{2}((|0\rangle-|2\rangle) -\left(\frac{1+i}{\sqrt{2}}\right)(|1\rangle-|3\rangle)),\nonumber\\
			&\frac{1}{2}((|0\rangle-|2\rangle) +\left(\frac{1+i}{\sqrt{2}}\right)(|1\rangle-|3\rangle))\},\nonumber\\
			&~\text{coming from}~ U_{1,1},\\
			\mathcal{E}_1^{(1)}&=\{\frac{1}{2}((|0\rangle+|2\rangle) -\left(\frac{1+i}{\sqrt{2}}\right)(|1\rangle-|3\rangle)),\nonumber\\
			&\frac{1}{2}((|0\rangle+|2\rangle) +i\left(\frac{1+i}{\sqrt{2}}\right)(|1\rangle-|3\rangle)),\nonumber\\
			&\frac{1}{2}((|0\rangle-|2\rangle) - \left(\frac{1-i}{\sqrt{2}}\right)(|1\rangle+|3\rangle)),\nonumber\\
			&\frac{1}{2}((|0\rangle-|2\rangle) +\left(\frac{1-i}{\sqrt{2}}\right)(|1\rangle+|3\rangle))\}\nonumber\\
			&~\text{coming from}~ U_{3,1}.
		\end{align}
		The arrangements have been written with convenience of constraints and quantum mechanical overlaps in mind. Quantum mechanical overlaps are easily
		seen to be $0,1/2,1/4,1$. This is in line with 2.4.2 above.
	\end{example}
	
	\begin{example} Let $d=p^2$ with $p$ an odd prime. There is no hope of mutual unbiasedness in this case as well as already indicated in Remark \ref{r3.5.5} Step (iv).
		We give quantum mechanical overlaps for the minimal system given in Theorem \ref{t3.6.2} and show that they have values $0,1,
		1/p, 1/p^2$.
		
		First we note a few immediate useful facts based on Remark \ref{r3.5.5} Step (iv) and parts (i)(d) and (ii) of proof of Theorem \ref{t3.6.2} above.
		Let $0\leq r,a,c,s,r^\prime,a^\prime,c^\prime,s^\prime,t,t^\prime$ $\leq p-1$ with $(r,a)\neq (r^\prime,a^\prime)$ and $0\leq b,b^\prime\leq p-1$ with
		$b\neq b^\prime$.
		\begin{align}
			&\text{(i)}~~\langle\eta(r,a;c,s)|\eta(r^\prime,a^\prime;c^\prime,s^\prime)\rangle
			= \sum_{t=0}^{p-1} \overline{\beta(r,a;c,s,t)}\beta(r^\prime, a^\prime;c^\prime,
			s^\prime,t)\langle\xi(r,c,t)|\xi(r^\prime,c^\prime,t)\rangle\nonumber\\
			&\text{simply because}~\langle \xi(r,c,t)|\xi(r^\prime,c^\prime,t^\prime\rangle=0~\text{for}~t\neq t^\prime.\nonumber\\
			& \text{(ii)}~\langle\eta(r,a;c,s)|\eta(r,a^\prime,c^\prime,s^\prime)\rangle =0\nonumber\\
			&\text{for}~ c\neq c^\prime ~\text{simply because}~\langle \xi(r,c,t)|\xi(r,c^\prime,t^\prime\rangle=0~\text{for}~ c\neq c^\prime.\nonumber\\
			&\text{(iii)}~ \langle \eta_{b;t,s}|\eta_{b^\prime;t^\prime,s^\prime}\rangle=0\nonumber\\
			&\text{for}~t\neq t^\prime~\text{simply because}~ \langle vp+t|v^\prime p+t^\prime\rangle =0~\text{for}~0\leq v,v^\prime
			\leq p-1~\text{and}~t\neq t^\prime.\nonumber
		\end{align}
		(iv) All other inner products like $\langle \eta|\eta^\prime\rangle$ for $\eta\neq \eta^\prime$ are likely to be non zero and our interest is in
		$\text{tr}(P_\eta P_{\eta^\prime})=|\langle \eta|\eta^\prime\rangle|^2$ and we proceed to compute such numbers i.e., non trivial quantum mechanical
		overlaps in the order we like.\\ \noindent
		(v) $|\langle\eta_{0;t,s}|\eta_{b^\prime;t,s^\prime}\rangle|^2=\frac{1}{p}$ for $b^\prime \neq 0$.\\
		(vi)  $|\langle\eta_{0;t,s}|\eta_{r,a;c,s^\prime}\rangle|^2=\frac{1}{p^2}$.\\
		(vii) We next consider $r\neq r^\prime$.
		\begin{enumerate}
		\item[(a)]
		\begin{align}
			\langle \xi(r,c,t)|\xi(r^\prime,c^\prime,t)\rangle & =
			\frac{1}{p}\sum_{v=0}^{p-1}e^{\left(\frac{2\pi i}{p}v(c-rt)- \frac{2\pi i}{p}v
				(c^\prime-r^\prime t)\right)}\nonumber\\
			&=   \frac{1}{p}\sum_{v=0}^{p-1}e^{\left(\frac{2\pi iv}{p}((c-c^\prime)-(r-r^\prime)t)\right)}
		\end{align}
		Because $r^{\prime\prime}=r-r^\prime  \neq 0$, $ \{r^{\prime\prime} t~(\text{mod}~p) : 1\leq t\leq p-1\}$ $= \{ u: 1\leq u\leq p-1\}$. So there
		is a unique $t$, say $t_{r^{\prime\prime}}$ with $0\leq t_{r^{\prime\prime}}\leq p-1$ such that $(c-c^\prime)-r^{\prime\prime}t =0 ~(\text{mod}~p)$. \\
		\item[(b)] For $t\neq t_{r^{\prime\prime}}$,
		\begin{align}
			\langle \xi(r,c,t)|\xi(r^\prime,c^\prime,t)\rangle & =  \frac{1}{p}\sum_{v=0}^{p-1}e^{\left(\frac{2\pi iv}{p}((c-c^\prime)-(r-r^\prime)t)\right)}=0,
		\end{align}
		simply because $0< (c-c^\prime)-(r-r^{\prime})t~(\text{mod}~p)\leq p-1$.\\
		\item[(c)] For $t=  t_{r^{\prime\prime}}$
		\begin{align}
			\langle \xi(r,c,t)|\xi(r^\prime,c^\prime,t)\rangle & =  \frac{1}{p}\sum_{v=0}^{p-1} 1= 1.
		\end{align}
		In particular, $ \xi(r,c,t_{r^{\prime\prime}})$ and $\xi(r^\prime,c^\prime,t_{r^{\prime\prime}})$ are multiples of each other by scalars of modulus one. Indeed, they both coincide with $\xi_{t_{r^{\prime\prime}}}$ as in (3.13) above.
		\\
		\item[(d)] By (b) and (c) just above and (i) above
		\begin{align}
			\langle\eta(r,a;c,s)|\eta(r^\prime,a^\prime; c^\prime,s^\prime)\rangle = \overline{\beta(r,a; c,s,t_{r^{\prime\prime}})}
			\beta(r^\prime,a^\prime; c^\prime,s^\prime,t_{r^{\prime\prime}})
		\end{align}
		and therefore
		\begin{align}
			|\langle\eta(r,a;c,s)|\eta(r^\prime,a^\prime; c^\prime,s^\prime)\rangle|^2=\frac{1}{p^2}
		\end{align}
		\end{enumerate}
		\noindent
		(viii) For $a\neq a^\prime$, by (i) above
		\begin{align}
			\langle\eta(r,a;c,s)|\eta(r,a^\prime; c,s^\prime)\rangle& = \sum_{t=0}^{p-1}\overline{\beta(r,a;c,s,t)}
			\beta(r,a^\prime;c,s^\prime,t)\nonumber\\
			&=\frac{1}{p}\sum_{t=0}^{p-1} e^{\left(\frac{2\pi i}{p}(a-a^\prime)\frac{(p-t)t}{2}\right)} e^{\left(\frac{2\pi i}{p}(s-s^\prime)t\right)}.
		\end{align}
		Without loss of generality, we may take $a>a^\prime$. Because $p$ is odd, $(p-t)t$ is even for $0\leq t\leq p-1$. So we may replace $a-a^\prime$
		by $a-a^\prime+p$, if the need be, in the above expression. Also $a-a^\prime$ or $a-a^\prime+p$ is even, say $2v$ with $1\leq v\leq p-1$. Thus the
		expression on the RHS becomes
		\begin{align}
			& \frac{1}{p}\sum_{t=0}^{p-1} e^{\left(\frac{2\pi i}{p}v(p-t)t\right)} e^{\left(\frac{2\pi i}{p}(s-s^\prime)t\right)}\nonumber\\
			&= \frac{1}{p}\sum_{t=0}^{p-1} e^{\left(\frac{-2\pi i}{p}vt^2\right)} e^{\left(\frac{2\pi i}{p}(s-s^\prime)t\right)}
			&=\frac{1}{p}\sum_{t=0}^{p-1} e^{\left(\frac{2\pi i}{p}((p-v)t^2+(s-s^\prime)t\right)}
		\end{align}
		Now
		\begin{align}
			|\langle\eta(r,a ;c,s)|\eta(r,a^\prime; c,s^\prime)\rangle |^2&=
			\frac{1}{p^2}\sum_{t,t^\prime=0}^{p-1} e^{\left(\frac{2\pi i}{p}v(t^{\prime 2}-t^2)\right)} e^{\left(\frac{2\pi i}{p}(s-s^\prime)(t-t^\prime)\right)}
			\nonumber\\
			&= \frac{1}{p^2}\sum_{t,t^\prime=0}^{p-1} e^{\left(\frac{2\pi i}{p}v(t^\prime+t)\right)(t^\prime-t)} e^{\left(\frac{2\pi i}{p}(s^\prime-s)(t^\prime-t)\right)}
			\nonumber\\
			&= \frac{1}{p^2}\sum_{\stackrel{0\leq t^\prime+t~(\text{mod}~p)\leq p-1}{0\leq t^\prime-t~(\text{mod}~p)\leq p-1}}^{p-1}
			e^{\left(\frac{2\pi i}{p}v(t^\prime+t)\right)(t^\prime-t)} e^{\left(\frac{2\pi i}{p}(s^\prime-s)(t^\prime-t)\right)}
		\end{align}
		Because $1\leq v ~ \leq p-1$, 
		there is a unique
		$u$ with $0\leq u~\leq p-1$ and $uv+(s^\prime-s)=0~(\text{mod}~p)$. So, the expression in (3.48) 
		
		\begin{align}
			&= \frac{1}{p^2}\Big{[}\sum_{\stackrel{0\leq t^\prime-t~(\text{mod}~p)\leq p-1}{ t^\prime +t=u~(\text{mod}~p)}}
			e^{\left(\frac{2\pi i}{p}(v(t^\prime+t)+(s^\prime-s))(t^\prime-t)\right)} \nonumber\\&+
			\sum_{\stackrel{0\leq t^\prime-t~(\text{mod}~p)\leq p-1}{ t^\prime +t\neq u~(\text{mod}~p)}} e^{\left(\frac{2\pi i}{p}(
				v(t^\prime+t)+(s^\prime-s))(t^\prime-t)\right)}\Big
			{]}
		\end{align}
		Now, for $t^\prime+t \neq u~(\text{mod}~ p)$ , $e^{\left(\frac{2\pi i}{p}(v(t+t^\prime)+(s^\prime-s))\right)}$ is a $p^{\text{th}}$ root of unity other than $1$, so
		\begin{align}
			\sum_{0\leq t^\prime-t ~(\text{mod}~p)~\leq p-1} e^{\left(\frac{2\pi i}{p}(v(t^\prime+t)+(s^\prime-s))(t^\prime-t)\right)}=0.
		\end{align}
		And for $t^\prime +t =u ~(\text{mod}~ p)$ , $e^{\left(\frac{2\pi i}{p}(v(t^\prime+t)+(s^\prime-s))(t^\prime-t)\right)}$ is 1 for
		$0\leq t^\prime -t ~(\text{mod}~p) \leq p-1$. So the first sum is $p$. Therefore $ |\langle\eta(r,a^\prime;c,s)|\eta(r,a^\prime; c,s^\prime)\rangle |^2=
		\frac{1}{p}$.
		\vskip2mm
		\noindent
		(ix) We now come to the case $0\neq b\neq b^\prime\neq 0$.
		
		(a) We first note that because $p$ is odd, either $b$ or $b+p$ is even and, therefore, has the form $2v$ with $0<v\leq p-1$. Now for
		$0\leq k\leq p-1$, $|kbp+t\rangle = |k(b+p)p+t\rangle$ and
		$\exp\left(\frac{2\pi i}{p}(\frac{k(k-1)}{2}b)\right)= \exp\left(\frac{2\pi i}{p}(\frac{k(k-1)}{2}(b+p)\right)$ as well. So we may
		replace $b$ by $2v$ in the expression for $\eta(b;t,s)$ in (ii)(c) of the proof of Theorem \ref{t3.6.2} above by
		\begin{align}
			\eta(2v;t,s)=\frac{1}{\sqrt{p}}\sum_{k=0}^{p-1}e^{\left(\frac{2\pi i}{p}(k-1)kv\right)}
			e^{\left(-\frac{2\pi i}{p}ks\right)}|2kvp+t\rangle
		\end{align}
		Similar arguments apply to $b^\prime$ with $v$ replaced by $v^\prime$ satisfying $0<v^\prime\leq p-1$ and $2v^\prime =b^\prime$ or $b^\prime+p$ according
		as $b^\prime$ is even or odd.
		
		(b) Because $1\leq v\neq v^\prime \leq p-1$,
		there is a unique $k_0$ with $1\leq k_0\leq p-1$ satisfying $v= k_0 v^\prime~(\text{mod}~p)$. Then for  $0\leq k\leq p-1$, $kv =kk_0v^\prime
		~(\text{mod}~p)$ and, therefore, $|2kvp+t\rangle=|2kk_0v^\prime p+t\rangle$. So
		\begin{align}
			\langle \eta(b;t,s)|\eta(b^\prime;t,s^\prime\rangle &=\langle \eta(2v;t,s)|\eta(2v^\prime;t,s^\prime\rangle\nonumber\\
			&=\frac{1}{p} \sum_{k=0}^{p-1} e^{\left(\frac{2\pi i}{p}(-(k-1)kv+(kk_0-1)kk_0 v^\prime +ks-kk_0 s^\prime)\right)}\nonumber\\
			&= \frac{1}{p} \sum_{k=0}^{p-1} e^{\left(\frac{2\pi i}{p}((k^2(k_0-1)v+ k(s-k_0 s^\prime )\right)}
		\end{align}
		The sum has the same form as that in (viii) above (with $p-v$ replaced by $(k_0-1)v$ and $s-s^\prime$ by $s-k_0s^\prime$). So
		$ |\langle \eta(b;t,s)|\eta(b^\prime,t,s^\prime\rangle|^2= \frac{1}{p}$.
		\vskip2mm
		\noindent
		(x) Finally, for $b\neq 0$, we consider the quantum mechanical overlap $\text{tr}(\mathcal{P}(r,a;c,s)$ $P(b;t,s^\prime))$.
		
		(a) We first note that for $t^\prime \neq t$, $\langle\xi(r;c,t^\prime)|\eta(b;t,s^\prime)\rangle=0$.
		
		(b)
		\begin{align}
			\langle\xi(r;c,t)|\eta(b;t,s^\prime)\rangle=\frac{1}{p}  \sum_{k=0}^{p-1} e^{\left(\frac{2\pi i}{p}kb(c-rt)\right)}
			e^{\left(\frac{2\pi i}{p}\frac{k(k-1)}{2}b\right)}e^{\left(\frac{2\pi i}{p}ks^\prime\right)}
		\end{align}
		We argue as in (ix) above and replace $b$ by $2v$ with $v=\frac{1}{2}b$ if $b$ is even and $\frac{1}{2}(b+p)$, if $b$ is odd. Then the above expression becomes
		\begin{align}
			&\frac{1}{p}\sum_{k=0}^{p-1} e^{\left(\frac{2\pi i}{p} 2kv(c-rt)\right)}
			e^{\left(\frac{2\pi i}{p} k(k-1)v\right)} e^{\left(\frac{2\pi i}{p} ks^\prime\right)}
			=\frac{1}{p}\sum_{k=0}^{p-1} e^{\left( k(\frac{2\pi i}{p} (2v(c-rt)+(k-1)v+s^\prime)\right)}
		\end{align}
		Because $0<v\leq p-1$, there is a unique $u$ with $0\leq u \leq p-1$ and $uv=s^\prime~(\text{mod}~p)$. So the above expression, viz.
		$\langle\xi(r;c,t)|\eta(b;t,s^\prime)\rangle$
		\begin{align}
			=\frac{1}{p}\sum_{k=0}^{p-1} e^{\left( k(\frac{2\pi i}{p}v (2(c-rt)+(k-1)+u)\right)}
		\end{align}
		So
		\begin{align}
			&|\langle\xi(r;c,t)|\eta(b;t,s^\prime)\rangle|^2
			=\frac{1}{p^2}\sum_{0\leq k,k_1 \leq p-1} e^{\left(\frac{2\pi i}{p}v(k-k_1)(2(c-rt)+(k+k_1)-1+u)\right)}\nonumber\\
			& =\frac{1}{p^2}\Big{[}\sum_{\stackrel{0\leq k-k_1~(\text{mod}~p)\leq p-1}{k+k_1+2(c-rt)-1+u =0~(\text{mod}~p)}}~1\nonumber\\
			&+ \sum_{\stackrel{0\leq k-k_1~(\text{mod}~p)\leq p-1 }{0\leq k+k_1~(\text{mod}~ p)}{\neq -2(c-rt)+1-u~(\text{mod}~p) \leq p-1}}
			e^{\left((k-k_1)(\text{mod}~p)(\frac{2\pi i}{p}v(2(c-rt)+(k+k_1)-1+u)\right)}\Big{]}\nonumber\\
			& = \frac{1}{p}
		\end{align}
		simply because $0<v(2(c-rt)+(k+k_1)-1+u)~(\text{mod}~p) \leq p-1$ in each term of the second summations and, therefore, arguments like those in (viii) above
		apply.
	\end{example}
	We may combine the above two examples into a single theorem below, if we like.
	\begin{theorem}\label{t3.6.5} The quantum mechanical overlaps $\text{tr}(\mathcal{P}_\eta \mathcal{P}_{\eta^\prime})$ arising from $\mathcal{P}$ as in
		part (iii) of Theorem \ref{t3.6.2} above for $d=p^2$, $p$ any prime number have values $0,{1}/{p},{1}/{p^2}, 1$.
	\end{theorem}
	\vskip2mm
	\subsubsection{\it Case d=6}~We now come to the simplest case of a non-square composite number of Example 2.3.
	\begin{example}\label{e3.6.6}
		We refer to Example \ref{e3.2.5} (b) and Figure 4. We first note that all the 12 ${\bf W}$-MASS's, are needed to cover ${\bf W}$. Next, there are
		two types of overlaps, viz., $\lambda_{(2,0)}=\lambda_{(4,0)}$, $\lambda_{(0,2)}=\lambda_{(0,4)}$, $\lambda_{(2,2)}=\lambda_{(4,4)}$,
		$\lambda_{(4,2)}=\lambda_{(2,4)}$ of the first type, and,  $\lambda_{(3,0)}, \lambda_{(0,3 )}, \lambda_{(3,3)}$ of the second type (with
		$(0,0)$ taken away from all). To be specific, the following Table in Figure 6 gives a good idea of the situation where we have written only convenient
		generators instead of the ${\bf W}$-MASS's or the overlaps.
		\begin{figure}[H]
			\centering
			\begin{tabular}{|c?c|c|c|} \hline
				& (3,0)&(0,3)&(3,3)\\  \thickhline
				(2,0)  & (1,0)       & (4,3)          & (1,3)\\
				or    &             &   or           &       \\
				(4,0) &             &  (2,3)         &      \\
				&$\bf{V}_1$     & $\bf{V}_8$       & $\bf{V}_6$   \\ \hline
				(0,2)  & (3,4)      & (0,1)          & (3,1)\\
				or    &  or         &                &       \\
				(0,4) & (3,2)       &                &      \\
				&$\bf{V}_9 $    & $\bf{V}_2 $      &$ \bf{V}_7 $  \\ \hline
				(2,2)  & (1,4)       & (4,1)          & (1,1)\\
				or   &   or        &                &       \\
				(4,4) &  (5,2)      &                &      \\
				&$\bf{V}_{10}$  & $\bf{V}_{11}$    & $\bf{V}_3$   \\ \hline
				(2,4)  & (1,2)       & (4,5)          & (1,5)\\
				or    &  or         &  or            &  or     \\
				(4,2) &  (5,4)      &  (2,1)         &  (5,1)    \\
				&$\bf{V}_4 $    & $\bf{V}_5$       &$ \bf{V}_{12}$  \\ \hline
				
			\end{tabular}
			\caption{}
		\end{figure}
	\end{example}
	We now develop the optimal simultaneous constraint technique in different steps.\vskip2mm\noindent
	(i) We start with the simplest unitary $U_{(1,0)}$ given by $U_{(1,0)}|k\rangle= e^{\frac{\pi i}{3}k}|k\rangle$ for $0\leq k\leq 5$.
	It generates ${\bf W}$-MASS ${\bf V}_1$.
	\begin{enumerate}
	\item[{(a)}] $U_{(1,0)}$ has unit eigenvectors $\{\xi_{1,k}=|k\rangle, 0\leq k\leq 5\}$ with corresponding eigenvalues $\{e^{\frac{\pi i}{3}k}, 0\leq k\leq 5\}$
	i.e., $\{1, -\omega^2, \omega,-1, \omega^2,-\omega\}$.
	\item[{(b)}] $U_{(2,0)}= U_{(1,0)}^2$ has eigenvalues $1, \omega=e^{\frac{2\pi i}{3}}, \omega^2 =e^{\frac{4\pi i}{3}}$ each of multiplicity 2, with
	the corresponding eigenspaces the span  $E_{1,0}$ of $\xi_{1,0}=|0\rangle$ and $\xi_{1,3}=|3\rangle$, the span  $E_{1,1}$ of $\xi_{1,1}=|1\rangle$
	and $\xi_{1,4}= |4\rangle$ and the span $E_{1,2}$ of $\xi_{1,2}=|2\rangle$ and $\xi_{1,5}=|5\rangle$ respectively
	\item[{(c)}] $U_{(3,0)}= U_{(1,0)}^3$ has eigenvalues $1$ and $-1$  each of multiplicity 3, with
	the corresponding eigenspaces the span  $G_{1,0}$ of $\xi_{1,0}=|0\rangle$,  $\xi_{1,2}=|2\rangle$ and $\xi_{1,4}=|4\rangle$, and
	the span  $G_{1,1}$ of $\xi_{1,1}=|1\rangle$, $\xi_{1,3}= |3 \rangle$ and $\xi_{1,5}=|5\rangle$ respectively.
	
	\item[{(d)}] Setting $P_{1,k} = |\xi_{1,k}\rangle \langle \xi_{1,k}|= |k\rangle\langle k|$ for $0\leq k\leq 5$ and $Q_{1,0}, Q_{1,1}, Q_{1,2}$ projections
	on $E_{1,0}, E_{1,1}, E_{1,2}$ we have $Q_{1,0}=P_{1,0}+P_{1,3}, ~Q_{1,1}=P_{1,1}+P_{1,4}$ and $Q_{1,2}=P_{1,2}+P_{1,5}$.
	
	\item[{(e)}] For the projections $R_{1,0}$ and $R_{1,1}$ on $G_{1,0}$ and $G_{1,1}$ respectively, we have $R_{1,0}=P_{1,0}+P_{1,2}+ P_{1,4}$
	and $R_{1,1}=P_{1,1}+P_{1,3}+ P_{1,5}$.
	
	\item[{(f)}] We may combine (d) and (e) into two equivalent useful representations in Figures 7 and 8.
	\begin{figure}[H]
		\centering
		\begin{tabular}{|c?c|c|c|}\hline
			&            &          & Row Sum\\ \thickhline
			& $P_{1,0}$ & $P_{1,3}$& $Q_{1,0}$\\ \hline
			&$P_{1,4}$ & $P_{1,1}$ & $Q_{1,1}$\\\hline
			&$P_{1,2}$ & $P_{1,5}$ & $Q_{1,2}$\\\hline
			Column sum & $R_{1,0}$ & $R_{1,1}$ & $I_6$\\\hline
		\end{tabular}
		\caption{}
	\end{figure}
	\begin{figure}[H]
		\centering
		\begin{tabular}{|c?c|c|c|} \hline
			&            &          &Row sum\\ \thickhline
			& $P_{1,0}$ & $P_{1,3}$& $Q_{1,0}$\\ \hline
			&$P_{1,2}$ & $P_{1,5}$ & $Q_{1,2}$\\\hline
			&$P_{1,4}$ & $P_{1,1}$ & $Q_{1,1}$\\\hline
			Column sum &$R_{1,0}$ & $R_{1,1}$ & $I_6$\\\hline
		\end{tabular}
		\caption{}
	\end{figure}

	\item[{(g)}] The elementary measurement $\mathcal{P}_1= \{P_{1,k}: 0\leq k\leq 5\}$ is ${\bf Q}_1$- and ${\bf R}_1$-constrained, where ${\bf Q}_1=
	\{Q_{1,0}, Q_{1,1},  Q_{12}\}$ and ${\bf R}_1= \{R_{1,0},R_{1,1}\}$.
	
	In view of the Tables in Figures 7 and 8, the simultaneous constraints for $\mathcal{P}_1$ have a nice useful relationship.

	\item[{(h)}] Figures 7 and 8 also show that $P_{1,k}$'s can be recovered from $Q_{1,r}$'s and $R_{1,r^\prime}$'s in the sense that
	$P_{1,0}=Q_{1,0}R_{1,0}$, $P_{1,1}=Q_{1,1}R_{1,1}$, $P_{1,2}=Q_{1,2}R_{1,0}$, $P_{1,3}=Q_{1,0}R_{1,1}$, $P_{1,4}=Q_{1,1}R_{1,0}$,
	and $P_{1,5}=Q_{1,2}R_{1,1}$,
	\end{enumerate}
	(ii) We now come to the unitary $U_{(0,1)}$ given by $U_{(0,1)}|k\rangle =|k+1\rangle$ for $0\leq k\leq 5$. It generates the ${\bf W}$-MASS
	${\bf V}_2$.
	\begin{enumerate}
	\item[{(a)}] Simple computations give that  $U_{(0,1)}$ has unit eigenvectors \\
	$\{ \xi_{2,k}= \frac{1}{\sqrt{6}}\sum_{s=0}^{5} e^{-\left(ks\frac{\pi i}{3}\right)}|s\rangle
	: 0\leq k \leq 5\}$ with corresponding eigenvalues \\
	$\{e^{\left(\frac{k\pi i}{3}\right)}: 0\leq k \leq 5\}$.
	
	\item[{(b)}] to (h)  Items (b) to (h) in (i) above now have obvious analogues with $\xi_{1,k}, P_{1,k}, E_{1,r}$, $ Q_{1,r}, G_{1,r^\prime}$ and $R_{1,r^\prime}$
	replaced by $\xi_{2,k}, P_{2,k}, E_{2,r}, Q_{2,r}, G_{2,r^\prime}$ and $R_{2,r^\prime}$ respectively
	for $0\leq k\leq 5$, $0\leq r\leq 2$, $0\leq r^\prime \leq 1$ and $\mathcal{P}_1, {\bf Q}_1, {\bf R}_1 $ respectively by
	$\mathcal{P}_2, {\bf Q}_2, {\bf R}_2 $ respectively.
	\end{enumerate}
	(iii) We consider the unitary $U_{(1,1)}$ given by $U_{(1,1)}|k\rangle =e^{\left(\frac{\pi i}{3}k\right)}|k+1\rangle$ for
	$0\leq k\leq 5$. It generates the ${\bf W}$-MASS
	${\bf V}_3$.
	\begin{enumerate}
	\item[{(a)}] Simple computations give that  $U_{(1,1)}$ has unit eigenvectors \\
	$\{ \xi_{3,k}= \frac{1}{\sqrt{6}}\sum_{s=0}^{5}
	e^{\left(-\frac{(2k+1)s\pi i}{6}\right)}  e^{\left( \frac{\pi i s(s-1)}{6}\right)}|s\rangle
	: 0\leq k \leq 5\}$ \\
	with corresponding eigenvalues $\{e^{\left(\frac{(2k+1)\pi i}{6}\right)} : 0\leq k \leq 5 \}
	= e^{\left(\frac{\pi i}{6}\right)}\{ e^{\left(\frac{k\pi i}{3}\right)} : 0\leq k \leq 5  \}$.
	
	\item[{(b)}] By Proposition \ref{p3.5.1} (or direct computation) $U_{(2,2)}=e^{\left(-\frac{\pi i}{3}\right)}U_{(1,1)}^2$. So $U_{(2,2)}$ has eigenvalues
	$ 1, \omega= e^{\left(\frac{2\pi i}{3}\right)}, \omega^2= e^{\left(\frac{4\pi i}{3}\right)} $ each of multiplicity 2 with corresponding
	eigenspaces the span $E_{3,0}$ of
	$\xi_{3,0}$ and $\xi_{3,3}$, the span  $E_{3,1}$ of $\xi_{3,1}$ and $\xi_{3,4}$, and the span of $E_{3,2}$ of $\xi_{3,2}$ and
	$\xi_{3,5}$ respectively.
	
	\item[{(c)}] By Proposition \ref{p3.5.1} (or, direct computations ), $U_{(3,3)}= - U_{(1,1)}^3$. So $U_{(3,3)}$ has eigenvalues $-i$ and $i$ each of multiplicity
	$3$ with corresponding eigenspaces the span  $G_{3,0}$ of $\xi_{3,0}$, $\xi_{3,2}$ and $\xi_{3,4}$, and the span $G_{3,1}$ of
	$\xi_{3,1}$, $\xi_{3,3}$ and $\xi_{3,5}$ respectively.
	
	\item[{(d)}] to (g) Items (d) to (g) in (i) above now have obvious analogues with $\xi_{1,k}$, $P_{1,k}$, $E_{1,r}$, $Q_{1,r}$, $G_{1, r^\prime}$
	and $R_{1,r^\prime}$ replaced by $\xi_{3,k}$, $P_{3,k}$, $E_{3,r}$, $Q_{3,r}$, $G_{3, r^\prime}$
	and $R_{3,r^\prime}$ respectively for $0\leq k\leq 5,~0\leq r\leq 2, ~0\leq r^\prime \leq 1$ and $\mathcal{P}_1$, ${\bf Q}_1$ and ${\bf R}_1$ replaced by
	$\mathcal{P}_3$, ${\bf Q}_3$ and ${\bf R}_3$ respectively.
	\end{enumerate}
	(iv) Now one constraint determined by $\lambda_{(4,2)} = \lambda_{(2,4)}$ remains to be explicitly expressed. The easiest way is
	to consider the unitary $U_{(1,2)}$ given by $U_{(1,2)}|k\rangle = e^{\left(\frac{\pi i k}{3}\right)} |k+2\rangle $ for $0\leq k\leq 5$.
	It generates the ${\bf W}$-MASS ${\bf V}_4$.
	\begin{enumerate}
	\item[{(a)}] Simple computations give that $U_{(1,2)}$ has unit eigenvectors
	\begin{align}
		\xi_{4,0} &=\frac{1}{\sqrt{3}}( |0\rangle +|2\rangle + e^{\left(\frac{2\pi i}{3}\right)}|4\rangle)\nonumber\\
		\xi_{4,2}& =\frac{1}{\sqrt{3}}( |0\rangle + e^{\left(-\frac{2\pi i}{3}\right)}|2\rangle + e^{\left(-\frac{2\pi i}{3}\right)}|4\rangle)\nonumber\\
		&=\frac{1}{\sqrt{3}}( |0\rangle + e^{\left(\frac{4pi i}{3}\right)}|2\rangle + e^{\left(\frac{4\pi i}{3}\right)}|4\rangle),\nonumber\\
		\xi_{4,4}&= \frac{1}{\sqrt{3}}( |0\rangle +e^{\left(\frac{2\pi i}{3}\right)}|2\rangle + |4\rangle)
	\end{align}
	with corresponding eigenvalues as $1, e^{\left(\frac{2\pi i}{3}\right)}, e^{\left(\frac{4\pi i}{3}\right)}$ respectively, on the one hand,
	and
	\begin{align}
		\xi_{4,1} &=\frac{1}{\sqrt{3}}( |1\rangle +|3\rangle + e^{\left(\frac{2\pi i}{3}\right)}|5\rangle)\nonumber\\
		\xi_{4,3}& =\frac{1}{\sqrt{3}}( |1\rangle + e^{\left(-\frac{2\pi i}{3}\right)}|3\rangle + e^{\left(-\frac{2\pi i}{3}\right)}|5\rangle)\nonumber\\
		\xi_{4,5}&= \frac{1}{\sqrt{3}}( |0\rangle +e^{\left(\frac{2\pi i}{3}\right)}|2\rangle + |5\rangle)
	\end{align}
	with corresponding eigenvalues
	$e^{\left(\frac{\pi i}{3}\right)},e^{\pi i }=-1,  e^{\left(\frac{5\pi i}{3}\right)}$ respectively, on the otherhand.
	
	\item[{(b)}]  By Proposition \ref{p3.5.1} (or, direct computations ), $U_{(2,4)}= e^{\left(-\frac{2\pi i}{3}\right)} U_{(1,2)}^2$,
	so $U_{(2,4 )}$ has eigenvalues $ e^{\left(\frac{4\pi i}{3}\right)}, 1$ and $ e^{\left(\frac{2\pi i}{3}\right)}$  each of multiplicity
	$2$ with corresponding eigenspaces the span of $E_{4,0}$ of $\xi_{4,0}$ and  $\xi_{4,3}$, the span $E_{4,1}$ of
	$\xi_{4,1}$ and  $\xi_{4,4}$, and the span $E_{4,2}$ of $\xi_{4,2}$ and $\xi_{4,5}$ respectively.
	
	\item[{(c)}]  By Proposition \ref{p3.5.1} (or, direct computations ), $U_{(3,0)}=  U_{(1,2)}^3$.
	So $U_{(3,0 )}$ has eigenvalues $1$ and $-1$  each of multiplicity
	$3$, with eigenspaces the span $G_{4,0}$ of $\xi_{4,0}, \xi_{4,2}$ and  $\xi_{4,4}$, and the span $G_{4,1}$ of
	$\xi_{4,1}, \xi_{4,3}$ and  $\xi_{4,5}$ respectively. This is in line with (i)(c) because $G_{1,0}=G_{4,0}$ and  $G_{1,1}=G_{4,1}$.
	
	\item[{(d)}] to (h) Items (d) to (h) in (i) above now have obvious analogues with changes as indicated in (iii) (d) to (h) above.
	
	\item[{(i)}] The choice of $U_{1,2}$ that generates ${\bf V}_4$ is easier than that of generators coming from any other in the fourth row.
	
	\item[{(j)}] We have an interesting alternative provided by the technique of determining $U_{(m,n)}$ from $U_{(pm,pn)}$ for $p$, a prime as
	indicated in Remark \ref{r3.5.5} or the proof of Theorem \ref{t3.6.2}. We will use the information in (i)(c) about $U_{(3,0)}$, which is equal to
	$U_{(1,2)}^3$  as well. We note that $U_{(1,2)}G_{1,0}=G_{1,0}$ and $U_{(1,2)}G_{1,1} = G_{1,1}$ . So
	$\left(U_{(1,2)}/G_{1,0}\right)^\ast = U_{(1,2)}^\ast/G_{1,0}$ and
	$\left(U_{(1,2)}/G_{1,1}\right)^\ast = U_{(1,2)}^\ast/G_{1,1}$. So $U_{(1,2)}$ restricted to $G_{1,0}$ and to $G_{1,1}$ have three
	eigenvalues each with unit eigenvectors  present in $G_{1,0}$ and $G_{1,1}$ respectively. Thus, we may solve for scalars
	$\alpha$, $\beta$, $\mu_k$'s and $\nu_k$'s that satisfy
	\begin{align}
		U_{(1,2)} \left( \sum_{k=0,2,4}\mu_k~|k\rangle\right)& = \alpha \left( \sum_{k=0,2,4}\mu_k~|k\rangle\right)\nonumber \\
		U_{(1,2)} \left( \sum_{k=1,3,5}\nu_k~|k\rangle\right)& = \beta \left( \sum_{k=1,3,5}\nu_k~|k\rangle\right)~\text{and}\nonumber, \\
		|\mu_0|^2+ |\mu_2|^2+|\mu_4|^2 &= 1=|\nu_1|^2+ |\nu_3|^2+|\nu_5|^2.
	\end{align}
	We obtain the system as in part (a) above for $\mu_0\geq 0, \nu_0 \geq 0$.
	\end{enumerate}
	We now have all the constraints in hand, It is as if $\mathcal{P}_1, \mathcal{P}_2, \mathcal{P}_3, \mathcal{P}_4 $ are working as anchors for them.
	
	\noindent
	(v) We emphasize the facts derived from Proposition \ref{p3.5.1} and Corollary \ref{c3.5.2} used in (i) to (iv) above for further use as well. We refer to
	the Table in Figure 6.
	\begin{enumerate}
	\item[{(a)}] For $(m,n)$ in the first two columns, $mn$ is even and, therefore, $U_{(3m,3n)}= U_{(m,n)}^3, U_{(m,n)}^6=I$ and the eigenvalues of $U_{(m,n)}$
	are $e^{\left(\frac{k\pi i}{3}\right)}, k=0, 1, \cdots, 5$ noted in this order; further, eigenvalues of
	$U_{(3m,3n)}$ are $1$ and $-1$ and occurring alternately in the corresponding listing.
	
	\item[{(b)}] For $(m,n)$ in the third column, $mn$ is odd and, therefore, $U_{(3m,3n)}=- U_{(m,n)}^3, U_{(m,n)}^6=-I$ and the eigenvalues of $U_{(m,n)}$
	are $ e^{\left(\frac{\pi i}{6}\right)}e^{\left(\frac{k\pi i}{3}\right)}, k=0, 1, \cdots, 5$ noted in this order; further, eigenvalues of
	$U_{(3m,3n)}$ are $-i$ and $i$ occurring alternately in the corresponding listing.
	
	\item[{(c)}] For all $ m,n \in \ZZ_6$, $U_{(2m,2n)}=  e^{\left(-\frac{\pi i}{3}mn\right)} U_{(m,n)}^2$. So we have the following cases.
	\begin{enumerate}
	\item[{($\alpha$)}] $U_{(2m,2n)}= U_{(m,n)}^2$ for $mn$ a multiple of $6$. i.e., $(m,n)$ in the first and  second row as well as column. So as noted
	in (a) above, the eigenvalues of $U_{(m,n)}$ for such $(m,n)$'s are  $e^{\left(\frac{k\pi i}{3}\right)}, k=0, 1, \cdots, 5$ noted in this order. So the eigenvalues
	of $U_{(2m,2n)}$ are $1, e^{\left(\frac{2\pi i}{3}\right)}, e^{\left(\frac{4\pi i}{3},\right)}, 1, e^{\left(\frac{2\pi i}{3}\right)},  e^{\left(\frac{4\pi i}{3}\right)}$
	in the corresponding listing . We note that $(2m,2n)=(2,0)~\text{or}~(0,2)$ according as $(m,n)= (4,3)~\text{or}~(3,4)$. This looks
	after ${\bf V}_8$ and ${\bf V}_9$.
	
	\item[{($\beta$)}] $U_{(2m,2n)}= - U_{(m,n)}^2$ for $mn$ an odd multiple of $3$, i.e., for $(m,n)=(1,3)$ or$ (3,1)$ which generate
	${\bf V}_6$ and ${\bf V}_7$ respectively.

	These are in the third column. So from (b) above, the eigenvalues of $U_{(m,n)}$ are $e^{\left(\frac{\pi i}{6}\right)}e^{\left(\frac{k\pi i}{3}\right)},
	~k=0, 1,\cdots, 5$ noted in this order, and, therefore, for $U_{2m,2n}$, the eigenvalues are
	$ e^{\left(\frac{4\pi i}{3}\right)}, 1, e^{\left(\frac{2\pi i}{3}\right)},  e^{\left(\frac{4\pi i}{3}\right)},1,e^{\left(\frac{2\pi i}{3}\right)}$
	for the corresponding listing. So a cyclic permutation is needed to arrive at the listing for $U_{(2,0)}$ and $U_{(0,2)}$ in (i) and (ii) above
	which happen to be $U_{(2m,2n)}$ for $(m,n)=(1,3)~\text{and}~(3,1)$ respectively.
	
	\item[{($\gamma$)}] $U_{(2m,2n)}= e^{\left(-\frac{4\pi i}{3}\right)} U_{(m,n)}^2 = e^{\left(\frac{2\pi i}{3}\right)}U_{(m,n)}^2$
	for $mn=4$, i.e., $(m,n)=(1,4)$ or $(4,1)$, which generate
	${\bf V}_{10}$ and ${\bf V}_{11}$ respectively. The eigenvalues of $U_{(m,n)}$ for these cases are $e^{\left(\frac{k\pi i}{3}\right)}, ~k=0, 1,\cdots, 5$ noted in this order
	and therefore, for $U_{(2m,2n)}$ are $e^{\left(\frac{2\pi i}{3}\right)}e^{\left(\frac{2k\pi i}{3}\right)}$, $k=0,1\cdots,5$, i.e,
	$ e^{\left(\frac{2\pi i}{3}\right)}, e^{\left(\frac{4\pi i}{3}\right)}, 1, e^{\left(\frac{2\pi i}{3}\right)}, e^{\left(\frac{4\pi i}{3}\right)},1$
	for the correspondig listing. Now $U_{(2m,2n)}=U_{(2,2)}$ for both these $(m,n)$ and the listing of eigenvalues for that as noted
	in (iii) above is $1, e^{\left(\frac{2\pi i}{3}\right)}, e^{\left(\frac{4\pi i}{3}\right)}, 1,
	e^{\left(\frac{2\pi i}{3}\right)}, e^{\left(\frac{4\pi i}{3}\right)}$. So a double cyclic permutation is called for.
	
	\item[{($\delta$)}] $U_{(2m,2n)}= e^{\left(-\frac{20\pi i}{3}\right)} U_{(m,n)}^2 = e^{\left(\frac{4\pi i}{3}\right)}U_{(m,n)}^2$
	for $mn=20$, i.e. for  $(m,n) =(4,5)$, which generates
	${\bf V}_{5}$. Eigenvalues of $U_{(m,n)}$ are  $e^{\left(\frac{\pi i}{6}\right)}e^{\left(\frac{k\pi i}{3}\right)}, ~k=0, 1,\cdots, 5$, noted in this
	order
	and therefore, for $U_{(2m,2n)}$, the eigenvalues are
	$ e^{\left(\frac{4\pi i}{3}\right)}, 1,  e^{\left(\frac{2\pi i}{3}\right)}, e^{\left(\frac{4\pi i}{3}\right)},1$, $e^{\left(\frac{2\pi i}{3}\right)}$
	for the corresponding listing. Now $U_{(2m,2n)}=U_{(2,4)}$ and the eigenvalues of $U_{(2,4)}$ are listed in the same order for $U_{(2,4)}$ in
	(iv) above.
	
	\item[{($\epsilon$)}] We now come to the only remaining case of $mn=5$, i.e., $(m,n)=(1,5)~\text{or}~(5,1)$, which generate ${\bf V}_{12}$. For this
	$U_{(2m,2n)}= e^{\left(-\frac{5\pi i}{3}\right)} U_{(m,n)}^2 = e^{\left(\frac{\pi i}{3}\right)}U_{(m,n)}^2$. By (b) above, eigenvalues of
	$U_{(1,5)}$ are $ e^{\left(\frac{\pi i}{6}\right)} e^{\left(\frac{k\pi i}{3}\right)}$, $k=0,1,\cdots,5$. So, for $U_{(2,4)}= U_{(2m,2n)}$ the
	eigenvalues are
	$e^{\left(\frac{2\pi i}{3}\right)}, e^{\left(\frac{4\pi i}{3}\right)}, 1$, $
	e^{\left(\frac{2\pi i}{3}\right)}, e^{\left(\frac{4\pi i}{3}\right)},1$ for the corresponding listing. But, as per the listing in (iv), the
	eigenvalues of $U_{(2,4)}$ are $e^{\left(\frac{4\pi i}{3}\right)}, 1,  e^{\left(\frac{2\pi i}{3}\right)},
	e^{\left(\frac{4\pi i}{3}\right)},1, e^{\left(\frac{2\pi i}{3}\right)}$. So a cyclic permutation is required.
	\end{enumerate}
	\end{enumerate}
	(vi) 
	\begin{enumerate}
	 \item[{(a)}] For
	 all the remaining $(m,n)$'s or, for that matter, ${\bf W}$-MASS's, ${\bf V}_j$ with $j\geq 5$, we decide to follow the listing of
	eigenvalues of $U_{(m,n)}$ for $(m,n)$ in the first two columns in Figure 6 as $e^{\left(\frac{k\pi i}{3}\right)}, k=0,1,\cdots,5$ and
	for $(m,n)$ in the third column in Figure 6 as $e^{\left(\frac{\pi i}{6}\right)}e^{\left(\frac{k\pi i}{3}\right)}, k=0,1,\cdots,5$. Take
	any such $(m,n)$ : let ${\bf V}_j$ be the corresponding ${\bf W}$-MASS generated by $(m,n)$. As done for $j=1,2,3,4$ in (i) to (iv) above,
	we can take orthonormal systems $\{\xi_{j,k}: k=0,1,\cdot,5\}$ for $U_{(m,n)}$ with corresponding listing of  eigenvalues. We can consider
	the analogues of $P_{1,k}$, $E_{1,r}$, $Q_{1,r}$, $G_{1, r^\prime}$
	and $R_{1,r^\prime}$ replaced by $P_{j,k}$, $E_{j,r}$, $Q_{j,r}$, $G_{j, r^\prime}$
	and $R_{j,r^\prime}$  by replacing $\xi_{1,k}$'s by $\xi_{j,k}$'s for $0\leq k\leq 5, 0\leq r\leq 2, 0\leq r^\prime \leq 1$. Next, we can go
	ahead with analogues of  $\mathcal{P}_j$, ${\bf Q}_j$ and ${\bf R}_j$  as well. But the difference lies in the eigenvalues or their listing
	as noted in (v) (c) and (b) above for analogues of (i) (b) and (i) (c)  respectively, though analogues of (d) to (h) in (i) are
	available.
	
	\item[{(b)}] As noted in the Table in Figure 6, each $\mathcal{P}_j$ is ${\bf Q}_{u_j}$- constrained for a unique $u_j$ with $u_j=1,2,3,\text{or}, 4$, viz.,
	the number of row it lies in, whereas each $\mathcal{P}_j$ is also ${\bf R}_{v_j}$- constrained for a unique value of $v_j$ with $v_j =1,2,\text{or}, 3$,
	viz., the number of column it lies in.
	\begin{enumerate}
	\item[{($\alpha$)}] In view of (v) (a) and (b), the listing $R_{j,0}, R_{j,1}$ of ${\bf R}_j$ also matches with the listing
	${ R}_{v_j,0}, { R}_{v_j,1}$ of ${\bf R}_{v_j}$.
	
	\item[{($\beta$)}] In view of (v)(c) $(\alpha)$ and $(\delta)$, the listing $Q_{j,0}, Q_{j,1}, Q_{j,2}$ of ${\bf Q}_{j}$ also matches with the listing
	$Q_{u_j,0}, Q_{u_j,1}, Q_{u_j,2}$ of ${\bf Q}_{u_j}$ for $j=8,9, \text{or}~ 5$, the corresponding $u_j$'s being 1,2, or 4 respectively.
	
	\item[{($\gamma$)}]
	In view of (v) $(c)$  $(\beta)$, the listing for ${\bf Q}_{6}$  differs from that of the corresponding
	${\bf Q}_1$ and that of ${\bf Q}_7$ differs from that of the corresponding ${\bf Q}_2$. In fact, $Q_{6,0}=Q_{1,2}$, $Q_{6,1}=Q_{1,0}$
	and $Q_{6,2}=Q_{1,1}$, and $Q_{7,0}=Q_{2,2}$, $Q_{7,1}=Q_{2,0}$ and, $Q_{7,2}=Q_{2,1}$.
	
	\item[{($\delta$)}]
	In view of (v) $(c)$ $(\gamma)$, the listings of ${\bf Q}_{10}$ and ${\bf Q}_{11}$   differ from that of the corresponding
	${\bf Q}_3$. In fact, for $j=10$ or $11$, $Q_{j,0}=Q_{3,1}$, $Q_{j,1}=Q_{3,2}$
	and $Q_{j,2}=Q_{3,0}$.
	
	\item[{($\epsilon$)}]
	In view of (v) $(c)$ $(\epsilon)$, the listings for  ${\bf Q}_{12}$ and the corresponding ${\bf Q}_{4}$   differ from each other.
	In fact, $Q_{12,0}=Q_{4,2}$, $Q_{12,1}=Q_{4,0}$ and $Q_{12,2}=Q_{4,1}$
	\end{enumerate}
	\item[{(c)}] Now the interesting point is that we need not actually compute $\mathcal{P}_j$ in full for $j\geq 5$. It is enough to just obtain $\xi_{j,0}$  and
	$\xi_{j,1}$ and the corresponding $P_{j,0}$ and $P_{j,1}$ for $j\geq 5$ using our simultaneous constraint technique and all the observations
	made above. We prefer to state it as a theorem :
	\end{enumerate}
	\begin{theorem}\label{t3.6.7}
		Simultaneous constraint technique for $d=6$ gives the ideal number $d^2=36$ of informationally complete rank one projections.
	\end{theorem}
	\begin{proof}
		We explain the technique and prove the theorem in different steps based on Example \ref{e3.6.6} above which we freely use together with notation,
		terminology and results there.
		\begin{enumerate}
		\item[{(i)}] We obtain the anchors $\mathcal{P}_1$, $\mathcal{P}_2$, $\mathcal{P}_3$ for ${\bf V}_1$, ${\bf V}_2$, ${\bf V}_3$ respectively as in 3.6.6 (i)
		to (iii),
		these have the trivial constraint $\{I_6\}$. So for estimation purposes, by the Constraint Technique 3.1.2, $\mathcal{P}_1\cup \mathcal{P}_2
		\cup \mathcal{P}_3$
		can be replaced by $\mathcal{P}_1\cup \mathcal{P}_2^\prime \cup \mathcal{P}_3^\prime$ with $\mathcal{P}_2^\prime=\{P_{2,k}: 0\leq k \leq 4\}$ and
		$\mathcal{P}_3^\prime=\{P_{3,k}: 0\leq k \leq 4\}$ ( or, by deleting any $P_{2,k_2}$ and any $P_{3,k_3}$ with $0\leq k_2,k_3\leq 5$, for that matter,
		and we arbitrarily choose $k_2=5=k_3$). It is as if we deleted $\xi_{2,5}$ and $\xi_{3,5}$ from the complete orthonormal bases for $U_{(0,1)}$
		and $U_{(1,1)}$ respectively obtained in Example \ref{e3.6.6} (ii) and (iii).
		
		\item[{(ii)}] We now come to $\mathcal{P}_4$ as obtained in Example \ref{e3.6.6} (iv). Then ${\bf R}_4= {\bf R}_1$ together with the listing as indicated in
		parts (i) and (iv) of Example \ref{e3.6.6}. So $\mathcal{P}_1$ and $\mathcal{P}_4$ are ${\bf R}_1$--constrained. By the Constraint Technique 3.1.2,
		$\mathcal{P}_4$
		can be replaced by $\mathcal{P}_4^\prime =\{P_{4,0}, P_{4,1}, P_{4,2}, P_{4,3}\}$ for estimation purposes. To elaborate, $P_{4,4}= R_{1,0}-P_{4,0}-P_{4,2}$
		and $P_{4,5}= R_{1,1}-P_{4,1}-P_{4,3}$ . ( Once again, the choice of deleting $P_{4,4}$ out of $\{P_{4,0}, P_{4,2}, P_{4,4}\}$ and of
		$P_{4,5}$ out of $\{P_{4,1}, P_{4,3}, P_{4,5}\}$ is arbitrary. We shall not repeat this kind of noting any more in this proof). It is as if we
		deleted $\xi_{4,4}$ and $\xi_{4,5}$ from the complete orthonormal basis $\{\xi_{4,k}: 0\leq k\leq 5\}$ of $U_{(1,2)}$ obtained in Example
		\ref{e3.6.6} (iv).
		
		\item[{(iii)}] We now have all the constraints ${\bf R}_1, {\bf R}_2, {\bf R}_3, {\bf Q}_1, {\bf Q}_2, {\bf Q}_3, {\bf Q}_4$ with us explicitly
		expressed in terms of $\mathcal{P}_1\cup \mathcal{P}_2^\prime\cup \mathcal{P}_3^\prime \cup \mathcal{P}_4^\prime$.
		
		\item[{(iv)}] We now come to $\mathcal{P}_j$ for $j\geq 5$. We claim that for each $j\geq 5$, for estimation purposes, $\mathcal{P}_j$ can be replaced by
		$\mathcal{P}_j^\prime =\{P_{j,0},P_{j,1}\}$. That is the gist of the simultaneous constraint technique which we explain.
		\begin{enumerate}
		\item[{(a)}] By Example \ref{e3.6.6} (vi)(b) there is a tuple $(0_j,1_j,2_j)$ of distinct numbers 0,1,2 in some order with $Q_{j,0}=Q_{u_j,0_j}, Q_{j,1}=Q_{u_j,1_j}$
		and $Q_{j,2}=Q_{u_j,2_j}$.
		
		By combining this with Example \ref{e3.6.6} (vi) (a), we obtain
		\begin{align}
			& P_{j,0} + P_{j,2}+P_{j,4}= R_{j,0} = R_{v_j,0}, \nonumber\\
			& P_{j,1} + P_{j,3}+P_{j,5}= R_{j,1} = R_{v_j,1}, \nonumber\\
			& P_{j,0} + P_{j,3}= Q_{j,0} = Q_{u_j,0_j}, \nonumber\\
			& P_{j,1} + P_{j,4}= Q_{j,1} = Q_{u_j,1_j},~~~~~~~~~~\text{and} \nonumber\\
			& P_{j,2} + P_{j,5}= Q_{j,2} = Q_{u_j,2_j}.
		\end{align}
		
		\item[{(b)}] We can solve for $P_{j,k}, ~k=2,3,4,5$ in terms of $P_{j,0}$, $P_{j,1}$, $Q_{u_j,0_j}$, $Q_{u_j,1_j}$, $Q_{u_j,2_j}$
		, $R_{v_j,0}$ and $R_{v_j,1}$ from (a) and obtain the following in succession
		\begin{align}
			&P_{j,3}=Q_{u_j,0_j} -P_{j,0},\nonumber\\
			&P_{j,4}=Q_{u_j,1_j}-P_{j,1},\nonumber\\
			&P_{j,2}= R_{v_j,0}- P_{j,0}-Q_{u_j,1_j}+P_{j,1},\nonumber\\
			&P_{j,5}= Q_{u_j,2_j}- R_{v_j,0} +P_{j,0}+ + Q_{u_j,1_j} - P_{j,1}.
		\end{align}
		For illustration,
		\begin{align},
			&P_{12,3}=Q_{4,2} -P_{12,0},\nonumber\\
			&P_{12,4}=Q_{4,0}-P_{12,1},\nonumber\\
			&P_{12,2}= R_{3,0}- P_{12,0}-Q_{4,0}+P_{12,1}, \nonumber\\
			&P_{12,5}= Q_{4,1}- R_{3,0} +P_{12,0}+ + Q_{4,0} - P_{12,1}.
		\end{align}
		
		\item[{(c)}] We give a pictorial representation of information in (a) and (b) above in Figures 9 and 10.
		\begin{figure}[H]
			\centering
			\begin{tabular}{|c?c|c|c|} \hline
				&            &          & Row sum\\ \thickhline
				& $P_{j,0}$ & $P_{j,3}$& $Q_{j,0}=Q_{u_j,0_j}$\\ \hline
				&$P_{j,4}$ & $P_{j,1}$ & $Q_{j,1}= Q_{u_j,1_j}$\\\hline
				&$P_{j,2}$ & $P_{j,5}$ & $Q_{j,2}=Q_{u_j,2_j}$\\\hline
				Column sum&$R_{j,0}=R_{v_j,0}$ & $R_{j,1}=R_{v_j,1}$ & $I_6$\\\hline
			\end{tabular}
			\caption{}
		\end{figure}
		\begin{figure}[H]
			\centering
			\begin{tabular}{|c?c|c|c|} \hline
				&            &          & Row sum\\ \thickhline
				& $P_{12,0}$ & $P_{12,3}$& $Q_{12,0}=Q_{4,2}$\\ \hline
				&$P_{12,4}$ & $P_{12,1}$ & $Q_{12,1}= Q_{4,0}$\\\hline
				&$P_{12,2}$ & $P_{12,5}$ & $Q_{12,2}=Q_{4,1}$\\\hline
				Column sum&$R_{12,0}=R_{3,0}$ & $R_{12,1}=R_{3,1}$ & $I_6$\\\hline
			\end{tabular}
			\caption{}
		\end{figure}
		\item[{(d)}] It is as if we deleted $\xi_{j,2},~ \xi_{j,3}, ~\xi_{j,4},~ \xi_{j,5}$ from the complete orthonormal system of $U_{m,n}$ chosen to
		generate ${\bf V}_j$. As a matter of fact, we need not compute them and manage with $\xi_{j,0}$ and $\xi_{j,1}$, with eigenvalues $1$ and
		$e^{\frac{\pi i}{3}}$, if $mn$ is even, or, $e^{\frac{i\pi}{6}}, e^{\frac{i\pi}{6}}e^{\frac{i\pi}{3}}=i$, if $mn$ is odd for
		the $(m,n)$ chosen to generate ${\bf V}_j$.
		\end{enumerate}
		\item[{(v)}] There is no hope of mutual unbiasedness as seen in the case of $d=p^2$ above in Theorem \ref{t3.6.5} . We note that
		\begin{align}
			\text{tr} (P_{j,k} P_{1,k^\prime})=|\langle \xi_{j,k},\xi_{1,k^\prime}\rangle|^2 =\frac{1}{6}~\text{for}~j=2~\text{or} ~3,
		\end{align}
		whereas
		\begin{align}
			\text{tr} (P_{4,k} P_{1,k^\prime})=|\langle \xi_{4,k},\xi_{1,k^\prime\rangle}|^2 =
			\left\{ \begin{array}{l} \frac{1}{3}~ \text{for}~k,~k^\prime ~\text{both odd or both even}\\
				\\0~\text{if one of}~k,k^\prime ~\text{is odd
					and the other even}\end{array}\right.
		\end{align}
		\begin{align}
			\text{tr}(P_{9,0} P_{2,0}) = |\langle \xi_{9,0},\xi_{2,0}\rangle |^2= |\langle \frac{1}{\sqrt{3}}\sum_{s=0,2,4}|s\rangle,\frac{1}{\sqrt{6}}\sum_{s=0}^{5}|s\rangle\rangle|
			= \frac{1}{2}
		\end{align}
		We shall elaborate more on this in Remark \ref{r3.6.8} (vi) below.
		
		\item[{(vi)}] The total number of rank one projections in $\mathcal{P}^\prime = \mathcal{P}_1\cup \mathcal{P}_2^\prime\cup \mathcal{P}_3^\prime\cup
		\mathcal{P}_4^\prime \cup \cup_{j=5}^{12} \mathcal{P}_j^\prime $ is $6+5+5+4+8\times 2=36$. Also $\mathcal{P}^\prime$ is complete in the sense
		that each $U_{jk}$ is a linear combination of members from $\mathcal{P}^\prime$ and thus $\mathcal{P}^\prime$ is a basis for $M_6$. So the
		technique gives the ideal numer 36 of such rank one projections.
		\end{enumerate}
	\end{proof}
	\begin{remark}\label{r3.6.8} It is clear from the discussion particularly, Figures 6 to 10 in Example \ref{e3.6.6} and the proof of Theorem \ref{t3.6.7} that there are many choices
		of choosing $\mathcal{P}^\prime$ of the ideal number 36 available. We note some of them :
		\begin{enumerate}
		\item[{(i)}] If we choose another generator of any $\lambda_{(m,n)}$ of full size, then the system of eigenvectors remains the same and only
		eigenvalues undergo a permutation and thus, the listing of new system, say, $\mathcal{S}_j=\{S_{j,k}: k=0,\cdots, 5\}$ will just
		be a permutation of $\mathcal{P}_j=\{P_{j,k}: k=0,\cdots, 5\}$ for $1\leq j\leq 12$.
		
		\item[{(ii)}] The situation is similar to that of (i) above for the overlaps by replacing eigenvectors by eigenprojections. So the listing of the
		new system, say, $\chi_j=\{\chi_{jk}, k=0,1,2\}$ will just be a permutation of ${\bf Q}_j=\{Q_{jk}:k=0,1,2\}$. We re-inforce that for $R_j$'s no change
		is needed.
		
		\item[{(iii)}] The care taken to match multiple $(2m,2n)$ by $2$ of $(m,n)$ made the matching more transparent and practical.
		
		\item[{(iv)}]  The net result is that the rows and columns in Figures 7 and 8 remain intact, though elements may undergo permutations and the sums in the
		last row or column may undergo permutations. In other words, we have cosets of $\{0,3\}$ and $\{0,2,4\}$ occurring in rows and columns
		respectively as subscripts. So, to figure out the sums of rows and columns, it is enough to  check for two different rank one projections
		in different cosets. In this Remark \ref{r3.1.3} comes in handy.
		
		\item[{(v)}] Now we obtain a basis for rank one projections for $M_6$. We need only two suitable such projections for each $j\geq 5$.

		(a) We can delete any row and any column and retain the rest. This gives us choices like $\{P_{j,0},P_{j,2}\}$ ,
		$\{P_{j,1},P_{j,3}\}$ etc..
		
		(b) Another one is to delete any row and then choose one each from the remaining rows. This gives us extra choices like $\{P_{1,0},P_{1,1}\}$ made
		in the proof of Theorem \ref{t3.6.7}. To go back to the starting point, we can  can start with any generator $U_{(m,n)}$
		of the ${\bf W}$-MASS ${\bf V}_j$, and we can
		take eigenvectors corresponding to any two eigenvalues with
		different squares and then take corresponding projections.
		
		(c) In fact, in view of analogue of (i)(h), all we need is, to figure out $u_j$ and $v_j$ and take any distinct $Q_{u_j,t_1}, Q_{u_j,t_2}$ and take
		products $Q_{u_j,t_1}R_{v_j,0}$ and $Q_{u_j,t_2}R_{v_j,0}$.
		
		\item[{(vi)}] The right choice in (v) above can reduce the number of quatum mechanical overlaps as is clear from Figures 9 and 10. We will list only a few.
		
		(a) For instance, in view of Example \ref{e3.6.6} (vi) (b)($\beta$) and $(\gamma)$ , $\text{tr}(P_{8,j} P_{6,k})=0$ for $j=0,1,3,4, ~k=0,3$ and $\text{tr}(P_{8,j} P_{6,k})=0,
		~\text{for}~j, k=4 ~\text{or}~1$.
		
		(b) On the other hand, we have $\text{tr}(P_{8,j} P_{l,k})=0,~\text{for}~l= 11,5,~j=0,2,4,~k=1,3,5, ~\text{or}~
		j=1,3,5,~\text{and}~ k=0,2,4$.
		\end{enumerate}
	\end{remark}
	\begin{remark}\label{r3.6.9} We see that the key is the table in Figure 6 and then the technique indicated by Figures 7 to 10. Next in line is the freedom
		provided by explanation in Remark \ref{r3.6.8} above. We can suitably modify all the arguments to suit for $d=2a$ with `$a$' an odd
		prime if we are able to give a right minimal set of $3(a+1)$ ${\bf W}$-MASS's and an analogue for the table in Figure 6 and display
		the technique by analogues of Tables in Figures 7 to 10.
		
		Same is true for $d=pa$ with `$p$' and `$a$' both distinct odd primes with $p<a$, say. In this certain arguments will be easier because
		now $d-1$ is even and, therefore, by Proposition \ref{p3.5.1} and Corollory \ref{c3.5.2}, $U_{(m,n)}^d=I_d$ and the set of eigenvalues of $U_{(m,n)}$ is
		$\{ \eta^s_d: s=0,1,\cdots,d-1\}$ for any generator $(m,n)$ of a full size $\lambda_{(m,n)}$ . This will impact the eigenvalues for
		$U_{(m,n)}^p$ and  $U_{(m,n)}^a$ to be $\{ \eta^s_a:s=0,1,\cdots a-1\}$ and $\{ \eta^t_p:t=0,1,\cdots p-1\}$ each of multiplicity $p$
		and $a$ respectively, as well.
	\end{remark}
	\subsubsection{\it The Case $d=2a$ with $a$ prime} As already indicated in Remark \ref{r3.6.9} above, we will develop analogues of the methods in
	\S3.6.2 above.
	\begin{theorem}\label{t3.6.10}
		Simultaneous constraints technique leads to the ideal number $d^2$ of a complete rank one projection system for $d=2a$ with `$a$' an odd prime.
	\end{theorem}
	\begin{proof}
		We divide the proof in a few parts.
		\begin{enumerate}
		\item[(i)] We begin with the following explicit system $\mathcal{M}$ of ${\bf W}$-MASS's consisting of parts $\mathcal{M}_1,\mathcal{M}_2, \mathcal{M}_3$ given below
		\begin{align}
			\mathcal{M}_1&= \{\lambda_{(1,0)}, \lambda_{(2j+1,2)}, 0\leq j\leq a-1\}, \nonumber \\
			\mathcal{M}_2&= \{\lambda_{(2,a)}, \lambda_{(2k,1)}, 0\leq k\leq a-1\}, \nonumber \\
			\mathcal{M}_3&= \{\lambda_{(1,a)}, \lambda_{(2g+1,1)}, 0\leq g \leq a-1\},
		\end{align}
		\item[(ii)] We claim that $\mathcal{M}$ covers $\ZZ_d\times \ZZ_d$.
		
		Consider any $(0,0)\neq (m,n)\in \ZZ_d\times \ZZ_d$.
		If $n=0,m=0$ or $m=n$, then $(m,n)=m(1,0)\in \lambda_{(1,0)}\in \mathcal{M}_1$, $(m,n)=n(0,1)\in \lambda_{(0,1)}\in \mathcal{M}_2$  and
		$(m,n)= m(1,1) \in \lambda_{(1,1)}\in \mathcal{M}_3$, respectively. Now consider the case $m\neq 0\neq n\neq m$ and let
		$h=\text{HCF}(m,n)$. Then $m=hm^\prime$ and $n=hn^\prime$ for $m^\prime,n^\prime$ with $\text{HCF}(m^\prime,n^\prime)=1$. So $(m,n)=
		h(m^\prime,n^\prime)$. Thus it is enough to consider the case when $\text{HCF}(m,n)=1$.
		\begin{enumerate}
		\item[(a)] Suppose $n$ is even. Then $m$ is odd. So $1\leq m\leq 2a-1$.
		
		Now $n=2^vk$ where $v\geq 1$ and $k$ is odd. So $1\leq 2^{v-1}k\leq a-1$. Set
		$k_1=k$ if $v=1$ and $2^{v-1}k+a$ if $v\geq 2$. So $\text{HCF}(k_1,2a)=1$. So there exists a unique $k_2$ with $1\leq k_2\leq 2a-1$ and
		$k_1k_2=1~(\text{mod}~2a)$. Further, $k_2$ is odd. So $m_1= mk_2~(\text{mod}~2a)$ is odd and $1 \leq m_1 \leq 2a-1$. Now $(m,n)= (m, 2\cdot 2^{v-1}k)$
		with respect to operations in the ring $(\ZZ_d,+,\cdot)$, i.e.,  addition and multiplication (mod $2a$), we have $k_1(m_1,2)= k_1( mk_2~(\text{mod}~ 2a),2)$
		$=(mk_1k_2~(\text{mod}~2a), 2k_1~(\text{mod}~2a)) = (m,n)$. So $(m,n)\in \lambda_{(m_1,2)}\in \mathcal{M}_1$.
		
		\item[(b)] Suppose $m$ is even. Then $n$ is odd. We first consider the case $n=a$. Then for any odd $v$, $nv= a~(\text{mod}~2a)$. We may interchange the
		roles of $m$ and $n$ in (a) above and obtain that $(m,n)=k_1(2,a)\in \lambda_{(2,a)}\in \mathcal{M}_2$. Now consider the case $n\neq a$. Then $\text{HCF}(n,2a)=1$
		. So as in (a) above, there is a unique $n\prime$ with $1\leq n^\prime \leq 2a-1$ and $nn^\prime =1 ~(\text{mod}~2a)$.
		Let $m_1 =mn^\prime ~(\text{mod}~2a)$. Because $m$ is even, we have that $m_1$ is even. So in the ring $\ZZ_d$, $(m,n)=n(m_1,1)\in \lambda_{(m_1,1)}
		\in \mathcal{M}_2$.
		
		\item[(c)] Now suppose that both $m$ and $n$ are odd. If $n=a$, then $(m,n)=m(1,a)\in \lambda_{(1,a)}$ , and if $m=a$, then
		$(m,n)=n(a,1)\in \lambda_{(a,1)}$. Both $\lambda_{(a,1)}$ and  $\lambda_{(1,a)}$ are in $\mathcal{M}_3$. Now, if $m\neq a\neq  n$, then
		let $n^\prime$ be as in (b) above. Then $(m,n)=n(m_1,1)$ where $m_1=mn^\prime~(\text{mod}~2a)$ is odd. So $(m,n)\in \lambda_{(m_1, 1)}\in \mathcal{M}_3$.
		
		Hence $\mathcal{M}$ covers $\ZZ_d\times \ZZ_d$ and therefore , $\{U_{(m,n)}: \{(m,n)\in \lambda_\sigma~ \\ \text{for some}~ \sigma~\text{in}
		~\mathcal{M}\}$ covers the unitary basis $\{U_{(m,n)}: (m,n)\in \ZZ_d\times \ZZ_d\}$.
		
		\item[(d)] To get an idea, for the case $a=3$, i.e. $d=6$, $ \mathcal{M}_1=\{V_1,V_4,V_9,V_{10}\}, \mathcal{M}_2=\{V_8,V_2,V_5,V_{11}\},
		\mathcal{M}_3=\{V_6,V_3,V_7,V_{12}\}$ occurring in the first, second and third column respectively in the Table in Figure 6.
		\end{enumerate}
		\item[(iii)] We now figure out overlapping members in the ${\bf W}$-MASS's, arising from $\mathcal{M}$. It will be convenient to write $(m,n)$ in place of $U_{(m,n)}$ and,
		at times, even for $\lambda_{(m,n)}$ it generates.
		\begin{enumerate}
		\item[(a)] For $(m,n)$ in $\mathcal{M}_1, \mathcal{M}_2~\text{or}~\mathcal{M}_3$, $a(m,n)$ is $(a,0), (0,a)~\text{or}~(a,a)$ respectively.
		\item[(b)] The situation for multiples by 2 is a bit different and we need suitable alternative generators and reorderings. So we start with noting them
		for the listing as given in (i) above for necessary adjustments in item (iv) below.
		\begin{align}
			{M}_1&= \{(2,0),(4j+2~ (\text{mod}~ 2a), 4), 0\leq j\leq a-1\}\nonumber\\
			&= \{(2,0),(4j+2,4), 0\leq j\leq \frac{a-3}{2},~\text{and} \nonumber\\
			&~~~~~~~~~(4j+2-2a,4)=(4(j-\frac{a-1}{2}),4), \frac{a-1}{2}\leq j\leq a-1\} \nonumber\\
			&= \{(2,0),(4v-2,4), 1\leq v=j+1\leq \frac{a-1}{2}, (4j^\prime,4), 0 \leq j^\prime\leq \frac{a-1}{2}\}, \nonumber\\
			{M}_2 &= \{(4,0),(4k~(\text{mod}~2a),2), 0\leq k \leq a-1\} \nonumber\\
			&= \left\{(4,0), (4k,2), 0\leq k \leq \frac{a-1}{2}, \right.\nonumber\\
			&\left. (4k-2a, 2)=(4(k-\frac{a-1}{2})-2,2), \frac{a+1}{2}\leq k \leq a-1\right\} \nonumber\\
			&= \{ (4,0),(4k,2), 0\leq k\leq \frac{a-1}{2}, (4k^\prime-2,2), 1\leq k^\prime\leq \frac{a-1}{2}\},\nonumber\\
			{M}_{3} &= \{(2,0), (4g+2,2), 0\leq g\leq a-1\}\nonumber\\
			&= \{(2,0), (4v-2,2), 1\leq v\leq \frac{a-1}{2}, (4g^\prime,2),0\leq g^\prime\leq \frac{a-1}{2}\}
		\end{align}
		\end{enumerate}
		\item[(iv)] For an application of the simultaneous constraint technique we will need alternative generators for the ${\bf W}$-MASS's set
		in (i) above which we shall now figure out. We deal in  the ring $\ZZ_d$ unless otherwise stated.
		\begin{enumerate}
		\item[(a)] If $1\leq \ell\leq 2a-1$ is odd and $\ell\neq a$, then $(\ell m, \ell n)$ is also a generator for full-size  $\lambda_{(m,n)}$ generated
		by $(m,n)$. To see this , $(\ell m, \ell n)\in \lambda_{(m,n)}$ and $s(\ell m,\ell n) = (0,0)$ if and only if $s \ell (m,n)=(0,0)$. But
		$\text{HCF}(\ell, 2a)= 1$. So $s\ell (m,n)=(0,0)$ if and only if $s =2a$. So $\lambda_{(\ell m,\ell n)}= \lambda_{(m,n)}$ and $(\ell m,\ell n)$ also
		generates it.
		
		\item[(b)] Let $\mu = \frac{a+1}{2}$, if it is odd and $a+\frac{a+1}{2}$ if $\frac{a+1}{2}$ is even. Then $\mu$ is odd and $\mu\neq a$. Further,
		$\text{HCF}(\mu,2a)=1$ and, therefore, $\{\nu\mu: 0\leq \nu \leq 2a-1\}= \{\alpha: 0\leq \alpha \leq 2a-1\}$, $\nu$ odd corresponds to $\alpha$
		odd, $\nu$ even corresponds to $\alpha$ even, $\nu= 0,2,4, a$ corresponds to $\alpha = 0,a+1,2,a$ respectively.
		
		\item[(c)] We replace $\mathcal{M}_1, \mathcal{M}_2, \mathcal{M}_3$  by $\lambda$'s generated by $(m,n)$'s or $(\mu m, \mu n)$'s as per our need which is evident in the format
		given below in three parts $\mathcal{M}_1^\prime, \mathcal{M}_2^\prime, \mathcal{M}_3^\prime$ respectively :
		\begin{align}
			\mathcal{M}_1^\prime &=\{(1,0), (2\alpha +1, a+1), ~0\leq\alpha \leq a-1\},\nonumber\\
			\mathcal{M}_2^\prime &=\{(a+1,a), (2k, 1), ~0\leq k \leq a-1\},\nonumber\\
			\mathcal{M}_3^\prime & =\{(1,a), (2g+1, 1), ~0\leq g \leq a-1\}.
		\end{align}
		Then $a(m,n)=(a,0), (0,a)~\text{or}~(a,a)$ for $(m,n) \in \mathcal{M}_1^\prime, \mathcal{M}_2^\prime, \mathcal{M}_3^\prime$ respectively.
		On the other hand, multiplication
		by 2 gives us
		\begin{align}
			{M}_1^\prime &=\{(2,0), (4\alpha +2, 2), ~0\leq\alpha \leq a-1\},\nonumber\\
			{M}_2^\prime &=\{(2,0), (4k, 2), ~0\leq k \leq a-1\},\nonumber\\
			{M}_3^\prime & =\{(2,0), (4g+2, 2), ~0\leq g \leq a-1\}.
		\end{align}
		So some reordering is called for. We prefer to first display it in the Table in Figure 11  in a more pleasing order for constraints
		rather than the ${\bf W}$-MASS's themselves.
		\begin{figure}[H]
			\centering
			\begin{tabular}{|c?c|c|c|} \hline
				&  $(a,0)$          & $(0,a)$         & $(a,a)$\\ \thickhline
				$(2,0)$ & $(1,0)$ & $(a+1,a)$& $(1,a)$\\ \hline
				$(0,2)$ &$(a,a+1)$ & $(0,1)$ & $(a,1)$\\\hline
				$(2,2)$  &$(1,a+1)$ & $(a+1,1)$ & $(1,1)$ \\\hline
				$(4,2)$ & $(a+2,a+1)$ & $(2,1)$ & $(a+2,1)$\\ \hline
				\vdots  & \vdots    & \vdots & \vdots \\\hline
				$(4b-2,2)$   & $(2b-1,a+1)$ & $(a+2b-1,1)$& $(2b-1,1)$\\ \hline
				$(4b,2)$   & $(a+2b,a+1)$ & $(2b,1)$& $(a+2b,1)$\\ \hline
				\vdots  & \vdots    & \vdots & \vdots \\\hline
				$(2a-4,2)$   & $(a-2,a+1)$ & $(a+(a-2),1)=(2a-2,1)$& $(a-2,1)$\\ \hline
				$(2a-2,2)$   & $(a+(a-1),a+1)=(2a-1,a+1)$ & $(a-1,1)$& $(2a-1,1)$\\ \hline
			\end{tabular}
			\caption{ $1\leq b \leq \frac{a-1}{2}$}
		\end{figure}
		Finally we replace $\mathcal{M}$ by $\mathcal{M}^{\prime\prime}$ which is the union of three parts:
		\begin{align}
			\mathcal{M}_1^{\prime\prime} &=\{(1,0), (j_t,a+1),~0\leq t \leq a-1\},\nonumber\\
			&\text{where}~j_t=t~\text{for}~t~\text{odd}~, ~\text{but}~a+t~\text{for}~t~\text{even}\nonumber\\
			&=\left\{(1,0), (a, a+1);(1,a+1), (a+2, a+1),\cdots, \right.\nonumber\\
			&\left.(2b-1,a+1),(a+2b,a+1),\cdots,(a-2,a+1),(2a-1,a+1)\right\},\nonumber\\
			\mathcal{M}_2^{\prime\prime} &=\{(a+1,a), (k_t,1),~0\leq t \leq a-1\},\nonumber\\
			&\text{where}~k_t=a+t~\text{for}~t~\text{odd}~, ~\text{but}~t~\text{for}~t~\text{even}\nonumber\\
			&=\left\{(a+1,a), (0, 1);(a+1,1), (2, 1),\cdots, (a+2b-1, 1),(2b,1), \right. \nonumber\\ & \left.~~~~~~~~~~~~~~~~~~~\cdots,(2a-2, 1),(a-1,1)\right\},\nonumber\\
			\mathcal{M}_3^{\prime\prime} &=\{(1,a), (g_t,1), ~0\leq t \leq a-1\},\nonumber\\
			&\text{where}~g_t=t~\text{for}~t~\text{odd}~, ~\text{but}~a+t~\text{for}~t~\text{even}\nonumber\\
			&=\left\{(1,a), (a, 1);(1,1), (a+2, 1),\right. \nonumber\\ &~~~~~~~~~~~~~~\left. \cdots; (2b-1, 1),(a+2b, 1),\cdots,(a-2, 1),(2a-1,1)\right\}.
		\end{align}
		Then $a(m,n)=(a,0),(0,a),~\text{or}~(a,a)$ according as $(m,n)\in  \mathcal{M}_1^{\prime\prime}, \mathcal{M}_2^{\prime\prime},
		~\text{or}~\mathcal{M}_3^{\prime\prime}$. Also $2(m,n)=(2,0), (0,2)$ or $(2t,2)$ for $1\leq t\leq a-1$ in the order as listed in
		$(m,n)\in  \mathcal{M}_1^{\prime\prime}, \mathcal{M}_2^{\prime\prime}, \mathcal{M}_3^{\prime\prime}$.
		\end{enumerate}
		\item[(iv)] We are now in a position to modify the proof of Example \ref{e3.6.6} and Theorem \ref{t3.6.7} by replacing $3$ at appropriate places by `$a$' and making other
		necessary changes. We begin by choosing anchors.
		\begin{enumerate}
		\item[(a)] Just as in the case of $d=6$, we run along the diagonal and then along the first column.That is, we label the ${\bf W}$-MASs's generated by
		$U_{(1,0)}$, $U_{(0,1)}$, $U_{(1,1)}$, $U_{(j_t, a+1)}$, $2\leq t\leq a-1$ by ${\bf V}_1$, ${\bf V}_2$, ${\bf V}_3$, ${\bf V}_{t+2}$, $2\leq t\leq a-1$
		respectively.
		
		\item[(b)] By Proposition \ref{p3.5.1}(iii) and Corollary \ref{c3.5.2}  all these unitaries in (a) above have simple eigenvalues which are $\{e^{\left(\frac{2\pi ik}{d}\right)}, 0\leq k\leq d-1\}$
		= $\{e^{\left(\frac{\pi ik}{a}\right)}, 0\leq k\leq 2a-1\}$ except for $U_{(1,1)}$  which has simple eigenvalues
		$e^{\left(\frac{\pi i}{d}\right)}\{e^{\left(\frac{2\pi ik}{d}\right)}, 0\leq k\leq d-1\}$ =
		$e^{\left(\frac{\pi i}{2a}\right)}\{e^{\left(\frac{\pi ik}{a}\right)}, 0\leq k\leq 2a-1\}$.
		
		\item[(c)] We choose the complete system of unit eigenvectors
		$ \{ |k\rangle : 0\leq k\leq 2a-1\}$ for $U_{(1,0)}$  indicated by the listing of eigenvalues in (b) above and the
		corresponding system $\mathcal{P}_1=\{ P_{1k} = |\xi_{1,k}\rangle\langle \xi_{1,k}| =|k\rangle\langle k|, : 0\leq k\leq 2a-1\}$ of rank one
		projections. It provides two constraints ${\bf Q}_1$ and ${\bf R}_1$ determined by $U_{(2,0)} = U_{(1,0)}^2$ and
		$U_{(a,0)} = U_{(1,0)}^a$ given by ${\bf Q}_1=\{Q_{1,0}, Q_{1,1},\cdots, Q_{1,a-1}\} $,
		${\bf R}_1=\{R_{1,0}, R_{1,1}\}$, where
		$Q_{1,r} =P_{1,r}+ P_{1,a+r} $ corresponding to the eigenvalue $e^{\left(\frac{2\pi i r}{a}\right)}$ of $U_{(2,0)}$ for $0\leq r\leq a-1$, whereas
		$R_{1,0}=\sum_{k=0}^{a-1} P_{1,2k}$ and  $R_{1,1}=\sum_{k=0}^{a-1} P_{1,2k+1}$ corresponding to the eigenvalues $1$ and $-1$ of $U_{(a,0)}$
		respectively. We give a useful pictorial representation in Figure 12.
		\begin{figure}[H]
			\centering
			\begin{tabular}{|c?c|c|c|} \hline
				&            &          & Row sum\\ \thickhline
				& $P_{1,0}$ & $P_{1,a}$& $Q_{1,0}$\\ \hline
				&$P_{1,a+1}$ & $P_{1,1}$ & $Q_{1,1}$\\\hline
				&$P_{1,2}$ & $P_{1,a+2}$ & $Q_{1,2}$\\\hline
				& \vdots    & \vdots & \vdots \\\hline
				& $P_{1,2a-2}$ & $P_{1,a-2}$& $Q_{1,a-2}$\\ \hline
				& $P_{1,a-1}$ & $P_{1,2a-1}$& $Q_{1,a-1}$\\ \hline
				Column sum&$R_{1,0}$ & $R_{1,1}$ & $I_{2a}$\\\hline
			\end{tabular}
			\caption{}
		\end{figure}
		
		\item[(d)] We choose complete system of unit eigenvectors $\{\xi_{2,k}=\sum_{s=0}^{2a-1} e^{\left(-\frac{ks\pi i}{a}\right)}|s\rangle : 0\leq k\leq 2a-1\}$
		for $U_{(0,1)}$ in the order indicated by the listing of eigenvalues as in (b) analogous to that for $d=6$ in Example \ref{e3.6.6} (ii) and then the corresponding
		system $\mathcal{P}_2=\{P_{2,k} =|\xi_{2,k}\rangle\langle \xi_{2,k}| : 0\leq k\leq 2a-1\}$ of rank one projections. We may use the trivial constraint
		$\{I_{2a}\}$ provided by $\mathcal{P}_1$ to reduce $\mathcal{P}_2$ to $\mathcal{P}_2^\prime$ by taking anyone, say, the last one
		$P_{2,2a-1}$ out. Furthermore, just as in (c) above, $\mathcal{P}_2$ provides constraints ${\bf Q}_2$ and ${\bf R}_2$ in an analogous manner.
		
		\item[(e)] The situation for $U_{(1,1)}$ or, for that matter, $U_{(m,n)}$ for any $(m,n)$ in the third column in Figure 9 foliates into two cases as follows
		because of the fact that $mn$ is odd.
		
		$(\alpha)$ $a+1$ is a multiple of $4$, i.e. $a=4v-1$ for some $v\geq 1$, and
		
		($\beta)$)  $a-1$ is a multiple of $4$, i.e. $a=4v+1$ for some $v\geq 1$.
		
		This is because, by Proposition \ref{p3.5.1} and the fact that $mn$ is odd, we have
		\begin{align}
			U_{(am,an)} = &e^{\left(-\frac{\pi i}{a}\frac{a(a-1)}{2} mn\right)}U_{(m,n)}^a\nonumber\\
			& e^{\left(-\frac{\pi i}{2}mn(a-1)\right)}U_{(m,n)}^a\nonumber\\
			=& \pm U_{(m,n)}^a,
		\end{align}
		according as $a-1$ or $a+1$ is a multiple of 4.
		
		Now the eigenvalues of $U_{(m,n)}$ are $e^{\left(\frac{\pi i}{2a}\right)}\{ e^{\left(\frac{\pi ik}{a}\right)}: 0\leq k\leq 2a-1\}$. So eigenvalues
		of $U_{(m,n)}^a$ are $i$ or $-i$ corresponding to $k$ even or odd respectively.
		Thus, the eigenvalues
		of $U_{(am,an)}$ are $-i$ or $i$ corresponding to $k$ even and odd respectively in case $(\alpha)$ ( in analogy with the case $d=6$ i.e., $a=3$ ) but the
		eigenvalues of $U_{(am,an)}$ are $i$ and $-i$ corresponding to $k$ even and odd respectively in case $(\beta)$, which is new.
		
		On the other hand, the analogy for $U_{(2m,2n)}$ with the case $d=6$ i.e., $a=3$, continues in the sense that the eigenvalues for
		$U_{(m,n)}^2$ are listed as $e^{\left(\frac{\pi i}{a}\right)}\{ e^{\left(\frac{2\pi ik}{a}\right)}: 0\leq k\leq 2a-1\}$
		and therefore for $U_{(2m,2n)}$ are listed as $e^{\left(2(k-\frac{mn-1}{2})\frac{\pi i}{a}\right)}, 0\leq k\leq 2a-1$.
		
		In this listing, the values for $k$ and $k+a$ coincide for $0\leq k\leq a-1$.
		
		In particular, for $U_{(2,2)}$ i.e., for $(m,n)=(1,1)$ the listing is the usual one as for $U_{(2,0)}$ or $U_{(0,2)}$ in (b) and (c)
		respectively above. But for other $(m,n)$'s, a cyclic permutation may be required.
		
		\item[(f)] In view of (e) above we can go ahead with the analogy with Example \ref{e3.6.6} (iii) for $U_{(1,1)}$ with a little extra care of
		dealing with two cases $(\alpha)$ and $(\beta)$ at due place. In other words we may obtain the complete system
		$\{\xi_{3,k}=\frac{1}{\sqrt{2a}}\sum_{s=0}^{2a-1} e^{\left(-\frac{(2k+1)s}{2a}\pi i\right)} e^{\left(\frac{\pi i}{a}\frac{s(s-1)}{2}\right)}|s\rangle,
		0\leq k\leq 2a-1\}$ of unit eigenvectors with corresponding eigenvalues as listed in (e) above and
		$\mathcal{P}_3=\{P_{3k}=|\xi_{3,k}\rangle\langle \xi_{3,k}| : 0\leq k\leq 2a-1\}$. Due to trivial constraint $\{I_{2a}\}$ provided by
		$\mathcal{P}_1$, $\mathcal{P}_3$ can be reduced by taking any one projection, say $P_{3,2a-1}$ out as in (c) above.
		$\mathcal{P}_3$ provides constraints ${\bf Q}_3$ and ${\bf R}_3$ in an analogous manner. But the corresponding eigenvalues for
		eigenprojectors  $R_{3,0}$ and $R_{3,1}$ are $-i$ and $i$ for case $(\alpha)$ and $i$ and -$i$ for case $(\beta)$.
		
		\item[(g)] For the remaining anchors $\{\mathcal{P}_{t+2} : 2\leq t\leq a-1\}$ we can use the constraint ${\bf R}_1$ as in (c) above. To elaborate,
		let $2\leq t\leq a-1$. As in Example \ref{e3.6.6}(i), $U_{(j_t,a+1)}$ restricted to $G_{1,0}$, the span of $\{|2k\rangle : 0\leq k\leq a-1\}$ has simple
		eigenvalues $\{e^{\left(\frac{2\pi i s}{a}\right)}:0\leq s\leq a-1\}$ and $U_{(j_t,a+1)}$ restricted to $G_{1,1}$, the span of
		$\{|2k+1\rangle : 0\leq k\leq a-1\}$ has simple
		eigenvalues $\{e^{\left(\frac{(2s+1)\pi i }{a}\right)}:0\leq s\leq a-1\}$. Pooling together, let $\{\xi_{t+2,k} :0\leq k\leq 2a-1\}$ be
		a complete orthonormal eigenvector system for $U_{(j_t,a+1)}$ corresponding to the eigenvalues
		$\{e^{\left(\frac{\pi i k}{a}\right)}:0\leq k\leq 2a-1\}$ and let $\mathcal{P}_{t+2} =\{P_{t+2,k}=|\xi_{t+2,k}\rangle\langle
		\xi_{t+2,k}| : 0\leq k \leq 2a-1\}$ be the set of corresponding rank one projections. These satisfy $R_{1,0}=\sum_{k=0}^{a-1} P_{t+2, 2k}$
		and $R_{1,1}=\sum_{k=0}^{a-1} P_{t+2, 2k+1}$. Also they give rise to the constraint ${\bf Q}_{t+2,r} = P_{t+2,r} +P_{t+2,a+r}$ for
		$0\leq r\leq a-1$. We give a pictorial representation in Figure 13.
		
		\begin{figure}[H]
			\centering
			\begin{tabular}{|c?c|c|c|} \hline
				&            &          & Row sum\\ \thickhline
				& $P_{t+2,0}$ & $P_{t+2,a}$& $Q_{t+2,0}$\\ \hline
				&$P_{t+2,a+1}$ & $P_{t+2,1}$ & $Q_{t+2,1}$\\\hline
				&$P_{t+2,2}$ & $P_{t+2,a+2}$ & $Q_{t+2,2}$\\\hline
				& \vdots    & \vdots & \vdots \\\hline
				& $P_{t+2,2a-2}$ & $P_{t+2,a-2}$& $Q_{t+2,a-2}$\\ \hline
				& $P_{t+2,a-1}$ & $P_{t+2,2a-1}$& $Q_{t+2,a-1}$\\ \hline
				Column sum&$R_{1,0}$ & $R_{1,1}$ & $I_{2a}$\\\hline
			\end{tabular}
			\caption{}
		\end{figure}
		$\mathcal{P}_{t+2}$ can be replaced by $\mathcal{P}_{t+2}^\prime$ by taking any projection from each column, say, $P_{t+2, 2a-2}$ and
		$P_{t+2, 2a-1}$ out.
		
		\item[(h)] The advantage of this choice of anchors above lies in the fact that the forms for $\xi_{t+2,k}$'s are stratified and the computation
		of quantum mechanical overlaps will thus be easier.
		\end{enumerate}
		\item[(vii)] We can label the remaining ${\bf W}$-MASS's the way we like. One choice could be to go column-wise. That is, let
		${\bf V}_{a+2}$, ${\bf V}_{a+3}$ and ${\bf V}_{a+4}$ be the ${\bf W}$-MASS's arising from $\lambda_{a,a+1}$, $\lambda_{1,a+1}$ and
		$\lambda_{a+1,a}$ respectively, ${\bf V}_{a+4+t}$ for $\lambda_{k_t,1}$, $1\leq t\leq a-1$, ${\bf V}_{2a+4}$ and ${\bf V}_{2a+5}$ for the
		${\bf W}$-Mass's arising from $\lambda_{1,a}$ and $\lambda_{a,1}$ respectively and, finally, ${\bf V}_{2a+4+t}$ for
		$\lambda_{g_t,1}$ for $2\leq t\leq a-1$. We continue with modification of proof for $d=6$. Let $\{\xi_{j,k} : 0\leq k\leq 2a-1\}$ be
		a complete  orthonormal system of eigenvectors and $\mathcal{P}_j =\{P_{j,k} = |\xi_{j,k}\rangle\langle \xi_{j,k}|: 0\leq k\leq 2a-1\}$ the corresponding
		rank one projections, the listing corresponding to simple eigenvalues $\{e^{\left(\frac{\pi i k}{a}\right)}:0\leq k\leq 2a-1\}$ of
		$U_{(m,n)}$ with $(m,n)$ in the first two columns giving rise to ${\bf W}$-MASS's ${\bf V}_j$ for $a+2\leq j\leq 2a+3$. For $(m,n)$ in the third
		column giving rise to ${\bf W}$-MASS's ${\bf V}_j$ for $2a+4\leq j\leq 3a+3$, the listing  of eigenvalues will change to
		$e^{\left(-\frac{\pi i}{2a}\right)}  \{e^{\left(\frac{\pi i k}{a}\right)}:0\leq k\leq 2a-1\}$ for the corresponding complete orthonormal
		system of eigenvectors $\{\xi_{j,k} : 0\leq k\leq 2a-1\}$ and $\mathcal{P}_j =\{P_{j,k} = |\xi_{j,k}\rangle\langle \xi_{j,k}|: 0\leq k\leq 2a-1\}$. Then
		each $\mathcal{P}_j$ is ${\bf Q}_j$-constrained and ${\bf R}_j$-constrained, where
		\begin{align}
			&{\bf Q}_j=\{Q_{j,0}, Q_{j,1},\cdots, Q_{j,a-1}\}, \nonumber\\ .
			&{\bf R}_j=\{R_{j,0}, R_{j,1}\}, \nonumber\\.
			&Q_{j,r} =P_{j,r}+ P_{j,a+r}, 0\leq r\leq a-1, \nonumber\\
			&R_{j,0}=\sum_{k=0}^{a-1} P_{j,2k},~ R_{j,1}=\sum_{k=0}^{a-1} P_{j,2k+1}.
		\end{align}
		Now $(m,n)$ lies in the $u_j^{\text{th}}$ row and $v_j^{\text{th}}$ column for some unique $1\leq u_j\leq a+1$ and $1\leq v_j\leq 3$. Then,
		as explained in (iv) above, as sets ${\bf Q}_j= {\bf Q}_{u_j}$  and ${\bf R}_j= {\bf R}_{v_j}$ but the listing is same for $(m,n)$ in the first
		two columns whereas there exists a permutaion $(0_j,1_j,\cdots, (a-1)_j)$ of $(0,1,\cdots,a-1)$ that satisfies the following for
		$(m,n)\neq (1,1)$ in the third column.
		\begin{align}
			Q_{j,s}=Q_{u_j,s_j}~\text{for}~0\leq s\leq a-1;
		\end{align}
		also in case $(\alpha)$, $R_{j,0} = R_{3,0}$ and $R_{j,1}=R_{3,1}$, but in case $(\beta)$, $R_{j,0} = R_{3,1}$ and $R_{j,1}=R_{3,0}$.
		We write $0_j^\prime =0, 1_j^\prime=1 $ for case $(\alpha)$ and $0_j^\prime =1, 1_j^\prime=0 $ for case $(\beta)$. Then the relationship
		amongst various $P_{j,k}$'s and $Q_{j^\prime,r}$'s, $R_{j^{\prime\prime},s}$'s can be displayed in a pictorial form as in Figure 14.
		
		\begin{figure}[H]
			\centering
			\begin{tabular}{|c?c|c|c|} \hline
				&            &          & Row sum\\ \thickhline
				& $P_{j,0}$ & $P_{j,a}$& $Q_{j,0}=Q_{u_j,0_j}$\\ \hline
				&$P_{j,a+1}$ & $P_{j,1}$ & $Q_{j,1}=Q_{u_j,1_j}$\\\hline
				&$P_{j,2}$ & $P_{j,a+2}$ & $Q_{j,2}=Q_{u_j,2_j}$\\\hline
				& \vdots    & \vdots & \vdots \\\hline
				& $P_{j,2a-2}$ & $P_{j,a-2}$& $Q_{j,a-2}=Q_{u_j,(a-2)_j}$\\ \hline
				& $P_{j,a-1}$ & $P_{j,2a-1}$& $Q_{j,a-1}=Q_{u_j,(a-1)_j}$\\ \hline
				Column sum&$R_{j,0}=R_{v_j,0_j^\prime}$ & $R_{j,1}=R_{v_j,1_j^\prime}$ & $I_{2a}$\\\hline
			\end{tabular}
			\caption{}
		\end{figure}
		
		\item[(viii)] For estimation purposes we may replace $\mathcal{P}_j$ by $\mathcal{P}_j^\prime$ by deleting any row and any column of $P_{jk}$'s in
		the above figure or even any row and then one each from each remaining row. One such simple instance is $\mathcal{P}_j^\prime
		=\{P_{j,k}:0\leq k\leq a-2\}$ and another simple one is $\mathcal{P}_j^{\prime\prime}
		=\{P_{j,2k}:0\leq k\leq a-2\}$. In fact, we need not compute the remaining projections of rank one.
		
		We may even formulate an analogue of Example \ref{e3.6.6}(i)(h) and just take $\mathcal{P}_j^\prime$=$\{Q_{u_j,t}R_{v_j,0}, 0\leq t\leq a-2\}$ instead of
		computing eigenvectors and corresponding rank one projections.
		
		\item[(ix)] The final step is to note that for estimation purposes
		\begin{align}
			\mathcal{P}^\prime = \mathcal{P}_1\cup  \mathcal{P}_2^\prime \cup \mathcal{P}_3^\prime  \cup \cup_{t=2}^{a-1} \mathcal{P}_{t+2}^\prime
			\cup \cup_{j=a+2}^{3a+3} \mathcal{P}_j^\prime
		\end{align}
		works fine and its cardinality is $2a+ (2a-1)+(2a-1)+(a-2)(2a-2)+(2a+2)(a-1)=4a^2=d^2$.
	\end{enumerate}
	\end{proof}
	\subsubsection{\it The case of $d=pa$ with $p, a$ odd prime numbers with $p<a$} We begin with a number theoretic remark.
	
	\begin{remark} We will use Remark \ref{r3.5.6} freely.
		\begin{enumerate}
		\item[(i)] $\ZZ_d$ is the disjoint union of cosets $\{K_b =\{ra+b: 0\leq r\leq p-1\}, 0\leq b\leq a-1\}$ of the subgroup
		$K_0 =\{ra:0\leq r\leq p-1\}$ of $\ZZ_d$.
		
		\item[(ii)] $\ZZ_d$ is the is the disjoint union of cosets $\{C_t =\{cp+t :0\leq c\leq a-1\}, 0\leq t\leq p-1\}$ of the subgroup
		$C_0 =\{cp :0\leq c\leq a-1\}$ of $\ZZ_d$.
		
		\item[(iii)] $ra+b=cp+t$ only if $rr_0+b=t~(\text{mod}~p)$, i.e., $r=(t-b~(\text{mod}~p))s_0~(\text{mod}~p)$, $~\text{if}~t\geq b~(\text{mod}~p)$ and
		$r=(p+t-b~(\text{mod}~p))s_0~(\text{mod}~p),~\text{if}~t< b~(\text{mod}~p)$. In particular, for $0\leq b\leq p-1$, the necessary condition
		becomes
		$ r=(t-b)s_0 (\text{mod}~p)~\text{if}~b\leq t~\text{and}~ r=(p+t-b)s_0 (\text{mod}~p)~\text{if}~b> t.$
		For $p\leq b^\prime\leq a-1$, $b^\prime$ has the form $b^\prime=b+jp$ for some suitable $j$ with $1\leq j\leq j_0$ and $0\leq b\leq p-1$.
		Then with the same $r$, $ra+b^\prime=c^\prime p+t$ with $c^\prime=c+j$. Arguments can be reversed with adjustment of the values of $c$, and, therefore, the
		condition is sufficient as well. We note this as a proposition below after setting notation.
		
		Let $c_t=ts_0~(\text{mod}~p)$ for $0\leq t\leq p-1$. We note that $c_0=0, c_1=s_0$ and $c_{p-t}= p-c_t$ for $0\leq t\leq p-1$.
		\end{enumerate}
	\end{remark}
	\begin{proposition}\label{p3.6.12} Let $0\leq t\leq p-1$ and $0\leq b\leq a-1$. Then $K_b\cap C_t=\{m_{b,t}\}$ with
		\begin{align}
			m_{b,t} =\left\{ \begin{array}{l} c_{t-b}a+b,~\text{for}~0\leq b\leq t, \\
				c_{p+t-b}a+b,~\text{for}~t< b\leq p-1, \\
				m_{b~(\text{mod}~p),t}+jp,~\text{for}~ p\leq b= jp+(b~(\text{mod}~p))\leq a-1
			\end{array}\right.
		\end{align}
		In other words there is a unique $m_{b,t}$ with $0\leq m_{b,t}\leq pa-1$ that satisfies $m_{b,t}= b~(\text{mod}~a)$
		and $m_{b,t}=t ~(\text{mod}~p)$.
	\end{proposition}
	
	\begin{corollary}\label{c3.6.13}
		\begin{enumerate}
		\item[(i)] $m_{0,1}=s_0a$,
		
		\item[(ii)] $m_{1,0}=fp=(p-s_0)a+1$,
		
		\item[(iii)] $m_{t,t} =t$ for $0\leq t\leq p-1$ and $m_{t+jp,t} = t+jp$ for
		$0\leq t\leq p-1$ and $p\leq t+jp\leq a-1$ i.e., $m_{b, b(\text{mod}~p)} =b$.
		\end{enumerate}
	\end{corollary}
	\begin{proof} This follows from Remark \ref{r3.5.6}(ii)(b) and (c).
	\end{proof}
	\begin{proposition}\label{p3.6.14} Let $0\leq t\leq p-1, 0\leq b\leq a-1$.
		\begin{enumerate}
		\item[(i)] $pm_{b,t}=bp$ and $am_{b,t}=ta$.
		
		\item[(ii)] $p(1,0)=(p,0),  p(m_{1,t},m_{0,1})=(p,0)$;
		
		\item[(iii)] $a(1,0)=(a,0),  a(m_{b,1},m_{1,0})= (a,0)$;
		
		\item[(iv)] $a(m_{1,t}, m_{0,1}) =(ta,a)$~ \text{and}~
		
		\item[(v)] $p(m_{b,1}, m_{1,0}) =(bp,p)$.
		\end{enumerate}
	\end{proposition}
	We present the information in Proposition \ref{p3.6.12}, Corollory \ref{c3.6.13} and Proposition \ref{p3.6.14} as a table in Figure 15. In fact, more is
	true that will help us apply our Simultaneous Constraint Technique.
	\begin{theorem}\label{t3.6.15} Let $d=pa$ with $p, a$ both odd primes with $p<a$.
		\begin{enumerate}
		\item[(i)] The set
		\begin{align}
			\mathcal{M}_{{\bf W}} &=\left\{\lambda_\sigma~:~\sigma=\sigma_1^0=(1,0), \sigma_{b,t}^1=(m_{b,t},1),
			\sigma_{t}^a=(m_{1,t},m_{0,1})=(m_{1,t}, s_0a),\right. \nonumber\\
			&\left. \sigma_{b}^p=(m_{b,1},,m_{1,0})=(m_{b,1}, fp), 0\leq b\leq a-1, 0\leq t\leq p-1\right\}
		\end{align}
		is a minimal cover of ${\bf W}$.
		
		\item[(ii)] The overlaps in $\lambda$'s in $\mathcal{M}_{{\bf W}}$ are as follows.
		\begin{enumerate}
		\item[(a)] $\lambda_{p,0}\subset \lambda_{1,0}$ and also $\lambda_{p,0}\subset \lambda_{m_{1,t},s_0a}$= $\lambda_{\sigma_t^a}$ for $0\leq t\leq p-1$.
		
		\item[(b)]  $\lambda_{a,0}\subset \lambda_{1,0}$ and also $\lambda_{a,0}\subset \lambda_{m_{b,1},m_{1,0}}$= $\lambda_{\sigma_b^p}$ for $0\leq b\leq a-1$.
		
		\item[(c)] For $0\leq b\leq a-1$,
		$\lambda_{bp,p}\subset \lambda_{{m_{b,1}} , m_{1,0}}=\lambda_{\sigma_b^p}$
		and also $\lambda_{bp,p}\subset \lambda_{m_{b,t},1}$= $\lambda_{\sigma_{b,t}^1}$ for $0\leq t\leq p-1$.
		
		\item[(d)] For $0\leq t\leq p-1$,  $\lambda_{ta,a}\subset \lambda_{m_{1,t},m_{0,1}} = \lambda_{\sigma_t^a}$  and also
		$\lambda_{ta,a}\subset \lambda_{m_{b,t},1} = \lambda_{\sigma_{b,t}^1}$ for $0\leq b\leq a-1$.
		\end{enumerate}
		\item[(iii)] The generators $U_\sigma$'s satisfy the following relations.
		\begin{enumerate}
		\item[(a)] $U_{p,0}= U_{1,0}^p$ and also $U_{p,0}= U_{m_{1,t},s_0a}^p$.
		
		\item[(b)] $U_{a,0}= U_{1,0}^a$ and also $U_{a,0}= U_{m_{b,1}, fp}^a$.
		
		\item[(c)] Let $\mu_b= a-\frac{p-1}{2}b~(\text{mod}~a)$ for $1\leq b\leq a-1$ .
		For $0\leq b\leq a-1$,
		\begin{align}
			U_{bp,p}=\left\{ \begin{array}{l} U_{m_{b,1},fp}^p ~\text{for}~b=0\\
				e^{\left(\frac{2\pi i}{a}\mu_b\right)}U_{m_{b,1},fp}^p ~\text{for}~b\geq 1
			\end{array}\right.
		\end{align}
		and, also for $0\leq t\leq p-1$,
		\begin{align}
			U_{bp,p}=\left\{ \begin{array}{l} U_{m_{b,t},1}^p ~\text{for}~b=0~\\
				e^{\left(\frac{2\pi i}{a}\mu_b\right)}U_{m_{b,t},1}^p ~\text{for}~b\geq 1.
			\end{array}\right.
		\end{align}
		
		\item[(d)] Let $\nu_t= p-\frac{a-1}{2}t~(\text{mod}~p)$ if $a-1$ is not a multiple of $p$ and  $1\leq t\leq p-1$. For $0\leq t\leq p-1$,
		\begin{align}
			U_{ta,a}=\left\{ \begin{array}{l} U_{m_{1,t},s_0a}^a ~\text{for}~t=0~\text{or}~a-1, a~\text{multiple of}~p,\\
				e^{\left(\frac{2\pi i}{p}\nu_t\right)}U_{m_{1,t},s_0a}^a  ~\text{for}~t \geq 1, a-1~\text{not a multiple of}~p
			\end{array}\right.
		\end{align}
		and, also for $0\leq b\leq a-1$,
		\begin{align}
			U_{ta,a}=\left\{ \begin{array}{l} U_{m_{b,t},1}^a ~\text{for}~t=0~\text{or}~a-1, a~ \text{multiple of} ~p,\\
				e^{\left(\frac{2\pi i}{p}\nu_t\right)}U_{m_{b,t},1}^a ~\text{for}~t\geq 1, a-1~\text{not a multiple of}~p.
			\end{array}\right.
		\end{align},
		\end{enumerate}
		\end{enumerate}
	\end{theorem}
	\begin{proof}  We note that $\mathcal{M}_{{\bf W}}$ has cardinality $1+d+p+a=(1+p)(1+a)$ in line with \cite{shvo1} and \cite{shvo2}.
	\begin{enumerate}
		\item[(i)] To see that  $\mathcal{M}_{{\bf W}}$ covers ${\bf W}$, all we have to do is to use Remark \ref{r3.5.6}.
		
		Consider any $(m,n) \in \ZZ_d\times\ZZ_d\smallsetminus \{(0,0)\}$. The cases $n=0, m=0, m=n $ are covered by $\lambda_\sigma$ for $\sigma =\sigma_1^0, \sigma_{0,0}^1, \sigma_{1,1}^1$ respectively. So it is enough to confine attention to the case $\text{HCF}(m,n)=1$, i.e.,
		$[m,n]=1 $. We consider different subcases.
		\begin{enumerate}
		\item[(a)] Remark \ref{r3.5.6} (i)(a) deals with the cases HCF $(n,pa) =1$, i.e., $[n,pa]=1$.
		
		\item[(b)] Let $n=n_1a$ with $[n_1,pa]=1$, then $[m,a]=1$. Let $m^\prime =\hat{n}_1s_0m$ so that $ m=n_1\hat{s}_0 m^\prime$. Then $(m,n)
		=n_1\hat{s}_0(m^\prime,s_0a)$. Also $[m^\prime, a]=1$. Let $t=m^\prime ~(\text{mod}~p)$. Then $m^\prime =bp+m_{1,t}$ for some
		$0\leq b\leq a-1$. Now $m_{1,t}=c_{t-1}a+1$, taking $c_{-1}=c_{p-1}=p-s_0$. So $(bp+1)(m_{1,t},s_0a)=(bp+c_{t-1}a+1,s_0a)=(m^\prime, s_0a)$. Therefore,
		$(m,n) =(bp+1)n_1\hat{s}_0(m_{1,t},s_0a)$. Hence $(m,n)\in \lambda_{m_{1,t},s_0a}=\lambda_{\sigma_t^a}$.
		
		\item[(c)] Let $n=n_1p$ with $[n_1,pa]=1$. Then $[m,p]=1$.
		Let $m^\prime =\hat{n}_1(ga+f)m$ so that $(m,n)=n_1(s_0a+p)(m^\prime,fp)$. Let $b=m^\prime~(\text{mod}~a)$. Then $m^\prime =ra+m_{b,1}$
		for some $0\leq r\leq p-1$. Now $m_{b,1}=cp+1$ for some $0\leq c\leq a-1$. So $ (ra+1)(m_{b,1},fp)=(ra+1)(cp+1,fp)$ =
		$(ra+cp+1,fp)=(m^\prime,fp)$. Therefore, $(m,n)=n_1(s_0a+p)(ra+1)(m_{b,1},fp)$. Hence $(m.n) \in \lambda_{(m_{b,1},fp)}= \lambda_{\sigma_b^p}$.

		\item[(d)] Let $n=n_1p^k$ for some $2\leq k\leq k_0+1 $ and $[n_1,pa]=1$. Then $[m,p]=1$. Now $n=n_1p^k =n_1p(p^{k-1}+a)$. We note that
		$[p^{k-1}+a,pa]=1$ and also $p+1\leq p^{k-1}+a<a+a=2a$. So $1\leq p^{k-1}+a<pa$. Thus by (i)(a), $(p^{k-1}+a)^\wedge$ is available.
		Now $(m,n) =(m,n_1p(p^{k-1}+a))= (p^{k-1}+a)(m(p^{k-1}+a)^\wedge,n_1p)$. But $(m(p^{k-1}+a)^\wedge,n_1p)$ comes under the subcase in (d)
		above.
		
		Hence $\mathcal{M}_{{\bf W}}$ covers ${\bf W}$.
		\end{enumerate}
		\item[(ii)] It follows from Proposition \ref{p3.6.14}.
		
		\item[(iii)] Putting $j=p$ and $j=a$ in Proposition \ref{p3.5.1}, we have the following for $(m,n)\in \ZZ_d\times\ZZ_d$.
		\begin{enumerate}
		\item[($\alpha$)] $U_{pm,pn} = e^{\left(-\frac{2\pi i}{a}\frac{p-1}{2}mn\right)}U_{m,n}^p$, and
		
		\item[($\beta$)] $U_{am,an} = e^{\left(-\frac{2\pi i}{p}\frac{a-1}{2}mn\right)}U_{m,n}^a$.
		\end{enumerate}
		\begin{enumerate}
		\item[(a)] follows immediately from the Master equations ($\alpha$) by taking $(m,n)= (1,0)$, $(m,n)=(m_{1,t},s_0a)$ because $mn$ is a multiple
		of $a$.
		
		\item[(b)] It follows immediately from $(\beta)$ by taking $(m,n)=(1,0)$, $(m_{b,1},fp)$ as $mn$ is a multiple of $p$.
		
		\item[(c)] and (d) The numbers $\mu_b$ and $\nu_t$ have been assigned the values so as to satisfy the desired equations.
		\end{enumerate}
		\end{enumerate}
	\end{proof}
	
	\begin{example}\label{e3.6.16} It is a good idea to understand the constraints provided by the overlaps in Theorem \ref{t3.6.15} just above in their own right
		and for further use.
		
		\def\Q{\mathbf{Q}}
		\begin{enumerate}
		\item[(i)] For $\sigma = (bp,p)$, $0 \leq b \leq a-1$ or $(p,0)$, the corresponding
		$U_\sigma$ has $a$ distinct eigenvalues,
		each one of multiplicity $p$. This will be explained shortly for
		$1 \leq b \leq a-1$, the other cases being easier.\
		The corresponding constraints, say $\Q^{(b)}$ for
		$\sigma=(bp,p)$ for $0 \leq b \leq a-1$ or $\Q^{(a)}$ for
		$\sigma =(p,0)$ will thus be uniform in the terminology of Remark \ref{r3.1.3}(iii).
		
		Let $1 \leq b \leq a-1$ and set $U=U_{bp,p}$. Then for $0 \leq k \leq d-1$,
		\begin{align}
			U|k\rangle
			&=e^{\left(\frac{2\pi i bpk}{d}\right)} |p+ k \rangle \nonumber\\
			&=e^{\left(\frac{2\pi i bk}{a}\right)} |p+k\rangle
		\end{align}
		Put $\beta =e^{\left(\frac {2 \pi i bp} {a} \right)}$.
		Fix $0\leq t\le p-1$, $0 \leq c \leq a-1$ and set
		\begin{align*}
			\gamma=e^{\left(\frac{2 \pi i bt} {a}\right)},\quad
			\delta=e^{\left(\frac{2 \pi i c} {a}\right)}.
		\end{align*}
		Then for $0\leq c^\prime \leq a-1$, we note that
		$U|c^\prime p+t\rangle =\beta^{c^\prime} \gamma|(c^\prime+1)p+t\rangle$ and
		put $l_{c^\prime} =\left(\frac{c^\prime(c^\prime+1)}{2}\right)$.
		
		We now set $\xi =\xi_{t,c}= \displaystyle\sum_{c^\prime=0}^{a-1} \beta^{l_{c^\prime}}\bar\delta^{c^\prime}|c^\prime p+t\rangle$.
		
		Then
		\begin{align*}
			U\xi
			&=\displaystyle\sum_{c^\prime=0}^{a-1}\beta^{l_{c^\prime}} \bar\delta^{c^\prime} \beta^{c^\prime} \gamma|(c^\prime+1)p+t\rangle\\
			&=\bar{\beta} \gamma \delta \displaystyle\sum_{c^\prime=0}^{a-1} \beta^{l_{c^\prime+1}} \bar\delta^{c^\prime+1} |(c^\prime+1)p+t\rangle\\
			&=\bar\beta \gamma\delta\xi.
		\end{align*}
		Now
		\begin{align*}
			\bar{\beta} \gamma \delta
			&=e^{\left(\frac{2\pi i} {a} (-bp+bt+c)\right)}\\
			&=e^{\left(\frac{2\pi i} {a} (c-b(p-t)) \right)}.
		\end{align*}
		For $0 \leq t \leq p-1$, we have $1 \leq p-t \leq p \leq a-1 $.
		Also $1\leq b \leq a-1$. So $\{b(p-t)~(\text{mod}~a),\ 0\leq t\leq p-1\}$
		consists exactly of $p$ elements from $\{1,2,\ldots,a-1\}$.
		So $c-b(p-t)(\mathrm{mod}\, \ a)$ varies over $\{0,1,\ldots,a-1\}~ p$-times as
		$c$ varies from 0 to $a-1$. For $0\le b^\prime\le a-1$, let
		$\tilde{S}^{b^\prime}=\{(t,c):0\le t\le p-1,\, 0\le c\le a-1,\, c-b(p-t)\equiv b^\prime~(\text{mod}\, a)\}$.
		Then $\#\tilde{S}^{b^\prime}=p$. Also for $(t,c) \in S^{\tilde{b^\prime}}$, corresponding
		$ \xi_{t,c}$'s are linearly independent. So their span $E_{b^\prime}$ is the $p$-dimensional
		eigenspace corresponding to the eigenvalue $e^{\left(\frac{2 \pi i b^\prime} {a}\right)}$ of $U=U_{bp,p}$.
		Set ${Q}_{b^\prime}=\text{the projection on}~E_{b^\prime}$. Then $\mathbf{Q}^{(b)} = \{ Q_{b^\prime} : 0\leq b^\prime \leq a-1\}$.
		
		\item[(ii)] Consider unitary operators $U_{a,0}$, $U_{ta,a}$ $(0\leq t\leq p-1)$
		coming from specified generators of
		$\lambda_{a,0}$, and $\lambda_{ta,a}$ $(0\leq t\leq p-1 )$
		respectively. Each of them has $p$ distinct eigenvalues of multiplicity $a$.\
		The corresponding constraint ${\bf R}$ is  uniform of size $a$ and has $p$ elements.
		We denote the constraint corresponding to $U_{a,0}$ by ${\bf R}^{(p)}$ and the constraint for $U_{ta,a}$ by ${\bf R}^{(t)}$ for $0 \leq t\leq p-1$.
		\end{enumerate}
		\begin{figure}[H]
			\begin{center}
				\includegraphics[width=0.9\linewidth]{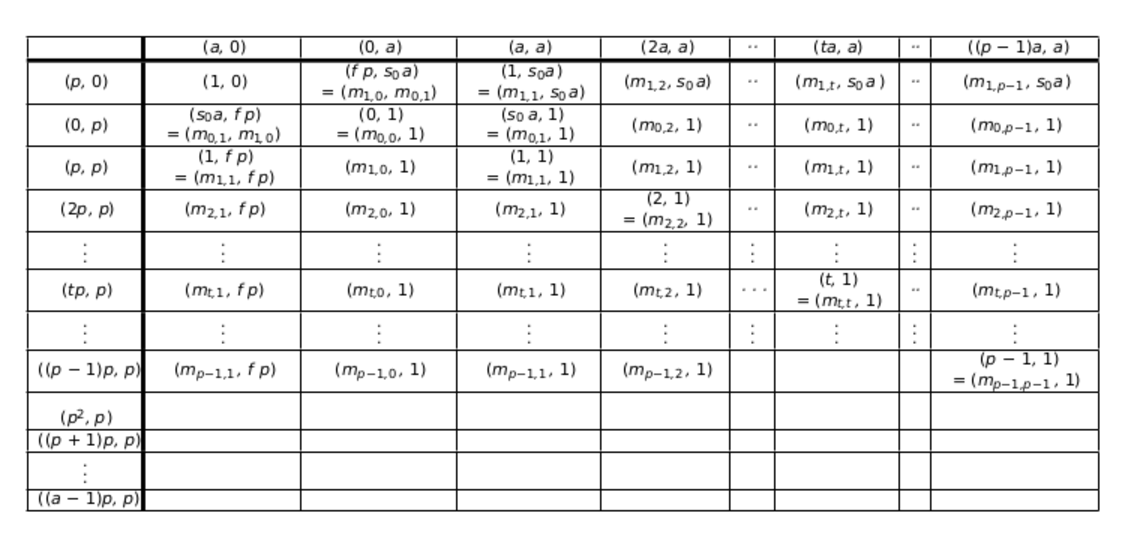}
				\caption[]{$c_t = ts_0~(\text{mod}~p), 0\leq t\leq p-1$;\hfill\break
                \begin{minipage}{\linewidth}
                \begin{equation*}
                	m_{b,t}=\begin{cases} c_{t-b}a+b, 0\leq b\leq t\\
                		c_{p+t-b}a+b, t< b\leq p-1  \\
                		m_{b(\text{mod}~ p),t}+b-b(\text{mod}~p), p\leq b\leq a-1;
                \end{cases}
                \end{equation*}
                \end{minipage}
					Note $c_0=0, c_1=s_0, c_{p-t}=p-c_t, m_{t,t}=t, m_{b,t}=b~\text{for}~t=b~(\text{mod}~p).$
					Block of size $p\times (p+1)$  consisting of horizontal lines with labelling $(0,p)$ to $((p-1)p,p)$
					has to be repeated by adding $(p^2, 0)$
					to the labelling and $(p,0)$ to every   element in the block.
					This action is repeated cumulatively till we reach the last block, which is truncated because the last labelling is $((a-1)p,p)$.}
			\end{center}
		\end{figure}
	\end{example}

	\begin{theorem}\label{t3.6.17} Simultaneous Constraint Technique leads to the ideal number $d^2$ of a complete rank one projection system for $d=pa$
		with $p,a$ distinct odd primes, say $p<a$.
	\end{theorem}
	\begin{proof} The ground is already set to a reasonable extent. We shall give two approaches. We start with $\mathcal{M}_{\bf W}$ as listed in
		Theorem \ref{t3.6.15} above and refer to the table in Figure 15 for help to understand. Consider any $\sigma=(m,n)$ listed in $\mathcal{M}_{\bf W}$.
		\begin{enumerate}
		
		\item[(i)] First of all, because $d=pa$ is odd, by Corollary \ref{c3.5.2}, the eigenvalues of $U_{m,n}$ are $\{\eta_d^k:0\leq k\leq p-1\}$=
		$\{e^{\left(\frac{2\pi i k}{pa}\right)}: 0\leq k\leq pa-1\}$ just because $U_{m,n}$ is a generator of full size ${\bf W}$-MASS.
		As a consequence, the eigenvalues of $U_{m,n}^p$ and $U_{m,n}^a$ are respectively $\{e^{\left(\frac{2\pi i b}{a}\right)}: 0\leq b\leq a-1\}$, each
		of multiplicity $p$ and $\{e^{\left(\frac{2\pi i t}{p}\right)}: 0\leq t\leq p-1\}$ each of multiplicity $a$.
		
		Let $\mathcal{P}_\sigma$ be a complete system of orthonormal unit eigenvectors $\{\xi_{\sigma,k}:0\leq k\leq pa-1\}$ corresponding to the
		listing $\{\eta_d^k: 0\leq k\leq pa-1\}$ of eigenvalues. For $0\leq k\leq pa-1$, let $P_{\sigma,k}=|\xi_{\sigma,k}\rangle\langle\xi_{\sigma, k}|$
		the rank one projections.
		
		Then $U_\sigma =\sum_{k=0}^{d-1}\eta_d^k P_{\sigma,k}$,
		\begin{align}
			U_\sigma^p = \sum_{j=0}^{a-1} \eta_a^j \left(\sum_{k\in K_j} P_{\sigma,k}\right) ~~~~~~~~~\text{and}~~~~~~~~
			U_\sigma^a = \sum_{s=0}^{p-1} \eta_p^s \left(\sum_{k\in C_s} P_{\sigma,k}\right).
			\label{0.47}
		\end{align}
		By Theorem \ref{t3.6.15} (iii), we have the following relations:
		\begin{align}
			&(a)~~~~~~ U_{p,0} = \sum_{j=0}^{a-1} \eta_a^j \left(\sum_{k\in K_j} P_{\sigma,k}\right), ~\text{for}~ \sigma = (1,0),
			(m_{1,t},s_0a), 0\leq t\leq p-1, \nonumber\\&~\text{i.e.},~\sigma\text{'s in the first row} \end{align}
			\begin{align}
			&(b)~~~~~~ U_{a,0} = \sum_{s=0}^{p-1} \eta_p^s \left(\sum_{k\in C_s} P_{\sigma,k}\right), ~\text{for}~ \sigma = (1,0),
			(m_{b,1}, fp), 0\leq b\leq a-1,~\nonumber\\ &\text{i.e.},~\sigma\text{'s in the first column} \end{align}
			\begin{align}
			&(c)~~~~~~ U_{0,p} = \sum_{j=0}^{a-1} \eta_a^j \left(\sum_{k\in K_j} P_{\sigma,k}\right), ~\text{for}~ \sigma = (0,1),
			(m_{0,t}, 1), 0\leq t\leq p-1, \nonumber\\ &~\text{i.e.},~\sigma\text{'s in the second row}\end{align}
			\begin{align}
			&(d)~~~~~~ U_{0,a} = \sum_{s=0}^{p-1} \eta_p^s \left(\sum_{k\in C_s} P_{\sigma,k}\right), ~\text{for}~ \sigma = (0,1),
			(m_{b,0}, 1), 0\leq b\leq a-1,\nonumber\\ & ~\text{i.e.},~\sigma\text{'s in the second column}. \end{align}
			\begin{align}
			(e)~~~~~~~~~~&~\text{For}~ 1\leq b\leq a-1, \nonumber\\
			&U_{bp,p} = \sum_{j=0}^{a-1} \eta_a^{(j+\mu_b)~(\text{mod}~a)} \left(\sum_{k\in K_j} P_{\sigma,k}\right),
			\nonumber\\&~\text{for}~ \sigma = (m_{b,1},fp),
			(m_{b,t}, 1), 0\leq t\leq p-1, \nonumber\\ & ~\text{i.e.},~\sigma\text{'s in the}~(b+2)^{\text{th}}~ \text{row labelled by} (bp,p).
			\end{align}
			\begin{align}
			(f)~~~~~&U_{ta,a} = \sum_{s=0}^{p-1} \eta_p^s \left(\sum_{k\in C_s} P_{\sigma,k}\right), \nonumber\\&~\text{for}~ \sigma= (m_{1,t}, s_0a),
			(m_{b,t}, 1), 0\leq b\leq a-1\nonumber\\ & ~\text{i.e.},~\sigma\text{'s in the}~(t+2)^{\text{th}}~\text{column labelled by}~(ta,a),\end{align}
			\begin{align}
			(g)~~~~~& ~\text{For}~a-1, \text{not a multiple of}~p, 1\leq t\leq p-1, \nonumber\\
			~~~~~~~~&U_{ta,a} = \sum_{s=0}^{p-1} \eta_p^{(s+\nu_t)(\text{mod}~p)} \left(\sum_{k\in C_a} P_{\sigma,k}\right), 
			\nonumber\\&~\text{for}~ \sigma= (m_{1,t}, s_0a),
			(m_{b,t}, 1), 0\leq b\leq a-1,\nonumber\\ & ~\text{i.e.},~\sigma\text{'s in the}~(t+2)^{\text{th}}~\text{column labelled by}~(ta,a).
		\end{align}
		\item[(ii)] For $h=1$ or $2$, $0\leq j\leq a-1$, let $Q_{h,j}$ be the projection on the eigenspace $E_{h,j}$ corresponding to the eigenvalue $\eta_a^j$ of
		$U_{p,0}$ and $U_{0,p}$ respectively. For $h=b+2$ , $1\leq b\leq a-1$,  $0\leq j\leq a-1$, let $Q_{h,j}$ be the projection on the
		eigenspace $E_{h,j}$ corresponding to the eigenvalue $\eta_a^{(j+\mu_b)~(\text{mod}~a)}$ of $U_{bp,p}$.  Then for $1\leq h\leq a+1$, $0\leq j\leq a-1$,
		$ Q_{h,j} =\sum_{k\in K_j} P_{\sigma,k} $ for $\sigma$'s in the $h^{\text{th}}$ row.
		
		Let $1\leq h\leq a+1$. Thus $\mathcal{P}_\sigma=\{P_{\sigma,k}: 0\leq k\leq pa-1\}$ is ${\bf Q}_{h}$-constrained with
		${\bf Q}_h=\{Q_{h,j}: 0\leq j\leq a-1\}$ with matching listing of cosets $K_j$'s occurring in $U_\sigma^p$ noted in $(\ref{0.47})$ in (i)
		for all $\sigma$'s in the $h^{\text{th}}$ row.
		
		\item[(iii)] For $r=1$ or $2$, $0\leq s\leq p-1$, let $R_{r,s}$ be the projection on the eigenspace $G_{r,s}$ corresponding to the
		eigenvalue $\eta_p^s$ of $U_{a,0}$ and $U_{0,a}$ respectively. For $r=t+2, 1\leq t\leq p-1, 0\leq s\leq p-1$, let $R_{r,s}$
		be the projection on the eigenspace 
		$G_{r,s}$ for $U_{ta,a}$ corresponding to the eigenvalue $\eta_p^s$ if $a-1$ is a multiple of $p$ and
		 $\eta_p^{(s+\nu_t)~(\text{mod}~p)}$ if $a-1$ is not
		a multiple of $p$. Then for $1\leq r\leq p+1$, $0\leq s\leq p-1$, $R_{r,s}=\sum_{k\in C_s} P_{\sigma,k}$ for $\sigma$'s in the $r^{\text{th}}$ column.
		Let $1\leq r\leq p+1$. Then $\mathcal{P}_\sigma =\{P_{\sigma,k} : 0\leq k\leq pa-1\}$ is ${\bf R}_r$-constrained with ${\bf R}_r=\{R_{r,s}:0\leq
		s\leq p-1\}$ with matching list of cosets $C_s$ occurring in $U_\sigma^a$ noted in
		$(\ref{0.47})$ in (i) for all $\sigma$'s in the $r^{\text{th}}$ column.
		
		\item[(iv)] By (ii) and (iii), for any given  $\sigma$ (in the list occurring in $\mathcal{M}_{\bf W}$), $\mathcal{P}_\sigma =
		\{P_{\sigma,k}: 0\leq k\leq pa-1\}$ is ${\bf Q}_h$-constrained as well as ${\bf R}_r$-constrained if and only if $\sigma$ lies in the
		$h^{\text{th}}$ row and  $r^{\text{th}}$ column in the Table in Figure 15. The constraints can be understood from the same prototype
		as in the Table in Figure 16. For each $\sigma$, the arrangement of $P_{\sigma,k}$ is identical with the part of Table in Figure 15 restricted to the
		first coordinates, $2^{\text{nd}}$ row and $2^{\text{nd}}$ columns onwards.
		
		\item[(v)] It is clear from the Table in Figure 16 that, if $Q$'s and $R$'s are known then, for
		estimation purposes, one full row and one full column of $P_{\sigma,k}$'s can be deleted
		as they are available linearly in terms of ${P}_{\sigma,k}$'s in the remaining rectangular block of size $(a-1)\times(p-1)$ . The same effect can be
		achieved by deleting $p+a-1$ elements from the rectangle so that each row and each column has at least one element removed but at most one
		element shares row and column with two more elements.
		\vspace{-0.3cm}
		\begin{figure}[H]
			\centering
			\begin{small}
				\begin{tabular}{|c?c|c|c|c|c|c|c|} \hline
					&                &                      &                &                      &         &                       & \begin{tabular}[x]{@{}c@{}}Row\\Sum\end{tabular}\\ \thickhline
					& $P_{\sigma,0}$ & $P_{\sigma,m_{0,1}}$ & $\cdots$       & $P_{\sigma,m_{0,t}}$ &$\cdots$ &$P_{\sigma,m_{0,p-1}}$ &$Q_{h,0}$\\ \hline
					& $P_{\sigma,m_{1,0}}$ & $P_{\sigma,1}=P_{\sigma,m_{1,1}}$ & $\cdots$       & $P_{\sigma,m_{1,t}}$ &$\cdots$ &$P_{\sigma,m_{1,p-1}}$ &$Q_{h,1}$\\ \hline
					& $ \vdots$      &    $\vdots$          &$\vdots$        & $\vdots$             & $\vdots$ & $\vdots$             &$\vdots$  \\ \hline
					& $P_{\sigma,m_{t,0}}$ & $P_{\sigma,m_{t,1}}$ & $\cdots$ & $P_{\sigma,t}=P_{\sigma,m_{t,t}}$ &$\cdots$ &$P_{\sigma,m_{t,p-1}}$ &$Q_{h,t}$\\ \hline
						& $ \vdots$      &    $\vdots$          &$\vdots$        & $\vdots$             & $\vdots$ & $\vdots$             &$\vdots$  \\ \hline
					&                &                      &    &                      &         &                       &          \\ \hline
					& $P_{\sigma,m_{p-1,0}}$ & $P_{\sigma,m_{p-1,1}}$ & $\cdots$ & $P_{\sigma,m_{p-1,t}}$ &$\cdots$ &$P_{\sigma,p-1}=P_{\sigma,m_{p-1,p-1}}$ &$Q_{h,1}$\\ \hline
					
					& $ \vdots$      &    $\vdots$          &$\vdots$        & $\vdots$             & $\vdots$ & $\vdots$             &$\vdots$  \\ \hline

					& $P_{\sigma,m_{b,0}}$ & $P_{\sigma,m_{b,1}}$ & $\cdots$ & $P_{\sigma,m_{b,t}}$ &$\cdots$ &$P_{\sigma,m_{b,p-1}}$ &$Q_{h,b}$\\ \hline
					
					& $ \vdots$      &    $\vdots$          &$\vdots$        & $\vdots$             & $\vdots$ & $\vdots$             &$\vdots$  \\ \hline

					& $P_{\sigma,m_{a-1,0}}$ & $P_{\sigma,m_{a-1,1}}$ & $\cdots$ & $P_{\sigma,m_{a-1,t}}$ &$\cdots$ &$P_{\sigma,m_{a-1,p-1}}$ &$Q_{h,a-1}$\\ \hline
					\begin{tabular}[x]{@{}c@{}}Column\\Sum\end{tabular}& $R_{r,0}$ & $R_{r,1}$ & $\cdots$ & $R_{r,t}$ &$\cdots$ &$R_{r,p-1}$ &$I_{pa}$\\ \hline
				\end{tabular}
			\end{small}
			\caption{$m_{b,t}$'s are as in Figure 15 that satisfy $m_{b,t}=b~(\text{mod}~a),~ m_{b,t}=t~(\text{mod}~p)$. $\sigma$ is in the $h^{\text{th}}$ row
				and $r^{\text{th}}$ column. First block of size $p\times p$ is repeated after adding $p$ to the second subscript of each element and then again after adding
				another $p$ and so on till we reach the last block being a truncated one of size $r_0\times p$. }
		\end{figure}
	
		To determine $Q_{h,j}$'s and $R_{r,s}$'s we first give the practical approach followed in this paper and indicate theoretical
		one in  Remark \ref{r3.6.18} after the proof.
		
		\item[(vi)] {\bf Anchors.} We choose the diagonals in the blocks in the Table in Figure 15 and label them by the row numbers $h=1,2,\cdots,a+1$ as
		anchors to determine the constraints. We make use of available constraints on cumulative basis for this purpose.
		
		To elaborate, we go in the following way stepwise.
		\begin{enumerate}
		\item[(a)] We start with the simplest one,
		$$\mathcal{P}_1=\mathcal{P}_{(1,0)} =\{P_{1,k}=P_{(1,0),k}=|k\rangle\langle k|: 0\leq k\leq pa-1\}.$$
		This gives us ${\bf Q}_1$ and ${\bf R}_1$ using (ii) and (iii). It also gives the trivial constraint $\{I_d\}$.
		
		\item[(b)] We now come to the cases $2\leq h\leq p+1$ viz., $h=b+2, 0\leq b\leq p-1$ and take $\mathcal{P}_h=\mathcal{P}_{b+2}=\mathcal{P}_{(b,1)}$
		=$\{P_{h,k}=P_{(b,1),k} : 0\leq k\leq pa-1\}$ the complete system of rank one projections listed as corresponding to the
		eigenvakues $\{\eta_d^k : 0\leq k\leq pa-1\}$ of $U_{b,1}$. This gives us ${\bf Q}_h$ and ${\bf R}_h$ using (ii) and (iii).
		
		The useful fact is that by Constraints Technique  3.1.2, for estimation purposes, $\mathcal{P}_h$ can be replaced by
		$\mathcal{P}_h^\prime$ obtained from $\mathcal{P}_h$ by deleting any one, say, $P_{h,pa-1}$, i.e.,
		$\mathcal{P}_h^\prime= \{P_{h,k}: 0\leq k\leq pa-2\}$. In practice, we need not compute $P_{h,pa-1}$.
		
		\item[(c)] we now come to the cases $p+2\leq h\leq a+1$, viz., $h=b+2, p\leq b \leq a-1$, and take $\mathcal{P}_h =\mathcal{P}_{b+2}
		=\mathcal{P}_{(b,1)}=\{P_{h,k}=P_{(b,1),k} : 0\leq k\leq pa-1\}$, the complete system of rank one projections listed as corresponding to
		the eigenvalues $\{\eta_d^k :0\leq k\leq pa-1\}$ of $U_{b,1}$. This gives us ${\bf Q}_h$ using (ii).
		
		Also,  $\mathcal{P}_h$ is ${\bf R}_{h~(\text{mod}~p)}$-constrained so the use of Constraint Technique 3.1.2 gives us that, for estimation
		purposes, $\mathcal{P}_h$ can be replaced by $\mathcal{P}_h^\prime$ after removing any row, or any element from each column,
		say $\{(a-1)p+s, 0\leq s\leq p-1\}$  so that $\mathcal{P}_h^\prime =\{P_{h,k}=P_{(b,1),k} : 0\leq k\leq p(a-1)-1\}$. In practice, we
		need not compute $P_{h,k}, pa-p\leq k\leq pa-1$.
		
		\item[(d)] Thus we have all the constraints in hand now using $\mathcal{P}_1, \mathcal{P}_h^\prime, 2\leq h\leq a+1$.
		\end{enumerate}
		\item[(vii)] {\bf Use of Simultaneous Constraint Technique.} We now deal with remaining $\sigma$'s in $\mathcal{M}_{{\bf W}}$. As said in (v) for
		estimation purposes, we can replace $\mathcal{P}_\sigma$ by $\mathcal{P}_\sigma^\prime$ by deleting one row and one column ( or other
		modes indicated in (v)), say, $\mathcal{P}_\sigma^\prime =\{P_{\sigma,k} : k=m_{b,t}, 0\leq b\leq a-2, 0\leq t \leq p-2\}$. We may label these 
		$\sigma$'s and accordingly, $\mathcal{P}^\prime_\sigma$'s the way we like by numbers $z$ from $a+2$ to $(p+1)(a+1)$.
		
		\item[(viii)] {\bf The final step.}  We collect $\mathcal{P}_1$, $\mathcal{P}^\prime_h$'s
		from (vi) and (vii) to form a set $\mathcal{P}^\prime$ of rank one projections and
		claim that it is a complete system of rank one projections of the ideal size $d^2=(pa)^2=p^2a^2$. We elaborate as follows.
		\begin{enumerate}
		\item[(a)] Because $\{U_{(m,n)}:(m,n)\in \ZZ_d\times\ZZ_d\}$ is a basis for $M_d=\mathcal{B}(\mathcal{H})$,  we have that $\cup\{\mathcal{P}_\sigma :
		\sigma\in \mathcal{M}_{\bf{W}}\}$ is informationallly complete system of rank one projections. Using (v) to (vii), this set can be reduced
		with the same capacity to be informationally complete by replacing it by
		$\mathcal{P}^\prime =\mathcal{P}_1\cup \cup_{h=2}^{p+1}\mathcal{P}_h^\prime \cup\cup_{h=p+2}^{a+1}\mathcal{P}_h^\prime\cup \cup_{z=a+2}^{(p+1)(a+1)}
		\mathcal{P}_z^\prime$.
		
		\item[(b)] \begin{align}
			\text{Size of}~\mathcal{P}_1 &= pa, \nonumber\\
			\text{Size of}~\mathcal{P}_h^\prime &= pa-1~\text{for}~2\leq h\leq p+1, \nonumber\\
			\text{Size of}~\mathcal{P}_h^\prime &= pa-p~\text{for}~p+2\leq h\leq a+1, \nonumber\\
			\text{Size of}~\mathcal{P}_z^\prime &= (p-1)(a-1) ~\text{for}~a+2\leq z\leq (p+1)(a+1). \nonumber
		\end{align}
		So, the size of $\mathcal{P}^\prime$ is $pa+p(pa-1)+(a-p)(pa-p)+p(a+1)(p-1)(a-1)=p^2a^2=d^2$.
		\end{enumerate}
		\end{enumerate}
	\end{proof}
	
	\begin{remark}\label{r3.6.18}
		\begin{itemize}
			\item[(i)] We can start with Example \ref{e3.6.16} and obtain the constraints ${\bf Q}^{(b)}$'s and ${\bf R}^{(t)}$'s
			and then relabel the projections involved keeping in mind discussion in (i)-(iii) of the proof of Theorem \ref{t3.6.17} above and obtain
			constraints ${\bf Q}_h, 1\leq h\leq a+1$ and ${\bf R}_r, 1\leq r\leq p+1$.
			\item[(ii)]
			Now it is clear from the Table in Figure 16 that for $\sigma$ listed in $\mathcal{M}_{\bf W}$,
			$P_{\sigma,m_{b,t}} = Q_{h,b}R_{r,t}$ for $\sigma$ in the $h^{\text{th}}$ row and $r^{\text{th}}$ column in Figure 15
			for $0\leq b\leq a-1$, $0\leq t\leq p-1$. Thus we obtain all $\mathcal{P}_\sigma$'s for $\sigma$'s listed in $\mathcal{M}_{\bf W}$. Now the
			arguments in parts (vi) to (viii) of the proof of Theorem \ref{t3.6.17} can be applied.
		\end{itemize}
	\end{remark}
	
	\subsection{Weyl operators and tensor product technique}
	
	The purpose of this subsection is  to give methods to obtain a complete system of rank one projections for any dimension $d\geq 2$
	together with quantum mechanical overlaps for some methods.
	
	We first recall that for $B_j$, a basis for a vector space $X_j,  1\leq j\leq k$ and $X=\otimes_{j=1}^{k} X_j$, the set
	$B=\{{\bf b}=\otimes_{j=1}^{k}b_j, b_j\in B_j~\text{for}1\leq j\leq k\}$ is a basis for $X$ and $\# B=\prod_{j=1}^k \# B_j$. We
	take $k\geq 2$ and dimension $X_j$ also $\geq 2$ for $1\leq j\leq k$.
	
	Next let $\mathcal{H}_j$ be a Hilbert space of dimension $d_j\geq 2$ and $X_j=\mathcal{B}(\mathcal{H}_j)$ and $B_j$, a basis for $X_j$,
	for $1\leq j \leq k$. Let $\mathcal{H}=\otimes_{j=1}^{k}\mathcal{H}_j$. If each $B_j$ consists of unitary operators (respectively, Hermitian operators, positive operators, rank one
	projections) on $\mathcal{H}_j$, then each ${\bf b}$ in $B$ is unitary ( respectively, Hermitian, positive, rank one projection).
	
	\subsubsection{\it Tensor product technique} For 
	$1\leq j \leq k$, let $B_j$ consist of rank one projections on the Hilbert space $\mathcal{H}_j$ of
	dimension $d_j\geq 2$ and $C_j=\{ \text{tr}(P_1P_2):P_1, P_2 \in B_j\}$, the set of quantum mechanical overlaps for $B_j$.
	Then $B=\{ {\bf b}=\otimes b_j, b_j\in B_j~\text{for}~1\leq j\leq k\}$ is a basis of rank one projections on the Hilbert space
	$\mathcal{H}=\otimes_{j=1}^{k}\mathcal{H}_j; \# B=d^2$ with $d=\prod_{j=1}^{k} d_j$. Further, the set of quantum mechanical overlaps
	for $B$ is $C=\{1,\prod_{j\in S}c_j, c_j\in C_j, c_j\neq 0~\text{or}~1, \emptyset \neq S\subset \{1,2,\cdots,k\}\}$ together with $0$, in case $0$ is in some $C_j$.
	This follows on noting that for ${\bf b}\neq {\bf b}^\prime$ in $B$,
	$S= \{j: b_j\neq b_j^\prime\}, \text{tr}({\bf b}{\bf b}^\prime)=\prod_{j\in S} \text{tr}(b_jb_j^\prime)$.
	
	\begin{example}\label{e3.7.1} {(Weyl Operators).} This is based on 2.2 above. (See \cite{3}). Let $d=\prod_{j=1}^{k} p_j^{s_j}, 2 \leq p_1<p_2\cdots <p_k,$ all
		primes and $s_j\geq 1$ for $1\leq j\leq k$. Set $d_j=p_j^{s_j}, \mathcal{H}_j=L^2(\mathbb{F}_{d_j})$, $X_j=B(\mathcal{H}_j)$ for $1\leq j\leq k$ and then $\mathcal{H}=\otimes_{j=1}^k \mathcal{H}_j$. Then dim($\mathcal{H})$ =$d$.
		\begin{enumerate}
		\item[(i)] Consider any $1\leq j\leq k$. Using 2.2.2 and the fact that
		$\{\{ P(a,y):~y\in \mathbb{F}_{d_j}\}, a\in \tilde{\mathbb{F}}_{d_j}\}$
		have trivial constraint $I_{d_j}$, we have that $B_j=\{P(\infty_j,y): y\in
		\mathbb{F}_{d_j}\}\cup \{\{P(a,y): y\in \mathbb{F}_{d_j}\setminus \{d_j-1\}\},
		a\in \mathbb{F}_{d_j}\}$ is a basis for $X_j$ and $C_j=\{0,1,1/d_j\}$.
		
		\item[(ii)] By Tensor product technique 3.7.1 above we have the following.
		\begin{enumerate}
		\item[(a)] $B=\{\otimes_{j=1}^{k} P(a_j,y_j), a_j\in \tilde{\mathbb{F}}_{d_j}, y_j\in
		\mathbb{F}_{d_j}~\text{in case}~ a_j=\infty_j~ \text{whereas}~
		y_j\in \mathbb{F}_{d_j}\setminus \{d_j-1\} ~\text{for}~a_j\in \mathbb{F}_{d_j}
		~\text{for}~1\leq j\leq k\}$ is a basis of rank one projections for
		$X=\mathcal{B}(\mathcal{H})$.
		
		\item[(b)] $\# B=d^2$.
		
		\item[(c)] The set $C$ of quantum mechanical overlaps for $B$ equals $\{0,1, \prod_{j\in S}
		\frac{1}{d_j}, \emptyset \neq S\subset \{1,2,\cdots,k\}\}$. In particular, $1/d\in C$.
		
		\item[(d)] $\#C= 2^k+1$ and, therefore, the set of quantum mechanical overlaps for $B$
		has cardinality $2^k+1$.
		
		\item[(e)] Hence we have a method of obtaining a complete system of rank one projections derived from Weyl operators of the ideal size $d^2$ and with $2^k+1$
		quantum mechanical overlaps.
		\end{enumerate}
		\end{enumerate}
	\end{example}
	\begin{example}\label{e3.7.2}{( Shift and multiply, cyclic group case)}
		Let $d=\prod_{j=1}^{k} p_j^{s_j}, 2 \leq p_1<p_2\cdots <p_k,$, all primes and
		$s_j\geq 1$ for $1\leq j\leq k$. Then $d$ is a product of $s$ primes, distinct or not, with $s=\sum_{j=1}^{k} s_j$.
		We consider different factorizations of $d$ in order to be able to combine the tensor product technique with basics in
		\S2 and Theorems \ref{t3.6.2}, \ref{t3.6.7}, \ref{t3.6.10}, and \ref{t3.6.17} obtained with the help of Optimal Constraint Technique and the Simultaneous Constraints Technique in \S 3.1.
		This will be done in Theorem \ref{t3.7.3} to obtain informationally complete systems of rank one projections of the ideal size
		$d^2$ where we also determine quantum mechanical overlaps in some cases using Theorem \ref{t3.6.5}.
		\begin{enumerate}
		\item[(i)] Consider any $1\leq j\leq k$. Then $s_j=1,2t_j~\text{or}~2t_j+1$ for
		some $t_j\geq 1$. Let $\ell_j=1,t_j~\text{or}~t_j+1$ respectively i.e.,
		$\ell_j=s_j/2$ if $s_j$ even and $(s_j+1)/2$ if $s_j$ is odd. So $p_j^{s_j}$
		has the form $p_j$, product of $p_j^2$ taken $t_j$ times  or  product of $p_j$
		and $p_j^2$ taken $t_j$ times respectively. We recall that empty product is to be taken as 1.
		
		Let ${\bf K} = \{j: s_j~\text{is odd}\}$, ${\bf K}^\prime = \{j: s_j~\text{is even}\}$. Then $p_j^{s_j}$ is the product of $d_{j,\ell}$'s, $1\leq
		\ell\leq \ell_j$, where for $j\in{\bf K}$, $d_{j,l_j} =p_j$ and $d_{j,l}=
		p_j^2$ for $\ell<\ell_j$, whereas for $j\in {\bf K}^\prime$, $d_{j,l}=p_j^2$
		for all $\ell\leq \ell_j$.
		\begin{enumerate}
		\item[(a)] $d= \prod_{j=1}^{k}\prod_{\ell=1}^{\ell_j} d_{j,\ell} =\prod_{u=1}^{\ell^\prime} d_u^\prime$, say with $\ell^\prime =\sum_{j=1}^{k} \ell_j
		=\frac{1}{2}(\sum_{j=1}^{k} s_j+ \#{\bf K})$ = $\frac{1}{2}(s+\#{\bf K})$. Here each $d_u^\prime$ is
		either some $p_j$ or its square.
		
		\item[(b)] Consider the case ${\bf K}\neq \emptyset$ and $\#{\bf K}$ is even, say
		${\bf K} =\{j_{2v-1},j_{2v}: 1\leq v\leq v^\prime\}$ with
		$v^\prime \geq 1$ .
		Let $p_v^\prime= p_{j_{2v-1}}p_{j_{2v}}$ for $1\leq v\leq
		v^\prime$. Then
		$$d= \prod_{v=1}^{v^\prime} p_v^\prime(\prod_{j\in {\bf K},\ell_j>1}\prod_{\ell=1}^{\ell_j-1}
		d_{j,\ell})\prod_{j\in {\bf K}^\prime}\prod_{\ell=1}^{\ell_j} d_{j,\ell} = \prod_{t=1}^{t^{\prime\prime}}
		d_t^{\prime\prime},$$ say with
		$t^{\prime\prime}= v^\prime+ \sum_{j\in {\bf K}}\frac{s_j-1}{2}
		+\sum_{j\in{\bf K}^\prime}\frac{s_j}{2} =\frac{1}{2}\sum_{j=1}^{k}s_j=\frac{1}{2}s$.
		
		Here each $d_{t}^{\prime\prime}$ is $p_j^2$ for some $j$ or $p_j p_{j^\prime}$
		for some $j\neq j^\prime \in {\bf K}$.
		
		\item[(c)]  ${\bf K}\neq \emptyset$ and $\#{\bf K}$ is odd, say
		${\bf K} =\{j_0, j_{2v-1},j_{2v}: 1\leq v\leq u\}$ with $u\geq 0$ .
		Let $p_v^\prime= p_{j_{2v-1}}p_{j_{2v}}$, $1\leq v \leq u$. Then
		$$d= p_{j_0} \prod_{v=1}^{u} p_v^\prime
		(\prod_{j\in {\bf K},\ell_j > 1}\prod_{\ell=1}^{\ell_j-1}
		d_{j,l})\prod_{j\in {\bf K}^\prime}\prod_{\ell=1}^{\ell_j} d_{j,\ell} = \prod_{r=1}^{r^{\prime\prime\prime}}
		d_r^{\prime\prime\prime},$$ say with
		$r^{\prime\prime\prime}= 1 + u+
		\sum_{j\in {\bf K}}(\ell_j-1)
		+\sum_{j\in {\bf K}^\prime}\ell_j =\frac{1}{2}(\sum_{j=1}^{k}s_j+1)=\frac{1}{2}(s+1)$.
		
		Here each $d_{r}^{\prime\prime}$ except for the first one, $p_{j_0}$, is
		either $p_j^2$ for some $j$ or $p_j p_{j^\prime}$
		for some $j\neq j^\prime \in {\bf K}$.
		\end{enumerate}
		\item[(ii)] For $p,a$ primes, $p<a$, let $\mathcal{P}^{(d)}$ be the complete set
		$\mathcal{P}^\prime$ of rank one projections obtained for $\mathcal{H}=\mathbb{C}^d$ with $d=p, p^2, pa $ as in Basics, Theorem \ref{t3.6.2},
		Theorems \ref{t3.6.7} and \ref{t3.6.10} (for $p=2$, $a$ an odd prime), Theorem \ref{t3.6.17} (for $p$ and $a$ both odd primes) of the ideal
		size $d^2$. We recall from Theorem \ref{t3.6.5} that the set of quantum mechanical overlaps for $d=p^2$ is $\{0, 1,\frac{1}{p},\frac{1}{p^2}\}$ and
		from Basics that for
		$d=p$, it is simply $\{0,1,\frac{1}{p}\}$.
		\end{enumerate}
	\end{example}
	\subsubsection{\it Finale} We have the finale for all $d$ by combining our Constraint techniques and tensor product technique as
	follows.
	
	\begin{theorem}\label{t3.7.3}{(Combination of tensor product and constraint techniques.)}
		Let $d=\prod_{j=1}^{k} p_j^{s_j}$, where $p_j$'s and $s_j$'s and other related
		symbols are as in Example \ref{e3.7.2} above. Set
		
		(a) $\mathcal{H}^{\prime}=
		\otimes_{u=1}^{\ell^{\prime}}\mathcal{H}_{d_{u}^{\prime}}$ ,
		
		(b) $\mathcal{H}^{\prime\prime}=
		\otimes_{t=1}^{t^{\prime\prime}}\mathcal{H}_{d_{t}^{\prime\prime}}$ , in case ${\bf K}\neq \emptyset$ and $\#{\bf K}$ even,
		
		(c) $\mathcal{H}^{\prime\prime\prime}=
		\otimes_{r=1}^{r^{\prime\prime\prime}}\mathcal{H}_{d_{r}^{\prime\prime\prime}}$, in case $\#{\bf K}$ odd.
		
		Then the Tensor product technique combined with constraint techniques in \S 3.1
		and 3.6 above gives a complete system of rank one projections in each of the scenario in (a), (b) and (c). Further, in case (a) the set of quantum
		mechanical overlaps is $\{0,1/d^\prime: d^\prime~\text{ is a factor of}~d~ \}$.
	\end{theorem}
	\begin{proof}
	\begin{enumerate}
		\item[(a)] Here we have to first apply Theorem \ref{t3.6.2} only to obtain
		$\mathcal{P}^{(d_u^\prime)}$ and Theorem \ref{t3.6.5} for quantum mechanical overlaps together with basics for mutually unbiased bases for prime dimensions.
		Thus $C_u=\{0,1/p_j,1/p_j^2,1\}$ for $d_u^\prime=p_j^2$ and $C_u=\{0,1, 1/p_j\}$ for $d_u^\prime =p_j$ . Then $C=\{0,1/d^\prime : d^\prime ~\text{ is a
			factor of}~ d~(\text{including}~1) \}$.
		
		\item[(b)] Here we apply Theorem \ref{t3.6.2} for $d_t^{\prime\prime}=p_j^2$ for some
		$j$ and Theorems \ref{t3.6.7}, \ref{t3.6.10} and \ref{t3.6.17} for $d_t^{\prime\prime}=p_jp_{j^\prime}$ with $j\neq j^\prime$ in ${\bf K}$.
		
		\item[(c)] Here, the situation is similar to (b) except for an extra factor $p_{j_0}$.
		\end{enumerate}
	\end{proof}
	
	\subsection{Generalized Gell-Mann basis technique}
	Gell-Mann \cite{Gel} gave a basis for $M_3$ consisting of 8 traceless $3\times 3$ Hermitian matrices $G_k: 1\leq k\leq 8$ together with $G_0=I_3$; they
	are also mutually orthogonal in the sense that $\text{tr}(G_jG_k)=0$ for $0\leq j\neq k\leq 8$. This generlization of Pauli matrices has been
	further generalized to arbitrary dimensions $2\leq d<\infty$ , say $\{G_j: 0\leq j\leq d^2-1\}$ by different people ( see, for instance, \cite{HE}) under different names like generators for $SU(d)$ or generalized Gell-Mann basis using different multiples of $G_j$'s to suit their
	purposes of studying different properties or motivating related concepts. For instance, \cite{KloHu} (see also \cite{AEHK},\cite{Klo}), confine attention to
	$\text{tr}(G_j^2)=d,~ 0\leq j\leq d^2-1$.
	
	\subsubsection{\it Generalized Gell-Mann Basis} Our purpose is to extract a complete system of rank one projections of the ideal size $d^2$ from them,
	so we take the form which
	has simple looking eigenvalues and use the notation followed in previous sections or subsections.
	\begin{enumerate}
	\item[(i)] Let for $0\leq j,k\leq d-1$, $J_{j,k}$ be the elementary matrix which has entry 1 at $j,k^\text{th}$ place and 0 otherwise. Let
	\begin{align}
		& T_{00}=I_d, ~T_{jj} =\sum_{m=0}^{j-1} J_{m,m}-jJ_{j,j}~~\text{for}~~ 1\leq j\leq d-1; \nonumber\\
		& T_{jk}= J_{j,k}+ J_{k,j}\text{for}~ 0\leq j<k\leq d-1,~\text{and}~\nonumber\\
		& T_{kj}= -i(J_{j,k}- J_{k,j})~ \text{for}~ 0\leq j<k\leq d-1.
	\end{align}
	
	\item[(ii)] Now $\{ T_{jj} :~0\leq j \leq d-1\} $ have a common orthonormal system of unit eigenvectors $\{|j\rangle : 0\leq j\leq d-1\}$ and the
	associated orthonormal system of rank one projections $\{P_j=|j\rangle\langle j|: 0\leq j\leq d-1\}$.

	For $0\leq j<k\leq d-1$, let $Q_{j,k}=P_j+P_k$.
	
	Consider any $0\leq j< k \leq d-1$. Set
	\begin{align}
		|\xi_{j,k;0}\rangle &= \frac{1}{\sqrt{2}}(|j\rangle + |k\rangle),  \xi_{j,k;1} = \frac{1}{\sqrt{2}}(|j\rangle - |k\rangle), \nonumber\\
		|\eta_{j,k;0}\rangle &= \frac{1}{\sqrt{2}}(|j\rangle + i|k\rangle),  \eta_{j,k;1} = \frac{1}{\sqrt{2}}(|j\rangle - i|k\rangle), \nonumber\\
		P_{j,k;\ell}&=  |\xi_{j,k;\ell}\rangle \langle \xi_{j,k;\ell}|, P_{k,j;\ell}=
		|\eta_{j,k;\ell}\rangle \langle \eta_{j,k;\ell}|, ~\ell=0,1
	\end{align}
	Then $T_{jk}=P_{j,k;0} - P_{j,k;1}$, and $T_{kj}=P_{k,j;0} - P_{k,j;1}$.
	\end{enumerate}
	\subsubsection{\it Modified constraint technique }
	We now draw in an analogy with Constraints and Constraint Technique in \S3.1.1 and \S3.1.2.
	\begin{enumerate}
	\item[(i)] We say that for $0\leq j<k\leq d-1$,
	$\{P_{j,k;\ell}:\ell=0,1\}$ and $\{P_{k,j;\ell}:\ell=0,1\}$ are $Q_{jk}$-consrained.
	
	\item[(ii)] We can obtain a complete system of rank one
	projections of the ideal size $d^2$ in several ways by deleting one each from
	$\{P_{j,k;\ell}:\ell=0,1\}$ and $\{P_{k,j;\ell}:\ell=0,1\}$ per $j,k$ with $0\leq j<k\leq d-1$. For instance,
	$\mathcal{P}^\prime =\{P_j:0\leq j\leq d-1\}\cup \{P_{j,k;0}, P_{k,j;0}, 0\leq j<k\leq d-1\}$ works fine.
	
	\item[(iii)] Indeed, for $0\leq j<k\leq d-1$, $P_{j,k}=P_{j,k;0}= \frac{1}{2}(P_j+P_k+ T_{jk})$ and
	$P_{k,j}=P_{k,j;0}= \frac{1}{2}(P_j+P_k+ T_{kj})$. Thus
	$\mathcal{P}^\prime =\{P_j:0\leq j\leq d-1\}\cup \{\frac{1}{2}(P_j +P_k + T_{jk}), \frac{1}{2}(P_j +P_k + T_{kj}), 0, \leq j<k\leq d-1\}$ .
	\end{enumerate}
	We consolidate the items above as a Theorem.
	\begin{theorem} ( Gell-Mann Basis). Let $\{T_{jk}: 0\leq j,k\leq d-1\}$ be the generalized Gell-Mann basis as in 3.8.1 (i) above,
		$P_{jj}=P_j$ for $0\leq j\leq d-1$
		and $\mathcal{P}^\prime = \{P_{j,k} :0\leq j,k \leq d-1\}$, the associated system of rank one projections as in 3.8.1 (ii) and 3.8.2 (iii) above. Then
		$\mathcal{P}^\prime$ is an informationally complete system of rank one projections with quantum mechanical overlaps $0,1,\frac{1}{2},
		\frac{1}{4}$.
	\end{theorem}
	
	\setcounter{section}{3}
	
	\section{Other Unitary Bases}

	We begin with a notion to be used in this section.
	\subsection{Collective unitary equivalence}


	\begin{definition}\label{def2.4} 
		Let $\mathcal{F},$ $\mathcal{G}$ be subsets of $\mathcal{B}(\mathcal{H}).$
		\begin{itemize}
			\item[(i)] We say that $\mathcal{F}$ is {\it collectively unitarily equivalent}
			to $\mathcal{G},$ in short, $\mathcal{F}$CUE$\mathcal{G},$ if
			there exists a
			$V \in \mathcal{U}(\mathcal{H})$ such that $\mathcal{G} = \{V^{\ast} AV: A \in
			\mathcal{F}\}.$ We may say $\mathcal{F}$CUE$\mathcal{G}$ via $V.$
			
			\item[(ii)] In case $\mathcal{F}$ and $\mathcal{G}$ are decomposed as
			$\mathcal{F} = \underset{\gamma \in \Gamma}{\cup} \mathcal{F}_{\gamma},$
			$\mathcal{G} = \underset{\gamma \in \Gamma}{\cup} \mathcal{G}_{\gamma}$
			respectively, then we may require that for $\gamma \in \Gamma,$
			$\mathcal{F}_{\gamma}$CUE$\mathcal{G}_{\gamma}$ via $V,$ and say that
			$\mathcal{F}_{\Gamma}$CUE$\mathcal{G}_{\Gamma},$ or, if no confusion arises
			$\mathcal{F}$CUE$\mathcal{G}.$
			
			\item[(iii)] In case $\mathcal{F}$ and $\mathcal{G}$ are indexed by a set
			$\Lambda$ as $\{A_{\alpha}:\alpha \in \Lambda \}$ and $\{B_{\alpha} : \alpha \in
			\Lambda \}$ respectively, then (as a special case of (ii) above) we may require
			that $B_{\alpha} = V^{\ast} A_{\alpha} V, \alpha \in \Lambda$ and say that
			$\mathcal{F}_{\Lambda}$CUE$\mathcal{G}_{\Lambda},$ or, if no confusion arises,
			$\mathcal{F}$CUE$\mathcal{G}.$
		\end{itemize}
	\end{definition}

	\begin{remark}\label{rem2.5} 
		\begin{itemize}
        
			\item[(i)] The operator $V$ in the definition above may not be unique. In
			fact, if $U^{\ast} \mathcal{G}U = \mathcal{G}$ for some $U \in
			\mathcal{U}(\mathcal{H})$ then $VU$ works fine too.
			
			\item[(ii)] $\mathcal{F}$CUE$\mathcal{G}$ if and only if
			$\mathcal{F}_{\Lambda}$CUE$\mathcal{G}_{\Lambda}$ for some indexing
			$\mathcal{F}_{\Lambda}$ and
			$\mathcal{G}_{\Lambda}$ of $\mathcal{F}$ and $\mathcal{G}$ respectively by the
			same index set $\Lambda.$ So we may fix some such indexing, if we like.
			
			\item[(iii)] The relation CUE (in both the senses in Definition \ref{def2.4}
			above) is an equivalence relation.
			
			\item[(iv)] It can be readily seen that if $\mathcal{F}$CUE$\mathcal{G}$ and
			$\mathcal{F}$ is a commuting family then so is $\mathcal{G}.$
			
			\item[(v)] If $\mathcal{F}_{\Lambda}$CUE$\mathcal{G}_{\Lambda}$ via $V,$ then
			for each $\alpha \in \Lambda,$ the spectrum $\sigma (A_{\alpha})=\sigma
			(B_{\alpha});$ and for $\alpha \in \Lambda,$ for an eigenvalue $\lambda$ of
			$A_{\alpha},$ $\xi$ is an eigenvector for $B_{\alpha}$ with eigenvalue $\lambda$
			if and only if $V\xi$ is an eigenvector for $A_{\alpha}$ with eigenvalue
			$\lambda.$
			
			\item[(vi)] If $\mathcal{F}= \{F_{\alpha}: 1 \leq \alpha \leq n\}$ is a
			commuting $n$-tuple of normal operators in $\mathcal{B}(\mathcal{H}),$ then
			there exists a $U \in \mathcal{U}(\mathcal{H})$ and an $n$-tuple $\mathcal{D}$
			of operators ($D_{\alpha}: 1 \leq \alpha \leq n $) represented by diagonal
			matrices $\{\widetilde{D}_{\alpha} : 1 \leq \alpha \leq n\}$ with respect to
			basis $\mathbf{e}$ such that $F_{\alpha} = U D_{\alpha} U^{\ast},$ $1 \leq
			\alpha \leq n.$ In other words, $\mathcal{F}$CUE$\mathcal{D}.$ The converse
			is
			also true.
		\end{itemize}
	\end{remark}
	
	\begin{remark}[{\it Application of CUE to construction of PPT
			matrices}]\label{rem2.6} 
		Garcia and Tener (\cite{GT}, Theorem 1.1) obtained a canonical
		decomposition for complex matrices $T$ which are UET, i.e., unitarily
		equivalent to their transposes $T^t,$ and developed a detailed analysis of such matrices.
		\begin{itemize}
			\item[(i)] In her paper \cite{AIS2} Ajit Iqbal Singh called
			a tuple $\mathcal{F} =
			(F_{\alpha} : 1 \leq \alpha \leq n)$ of $d \times d$ matrices {\it collectively
				unitarily equivalent} to the respective {\it transposes,} in short, CUET,
			if
			$ \mathcal{F}$CUE$\mathcal{F}^t,$ where $\mathcal{F}^t = (F_{\alpha}^t : 1 \leq
			\alpha \leq n).$
			She gave a few examples and results and utilized them to obtain block
			matrices that are positive under partial transpose, in short, PPT.
			
			

			(\cite{GT}, items 8.3, 8.4 and 8.5) tell us how to construct CUET
			tuples.
			
			\item[(ii)] Tadez and Zyczkowski (\cite{TZ}, 4.1) indicate a general method to
			construct a class of CUET tuples of matrices. Let $P$ be the permutation
			matrix given in column form as $[e_1, e_d, e_{d-1}, \ldots, e_2].$ Let ${\xi} =
			({\xi}_j) \in \mathbb{C}^d.$ Set $C_{\xi} = [C^{\xi}_{jk}]$ to be the $d \times
			d$ circulant matrix given by $C^{\xi}_{jk} = \xi_{j-k~
				(\text{mod}\, d)}.$ Then
			$C_{\xi}^t = P^t C_{\xi} P.$ Because $P$ is a real unitary matrix, this gives
			that any tuple consisting of $C_{\xi}$'s is CUET.
		\end{itemize}

	\end{remark}

	\subsection{Fan systems as invariants for unitary bases}
	The main purpose here is to classify unitary bases up to equivalence.
	
	\begin{definition}
		A {\it tagged unitary system,} in short, TUS, is a triple
		$\mathbf{T}=(x_0, U_{x_{0}}, \mathbf{W})$ where, $x_0 \in X,$ $U_{x_{0}} \in
		\mathcal{U} (\mathcal{H}),$ $\mathbf{W} = \{W_y : y \in X, y \neq x_0\}$ is a
		unitary system. We say $\mathbf{T}$ is tagged at $x_0.$
	\end{definition}
	
	\begin{remark}\label{re2.2} 
		Let $\mathbf{T}$ be a tag at $x_0.$ Set $W_{x_{0}} = I.$ We set
		$\mathbf{U}=\{U_x = U_{x_{0}} W_x: x \in X\}.$ Then $\mathbf{U}$ satisfies
		$W_x^{\ast} W_y = U_x^{\ast} U_y$ for $x,$ $y \in X,$ and, therefore,
		$\mathbf{U}$ is a UB. We call $\mathbf{U}$ the {\it $\mathbf{T}$-associated UB.}
		On the other hand, given a UB $\mathbf{U},$ for $x_0 \in X,$ setting
		$\mathbf{W} = \{W_x = U_{x_{0}}^{\ast} U_x, x \in X, x \neq x_0 \},$ we have
		$\mathbf{T}= (x_0, U_{x_{0}}, \mathbf{W})$ is a tag at $x_0.$ It satisfies
		$W_x^{\ast} W_y = U_x^{\ast} U_y$ for $x,$ $y \in X,$ where $W_{x_{o}}$ has been
		taken as $I.$ We call $\mathbf{T}$ the $\mathbf{U}$-{\it associated tag} at
		$x_0.$ We note that in both cases, for $x, y \in X,$ $W_y W_x^{\ast} =
		U_{x_{0}}^{\ast} U_y U_x^{\ast} U_{x_{0}}.$ Further, for $x,$ $y \in X,$ $W_x
		W_y = W_y W_x$ if and only if $U_x U_{x_{0}}^{\ast} U_y= U_y U_{x_{0}}^{\ast}
		U_x.$ We denote this condition by $\mathbf{T}(x, x_0, y)$ and call it by {\it
			Twill}. Really speaking, both the
		associations are on the left and have obvious right versions as well.
	\end{remark}

	\begin{theorem}\label{thm2.7} 
		Let $\mathbf{U},$ $\mathbf{U}^{\prime}$ be unitary bases for $\mathcal{H}.$ Then
		the following are equivalent.
		\begin{itemize}
			\item[(i)] $\mathbf{U}$ is equivalent to $\mathbf{U}^{\prime},$
			\item[(ii)] for some $\mathbf{U}$-associated tag $\mathbf{T}$ and some
			$\mathbf{U}^{\prime}$-associated tag $\mathbf{T}^{\prime},$ \\ $\mathbf{W}$CUE
			$\mathbf{W}^{\prime}.$
			
			\item[(iii)] for each $\mathbf{U}$-associated tag $\mathbf{T},$ there is a
			$\mathbf{U}^{\prime}$-associated tag $\mathbf{T}^{\prime}$ such that
			$\mathbf{W}$CUE$\mathbf{W}^{\prime}.$
		\end{itemize}
	\end{theorem}
	
	\begin{proof}
		(i) $\Rightarrow$ (iii), Suppose $\mathbf{U} \sim \mathbf{U}^{\prime}.$ Then
		there exist $V_1, V_2 \in \mathcal{U}(\mathcal{H})$ and a relabelling $x
		\rightarrow x^{\prime}$ of $X$ such that $U_{x^{\prime}}^{\prime} = V_1 U_{x}
		V_2$ for $x \in X.$ Consider any $x_0 \in X$ and let $\mathbf{T} = (x_0,
		U_{x_{0}},
		\mathbf{W})$ be the $\mathbf{U}$-associated tag at $x_0$ and
		$\mathbf{T}^{\prime}= (x_0^{\prime}, U_{x_{0}^{\prime}}^{\prime},
		\mathbf{W}^{\prime}),$ the $\mathbf{U}^{\prime}$-associated tag at
		$x_0^{\prime}.$ Set $W_{x^{\prime}_{0}}^{\prime} = W_{x_{0}} = I. $ Then
		$U_{x_{0}^{\prime}}^{\prime} = V_1 U_{x_{0}} V_2$ and, therefore, for $x \in X,$
		$W_{x^{\prime}}^{\prime} = U_{x_{0}^{\prime}}^{\prime^{\ast}}
		U_{x^{\prime}}^{\prime}=V_2^{\ast} U_{x_{0}}^{\ast} V_1^{\ast} V_1 U_x V_2 =
		V_2^{\ast} W_x V_2.$ So $\mathbf{W}$CUE$\mathbf{W}^{\prime}$ via $V_2.$\\
		(iii) $\Rightarrow$ (ii) is trivial.\\
		(ii) $\Rightarrow$ (i), Let $\mathbf{T} = (x_0, U_{x_{0}}, \mathbf{W})$ and
		$\mathbf{T}^{\prime} = (x_0^{\prime}, U_{x_{0}^{\prime}}^{\prime},
		\mathbf{W}^{\prime})$ be the $U$-associated tag at $x_0$ and
		$\mathbf{U}^{\prime}$-associated tag at $x_0^{\prime}$ respectively with
		$\mathbf{W}$CUE$\mathbf{W}^{\prime}.$ Then by Remark \ref{rem2.5}(ii), there
		exist a $V \in \mathcal{U}(\mathcal{H})$ and a bijective function on $X
		\backslash \{x_0\}$ onto $X \backslash \{x_0^{\prime}\},$ say $x \rightarrow
		x^{\prime}$ such that $W_{x^{\prime}}^{\prime} = V^{\ast} W_x V$ for $x \in X
		\backslash \{x_0\}.$
		
		Set $V_1 = U_{x_{0}^{\prime}}^{\prime} V^{\ast} U_{x_{0}}^{\ast}.$ Then $V_1 \in
		\mathcal{U}(\mathcal{H}).$ Also $U_{x_{0}^{\prime}}^{\prime} = V_1 U_{x_{0}} V.$
		Further, for $x \in X \backslash \{x_0\}$
		\begin{align*}
			U_{x^{\prime}}^{\prime} &= U_{x_{0}^{\prime}}^{\prime} W_{x^{\prime}}^{\prime}
			= U_{x_{0}^{\prime}}^{\prime} (V^{\ast} W_x V)\\
			&= (U_{x_{0}^{\prime}}^{\prime} V^{\ast} U_{x_{0}}^{\ast})(U_{x_{0}} W_{x})V
			\\
			&= V_1 U_x V.
		\end{align*}
		So $\mathbf{U}$ is equivalent to $\mathbf{U}^{\prime}.$
	\end{proof}

	\begin{theorem}\label{thm2.8}
		Let $\mathbf{W}$ be a unitary system.
		
		Let $\mathbf{W}^{\prime}$ be a unitary system and
		$\mathcal{V}_{\mathbf{W}^{\prime}} = \{\mathbf{V}_{\beta}^{\prime}: \beta \in
		\Lambda^{\prime}\},$ the family of $\mathbf{W}^{\prime}$-MASS's as in Theorem~\ref{thm3.2.1} (i).\
		Then $\mathbf{W}$CUE$\mathbf{W}^{\prime}$ via $V$ iff there is a bijective map on
		$\Lambda$ to $\Lambda^{\prime},$ say, $\alpha \rightarrow \alpha^{\prime},$ such
		that $\mathbf{V}_{\alpha}$CUE$\mathbf{V}^{\prime}_{\alpha^{\prime}}$ via $V.$
	\end{theorem}
	
	\begin{proof}
		
		Suppose $\mathbf{W}$CUE$\mathbf{W}^{\prime}$ via $V.$ Then
		$\{\mathbf{\widehat{V}_{\alpha}} =\{V^{\ast} AV : A \in \mathbf{V}_{\alpha}
		\}, \alpha
		\in \Lambda \}$ is a maximal family of $\mathbf{W}^{\prime}$-MASS's with
		$\mathbf{W}^{\prime} =
		\underset{\alpha \in \Lambda}{\cup} \widehat{\mathbf{V}}_{\alpha}.$ So, by
		uniqueness, each $\widehat{\mathbf{V}}_{\alpha}$ is some unique
		$\mathbf{V}_{\beta}^{\prime}.$ Set $\beta = \alpha^{\prime}.$ On the other hand,
		each $\mathbf{V}_{\beta}^{\prime}$ is some unique
		$\widehat{\mathbf{V}}_{\alpha}.$ So the map $\alpha \rightarrow
		\alpha^{\prime}$ is bijective on $\Lambda$ to $\Lambda^{\prime}.$ The converse
		part is trivial.
	\end{proof}

	\begin{definitiondis}[Fan representation \& Hadamard fans]\mbox{}
		\begin{itemize}
			
			\item[(i)] The family $H_{\mathbf{W}} = \left \{H_{\alpha} : \alpha \in
			\Lambda \right \}$ facilitated as in Remark~\ref{r3.2.2}(i) above, will be called
			the {\it Hadamard fan} of $\mathbf{W}.$ We
			note that if $\mathbf{W}$ and $\mathbf{W}^{\prime}$ are unitary systems with
			$\mathbf{W}$CUE$\mathbf{W}^{\prime}$ then their Hadamard fans are the same to
			within a labelling of $\Lambda$ and permutation of rows and columns of
			$H_{\alpha}.$
			
			\item[(ii)] Let $\mathbf{U}$ be a UB and $\mathcal{W} = \{\mathbf{W} :
			\mathbf{T}= (x_0, \mathbf{U}_{x_{0}}, \mathbf{W})$ is the tag of
			$\mathbf{U}\,\, \mbox{at}\,\,x_0 \}$ and $\mathcal{V} = \left \{
			\mathcal{V}_{\mathbf{W}} : \mathbf{W \in \mathcal{W}}\right \}.$ Then
			$\mathcal{V}$ will be called the {\it fan system} of $\mathbf{U}.$ We note that
			it follows from Theorem \ref{thm2.7} and Theorem \ref{thm2.8} that $\mathcal{V}$ is an invariant
			for $\mathbf{U}$ in the sense that if $\mathbf{U}^{\prime}$ is a UB and
			$\mathcal{V}^{\prime}$ is its fan system then $\mathbf{U} \sim
			\mathbf{U}^{\prime}$ if and only if to within CUE $\mathcal{V} \cap
			\mathcal{V}^{\prime} \neq
			\phi$ if and only if to within CUE $\mathcal{V} = \mathcal{V}^{\prime}.$
			
			\item[(iii)] Let $\widetilde{H}_{\mathbf{W}} = \{\widetilde{H}_{\alpha} : \alpha
			\in \Lambda\}.$ We call $\widetilde{\mathbf{H}} = \{\widetilde{H}_{\mathbf{W}} :
			\mathbf{T} = (x_0, U_{x_{0}}, \mathbf{W})$ is the tag of
			$\mathbf{U}\,\,\mbox{at}\,\, x_0 \}$ the {\it Hadamard fan} of
			$\mathbf{W}.$ We note that if $\mathbf{U}$ and $\mathbf{U}^{\prime}$ are UB's
			with
			Hadamard fan systems $\mathbf{H}$ and $\mathbf{H}^{\prime}$ respectively, and if
			$\mathbf{U} \sim \mathbf{U}^{\prime}$ then $\mathbf{H} = \mathbf{H}^{\prime}.$
			The converse does not hold.
		\end{itemize}
	\end{definitiondis}

	\subsection{Maximally entangled state bases}
	The question that triggered this
	paper, in fact, is the following
	one in the context of maximally entangled states (MES) with phases. How to
	distinguish pairwise orthogonal systems of MES using local quantum operations
	supplemented by classical communication?
	
	If one can figure out sets of pairwise orthogonal MES, locally unitarily
	connected up to global phases to the Bell basis \cite{Be,BBCJPW} then the task
	of distinguishing the states from the aforesaid sets is equivalent to that of
	distinguishing locally the Bell states. We elaborate as follows.
	
	\subsubsection{\it The question of distinguishing MES}
	We now put the question in the language used in the second section. Let
	$\{|\Psi_x \rangle : x \in X\}$ be an orthonormal basis in $\mathcal{H} \otimes
	\mathcal{H}$ consisting of MES only. Do there exist unitaries $V_1, V_2 \in
	\mathcal{U}(\mathcal{H}),$ a bijective function $g$ on $X$ to itself and a
	function $f$ on
	$X$ to $S^1$ such that $|\psi_{g(x)}\rangle = f(x) (V_1 \otimes V_2)(U_{x}
	\otimes I) \Omega,$ $x \in X$ where $\{U_x : x \in X\}$ is the basis $\{U_{mn}
	: m, n \in Z_d\}$ as explained in \S2.3 ?
	
	\subsubsection{\it Unitary basis version of the Question of distinguishing MES bases}.
	In view of the \S 2.1.1(iv)(a), there exists a system $\{T_x : x \in X \}$
	of mutually orthogonal unitaries in $\mathcal{U}(\mathcal{H})$ such that
	$|\psi_x \rangle = (T_x \otimes I) \Omega$ for $x \in X$ and further, by \S
	2.1.1(iv), for $x \in X,$ $(V_1 \otimes V_2)(U_x \otimes I) \Omega$
	\begin{align*}
		&= (V_1 \otimes I) (I \otimes V_2) (U_x \otimes I) \Omega\\
		&= (V_1 \otimes I) (U_x \otimes I) (I \otimes V_2) \Omega\\
		&= (V_1 \otimes I) (U_x \otimes I) (V_2^t \otimes I) \Omega \\
		&= (V_1 U_x V_2^t
		\otimes I) \Omega.
	\end{align*}
	Now $\widetilde{V}_2 = V_2^t$ is a unitary if $V_2$ is so. So the question
	reduces to: Do there exist unitaries $V_1, \widetilde{V}_2 \in
	\mathcal{U}(\mathcal{H})$, a bijective function $g$ on $X$ to itself and a function $f:X \rightarrow S^1$ such that
	$T_{g(x)} = f(x) V_1 U_x \widetilde{V}_2,$ $x \in X ?$
	
	In the terminology of \S2.1.1(ii), the question takes the form : Does there exist a function $\widetilde{f}$
	on $X$ to $S^1$ such that $\{\widetilde{f}(x) T_x : x \in X\}$ is equivalent to
	$\{U_x : x \in X\}?$
	
	We shall utilise the results and methods given above to
	answer this.
	
	\begin{definition}
		We call two unitary bases $\mathbf{U}$ and $\mathbf{U}^{\prime}$ {\it
			phase-equivalent} if there exists a function $\widetilde{f}$ on $X$ to $S^1$
		such that $\{\widetilde{f}(x) U_x^{\prime} : x \in X \}$ is equivalent
		to $\{U_x : x \in X \}.$
	\end{definition}

	\begin{definition}
		For subsets $\mathcal{F}$ and $\mathcal{G}$ of $\mathcal{B}(\mathcal{H})$ we
		say $\mathcal{F}$ is {\it phase-collectively-unitarily equivalent} to
		$\mathcal{G}$ and write $\mathcal{F}$PCUE$\mathcal{G}$ if there exists a
		function $f$ on $\mathcal{F}$ to $S^{1}$ such that $f\mathcal{F}$ CUE
		$\mathcal{G}$ where, $f \mathcal{F} = \{f(A) A : A \in \mathcal{F}\}.$
	\end{definition}
	
	\begin{remark} Let $\mathbf{W} = \{ W_y : y \in Y\}$ be a
		unitary system and $h : Y \rightarrow
		S^{1}$ be a function. Then
		\begin{itemize}
			\item[(i)] $h \mathbf{W} = \{h(y) W_y : y \in W \}$ is a unitary system,
			$\mathbf{W}$ is abelian if and only if $h\mathbf{W}$ is abelian,
			\item[(ii)] $\mathbf{V} = \{W_y : y \in Z \}$ with $Z \subset Y$ is a
			$\mathbf{W}$-MASS if and only if $(h|Z) \mathbf{V}$ is a $\mathbf{W}$-MASS, and
			
			\item[(iii)] $h \mathcal{V}_{\mathbf{W}} = \mathcal{V}_{h\mathbf{W}}.$
		\end{itemize}
		We can now have the obvious generalizations of Theorems and items in \S4.2
		with obvious modifications of the corresponding proofs. Here
		is an illustration which will be strengthened further by examples that follow.
	\end{remark}
	
	\begin{theorem}
		Let $\mathbf{U}, \mathbf{U}^{\prime}$ be unitary bases for $\mathcal{H}.$ Then
		the following are equivalent.
		\begin{itemize}
			\item[(i)] $\mathbf{U}$ is phase equivalent to $\mathbf{U}^{\prime}.$
			\item[(ii)] For some $\mathbf{U}$-associated tag $\mathbf{T}$ and some
			$\mathbf{U}^{\prime}$-associated tag $\mathbf{T}^{\prime},$\\
			$\mathbf{W}$PCUE$\mathbf{W}^{\prime}.$
			\item[(iii)] For each $\mathbf{U}$-associated tag $\mathbf{T},$ there
			is a $\mathbf{U}^{\prime}$-associated tag $\mathbf{T}^{\prime}$ such that
			$\mathbf{W}$ PCUE $\mathbf{W}^{\prime}.$
			\item[(iv)] $\mathbf{U}$ and $\mathbf{U}^{\prime}$ have the same fan systems to
			within PCUE.
		\end{itemize}
	\end{theorem}
	The purpose of the rest of this section is to illustrate concepts and results by
	more examples and throw more light on them.
	
	\subsection{Example (shift and multiply)} The emphasis is on various subcases of The Shift and Multiply Example as described in \S \ref{e2.1.4}.
	based on (\cite{13}, III.A) or (\cite{14}, Proposition 9)
	beyond the special case considered in \S2 and \S3 at various places. We distinguish them for their own sake, or for that matter,
	the corresponding MES bases.
	\subsubsection{\it Conditions for Maximal abelian subsystems}. We figure them out for various set-ups.
	\begin{itemize}
		\item[(i)]
		For $(m,n),(m^{\prime},n^{\prime}) \in X$ we say $(m,n)$
		{\it commutes }
		with $(m^{\prime},n^{\prime})$ and write it as $(m,n) \Delta
		(m^{\prime},n^{\prime})$ if $U_{m,n}$ commutes with $U_{m^{\prime},n^{\prime}}.$
		We now proceed to obtain maximal commuting subsets of $\mathbf{U}$ (to be
		called $\mathbf{U}$-MASS's) or,
		equivalently, of $X.$
		\item[(ii)]
		Let $(m,n),(m^{\prime},n^{\prime}) \in X.$ Then $U_{mn} U_{m^{\prime}
			n^{\prime}} = U_{m^{\prime}n^{\prime}} U_{mn}$ if and only if for $k \in Y_d,$
		$U_{mn}\left (H_{m^{\prime}k}^{n^{\prime}} |\lambda(n^{\prime},k) \rangle
		\right ) = U_{m^{\prime}n^{\prime}}\left (H_{mk}^{n} |\lambda(n,k) \rangle
		\right ) $ if and only if for $k \in Y_d,$ $ H_{m, \lambda(n^{\prime},k)}^{n}
		H_{m^{\prime}k}^{n^{\prime}} | \lambda (n, \lambda (n^{\prime},k) )\rangle
		\!\!= \!\! H_{m^{\prime}, \lambda(n,k)}^{n^{\prime}}$ $H_{mk}^n | \lambda
		(n^{\prime}, \lambda(n,k)) \rangle $ if and only if
		\begin{equation}\lambda (n,\lambda (n^{\prime}, k)) = \lambda (n^{\prime},
			\lambda (n,k)), \,\,k \in Y_d \,\,\,\,\,\,\,,~\text{in~ short},\left [\lambda_{n,n^{\prime}} \right
			]\end{equation}
		and
		\begin{equation}
			~~~~H_{m, \lambda(n^{\prime}, k)}^{n} H_{m^{\prime}k}^{n^{\prime}} =
			H_{m^{\prime}, \lambda(n, k)}^{n^{\prime}} H_{mk}^{n}, ~~k \in Y_d \,,\text{in short},
			\left [H_{(m,n),(m^{\prime}, n^{\prime})}\right ].
		\end{equation}
		We call these conditions {\it Latin criss-cross} and {\it Hadamard
			criss-cross} respectively.
		\item[(iii)] {\bf Latin squares.\,\,} A latin square $\lambda$ may be called a
		quasigroup $L$ in the
		sense that the binary operation `.' on $L$ given by $a.b=\lambda(a,b)$
		satisfies the condition that, given $s,t \in L,$ the equations $x.s=t$ and
		$s.y=t$ have unique solutions in $L;$ one may see, for instance, the book by
		Smith \cite{jdhs} for more details. Keeping this in mind we introduce a few
		notions for $\lambda.$
		\begin{enumerate}
		\item[(a)] An element $a$ of $L$ will be called a {\it left identity} if $a.b=b$ for
		$b$ in
		$L.$ We note that a left identity, if it exists, is unique. Similar remarks
		apply to the notion and uniqueness of {\it right identity.}
		\item[(b)] $\lambda$ is called {\it associative} if `$\cdot$' is associative.
		\item[(c)] Elements $a,b$ in $L$ will be said to be {\it commuting} if $a.b=b.a.$
		\item[(d)] The {\it centre} $Z(L)$ of $L$ is $\{a \in L: a.b=b.a \,\,\mbox{for
			each}\,\, b\,\, \mbox{in}\,\, L\}.$
		
		We shall mainly consider latin squares arising from a group $G$ (with
		multiplication written as juxtaposition and identity written as $e$) or {\it
			right
			divisors} or {\it left divisors} in the group $G$ as follows:
		\item[(e)] $a.b=ab,$
		\item[(f)] $a.b=ab^{-1},$
		\item[(g)] $a.b=a^{-1}b,$ for $a,b$ in $G.$
		
		Direct computations give the following. 
		\item[(h)] A right (respectively, left) divisor latin square has $e$ as right
		(respectively, left) identity. Further, any such latin square has both right
		and left identity if and only if $a^2 = e$ for each $a$ in
		$G$ if and only if $L$ is the same as $G.$
		\item[(i)] Any such $\lambda$ is associative if and only if $G$ and $L$ coincide.
		\item[(j)] Elements $a,b$ in any such $L$ commute if and only if $(ab^{-1})^2 = e.$
		\item[(k)] In particular, if the number of elements in $G,$ $\#G$ is an odd number $\ge
		3$ then no two distinct
		elements in any such $L$ commute.\\
		We may have {\it twisted version} of (e), (f) and (g) as follows and then draw
		the same conclusions as above for them.
		\item[(l)] $a.b=ba,$
		\item[(m)] $a.b=b^{-1} a,$
		\item[(n)] $a.b=ba^{-1},$ for $a,b$ in $G.$
		Let $\lambda^{-1}$ be the latin square $\mu$ defined by $\mu (a, \lambda
		(a,b))=b$ for $a,b$ in $L.$
		\item[(o)] Direct computations give that $\mu^{-1}=\lambda.$ We may say that
		$(\lambda, \mu)$ is an {\it inverse-pair}. We note that latin squares listed
		above may then be inverse-paired as $((e),(g)),$ $((f),(m))$ and $((l), (n)).$
		\end{enumerate}
		\item[(iv)] {\bf Latin criss-cross.\,\,} Item (iii) above immediately gives the
		following facts.
		
		\begin{enumerate}
		\item[(a)] Latin criss-cross for an associative latin square reduces to
		$n.n^{\prime} = n^{\prime}. n.$
		\item[(b)] If $\lambda$ has a right identity then Latin criss-cross implies that
		$n.n^{\prime}=n^{\prime}.n.$ In particular, if $\lambda$ is a right divisor
		latin square and
		$\#L$ is an odd number $\ge 3,$ then Latin criss-cross reduces to
		$n=n^{\prime}.$
		\item[(c)] Direct computations give that if $\lambda$ is a left divisor latin square
		then Latin criss-cross reduces to $nn^{\prime}=n^{\prime}n.$ 
		\item[(d)] Suppose $\lambda$ is a right divisor latin square. Then Latin criss-cross
		holds if and only if $(n^{\prime} n^{-1})^2 =e,$
		$n^{\prime}n=nn^{\prime}$ and $n^{\prime}n^{-1} \in Z(G).$ For the sake of
		illustration we give details. We first note
		that Latin criss-cross holds if and only if $n^{\prime} kn^{-1} =
		nkn^{\prime -1}$
		for all $k.$ Taking $k=e, n, n^2$ this implies $n^{\prime} n^{-1}=nn^{\prime
			-1},$
		$n^{\prime} = n^2 n^{\prime -1}$ and $n^{\prime}n=n^3 n^{\prime -1}$
		which, in turn,
		implies $(n^{\prime} n^{-1})^2=e,$ $n^{\prime 2} = n^2$ and $n^{\prime}n = n^3
		n^{\prime -1}$ and thus, $(n^{\prime} n^{-1})^2=e,$ $n^{\prime 2} = n^2,$
		$n^{\prime}n=nn^{\prime}.$ This is equivalent to $n^{\prime}n = nn^{\prime},$
		$n^2 = n^{\prime 2}$ as also to $(n^{\prime} n^{-1})^2=e,$ $n^{\prime}n =
		nn^{\prime}.$ Therefore, Latin criss-cross is equivalent to $n^{\prime}n =
		nn^{\prime},$ $n^{\prime} n^{-1} = nn^{\prime -1},$ $n^{\prime} kn^{-1} =
		nkn^{\prime -1}$
		for all $k,$ i.e., $n^{\prime}n =nn^{\prime},$ $n^{\prime} n^{-1} = nn^{\prime
			-1} =
		n^{\prime -1}n,$ $n^{\prime} n^{-1}(nk) = (nk) n^{\prime -1} n$ for all $k,$
		\,\,
		i.e., $n^{\prime} n=nn^{\prime},$ $n^{\prime} n^{-1}=nn^{\prime -1},$
		$(n^{\prime}
		n^{-1})\ell = \ell (n^{\prime} n^{-1})$ for all $\ell$
		i.e., $n^{\prime}n = nn^{\prime},$ $(n^{\prime} n^{-1})^2=e,$ $n^{\prime}
		n^{-1} \in Z(G).$
		\item[(e)] If $\lambda$ is a right divisor latin squre and $Z(G)$ consists of the
		identity then Latin criss-cross reduces to $n=n^{\prime}.$
		\item[(f)] Suppose $G$ is an abelian group with identity written as $0$ and $\#G >2.$
		Then (c)
		above gives that Latin criss-cross is satisfied automatically for the left
		subtraction latin square. Also (d) above gives that Latin criss-cross for the
		right subtraction latin square is satisfied if and only if $n+n = n^{\prime}
		+n^{\prime}.$ As already noted in (b) above, it is possible for $n \neq
		n^{\prime}$ only if $\#G$ is even and in that case, for $0 \neq g \in G,$
		order of $g$ even, say, $2r,$ $n=0,$ $n^{\prime} = rg
		(=\underbrace{g+\cdots+g}_{r \,\,\text{times}} )$ satisfy the
		requirement. We now assume that $\#G$ is even. We can divide $L$ into mutually
		disjoint equivalence classes $L_h$, indexed by the set $S=\{h=g+g\,\,:\,\, g \in
		G\}$, given by $L_h=\{n \in G: n+n=h\}.$ We note that $L_h = \cup \{ L_0 + g :
		g+g=h\}$ and Latin criss-cross is satisfied if and only if $n,n^{\prime}$ both
		belong to some $L_h.$ In case $\exp G=2,$ we have $L=G,$ $S=\{0\},$ $L_0=L$ and,
		therefore, Latin criss-cross is automatically satisfied. On the other hand, if
		$\exp G > 2,$ then $\{0\} \subsetneqq S, \{0\} \subsetneqq L_0 \subsetneqq L$
		and $\#L_0 \leq \#L_h $ for $h \in S.$. Here, $\exp G$ means the minimum positive number $n$ such that for each $g \in G, ng=e$.
		\end{enumerate}
		\item[(v)] {\bf Hadamard criss-cross.}
		\begin{enumerate}
		\item[(a)] We first consider the case $n=n^{\prime}.$ Hadamard criss-cross becomes
		\begin{align}
			H_{m,n.k}^{n} H_{m^{\prime},k}^{n} &= H_{m^{\prime}, n.k}^{n}
			H_{m,k}^{n}\,\,\mbox{for each}\,\,k,\nonumber\\
			&i.e., \frac{H_{m^{\prime},n.k}^{n}}{H_{m, n.k}^{n}} =
			\frac{H_{m^{\prime},k}^{n}}{H_{m,k}^{n}}~
			\text{for each}~ k.
		\end{align}
		
		\item[(b)] Suppose $L$ is a left divisor latin square. Then for $n=n^{\prime}=e,$
		Hadamard criss-cross becomes an identity and
		therefore,
		$m, m^{\prime}$ can be arbitrary. This gives rise to a full-size
		$\mathbf{U}$-MASS.
		
		\item[(c)] We now consider the case when $n=n^{\prime}$ is a must. For $j, k \in
		\mathbb{Z}_d,$ let $H_{j,k}^{n} = (\eta_d)^{jk},$ where $\eta_d = e^{\left(\frac{2 \pi i}{d}\right)}.$ Hadamard criss-cross is equivalent to
		$(m-m^{\prime})(n.k-k)=0~ (\mbox{mod}\,d)$ for all $k.$ If $(n.k - k)$ is co-prime
		to $d$ for some $k,$ then Hadamard criss-cross
		holds if and only if $m=m^{\prime}.$ This means $(m,n)$ belongs to the unique
		$\mathbf{U}$-MASS $\{(m,n)\}$ containing it. In particular, it is so if $n$
		is not the left identity for $\lambda$ and $d$ is a prime. We record important
		consequences.
		
		\item[(d)] {\bf Singleton $\mathbf{U}$-MASS's example.\,\,} We consider the
		right subtraction latin square coming from
		$\mathbb{Z}_3$ and take, for $n=0,1,$ or $2$, 
		\begin{align*}
		H^n=\left [\begin{array}{ccc} 1 & 1
			&
			1
			\\ 1 & \omega & \omega^2 \\ 1 & \omega^2 & \omega \end{array} \right ].
			\end{align*}
		
		By (iv)(b), Latin criss-cross is satisfied if and only if $n^{\prime}=n.$ By (v)
		above, Hadamard criss-cross is satisfied if and only if $\frac{H_{m^{\prime},
				n-k}^{n}}{H_{m, n-k}^{n}} = \frac{H_{m^{\prime},k}^{n}}{H_{m,k}^{n}}$ for each
		$k.$ For $n=1$ or 2, taking $k=0$ this forces
		$\frac{H_{m^{\prime},n}^{n}}{H_{m,n}^{n}}=1$ i.e. $H_{m^{\prime},n}^{n} =
		H_{m,n}^{n},$ which, in turn, forces $m^{\prime}=m.$ So $\mathbf{U}$-MASS's are
		all of size~1.
		
		\item[(e)] We consider the non-abelian group $G=S_3$ of permutations on $\{a, b, c\}.$
		We label the elements of $G$ in any manner by $\{0,1,2,3,4,5\}$ but with $0=e,$
		$1= \,\,\mbox{the cycle}\,\, (ab).$ Set $\mathbf{W}= \mathbf{U} \backslash
		\{(0,0)\}.$ Then $n=1=k$ satisfy the requirement that $(n.k-k)$ is co-prime to $d.$
		So for $0 \leq m \leq 5,$ $\{(m,1)\}$ is a $\mathbf{W}$-MASS. We
		can label the
		cycle $(ac)$ as 3 and cycle $(bc)$ as 5, then
		labelling $(ab)(bc) = (abc)$ as 2 and $(ab)(ac)=(acb)$ as 4
		similar arguments give that for $0
		\leq m \leq 5,$ $\{(m,3)\}$ and $\{(m,5)\}$ are $\mathbf{W}$-MASS's. Thus, we
		have 18
		$\mathbf{W}$-MASS's of size one each, say, $F_{1}, F_{2}, \ldots, F_{18}.$
		we
		have four full-size
		$\mathbf{W}$-MASS's viz., $F_0=\{(m,0): 1 \leq m \leq 5\},$ $F_{19} = \{(0,2),
		(3,2), (0,4), (3,4), (3,0) \},$ $F_{20} = \{(1,2), (4,2), (2,4), (5,4),
		(3,0)\}$ and $F_{21} = \{(2,2), (5,2), (1,4),(4,4),
		(3,0) \}.$ The element $(3,0)$ is present in all these four
		$\mathbf{W}$-MASS's. All other elements belong to a unique $\mathbf{W}$-MASS.
		Figure 17 gives an idea.
		\end{enumerate}
		\begin{figure}[H]
			\centering
			\includegraphics[width=0.6\linewidth]{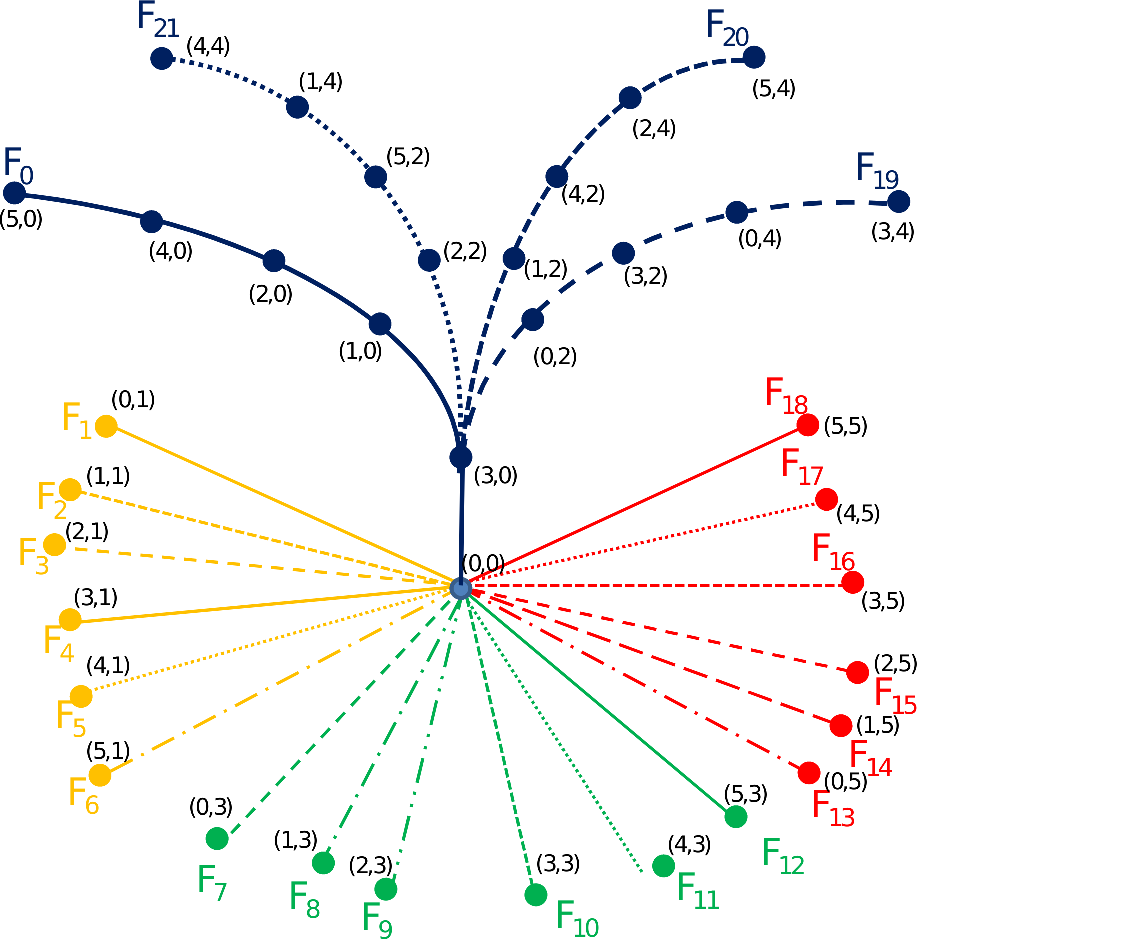} 
			\caption{}
		\end{figure}
		
		\item[(vi)] {\bf Tags and Twills \,\,} Contents of \S~4.2 tell us that it
		is $\mathbf{W}$-MASS's for tags of $\mathbf{U}$ that really help us. And
		$\mathbf{U}$-MASS's help directly if (a scalar multiple of) $I$ is in
		$\mathbf{U}$ simply because then apart from (the scalar multiple of) $I$
		occurring in all $\mathbf{U}$-MASS's, $\mathbf{W}$-MASS's and
		$\mathbf{U}$-MASS's
		are same. Let $\mathbf{T} = (x_0, U_{x_{0}}, \mathbf{W})$ be a tag of
		$\mathbf{U}.$ As noted in Remark~\ref{re2.2}, for $x,y \in X \backslash \{x_0\},$
		$W_x W_y = W_y W_x$ if and only if Twill $\mathbf{T}(x, x_0, y)$ viz., $U_x
		U_{x_{0}}^{\ast}
		U_y = U_y U_{x_{0}}^{\ast} U_x$ is satisfied. We now figure these out for some
		of the
		cases considered above.
		\begin{enumerate}
		\item[(a)] Let $x_0 = (m_0, n_0).$ Then for $k \in Y_d,$
		\begin{align*}
		U_{x_{0}}^{\ast}
		|k \rangle = U_{x_{0}}^{-1} |k \rangle = \overline{H_{m_{0},
				\mu (n_{0}, k)}} | \mu (n_0, k) \rangle.
		\end{align*}
		So the Twill is equivalent to \\$\left [\lambda (n, n_{0}, n^{\prime}) \right ]$ and $\left [H (m,n), (m_0, n_0), (m^{\prime}, n^{\prime})
		\right ]$, given by
		\begin{align}\lambda (n, \mu (n_0, \lambda (n^{\prime}, k)) = \lambda (n^{\prime}, \mu
			(n_{0}, \lambda (n,k)), k \in Y_d,
		\end{align}
		and
		\begin{align}
			& H_{m, \mu (n_{0}, \lambda (n^{\prime}, k))}^{n} \overline{H_{m_{0},
					\mu (n_{0}, \lambda (n^{\prime}, k))}^{n_{0}}} H_{m^{\prime},k}^{n^{\prime}}
			\nonumber\\
			&= H_{m^{\prime}, \mu (n_{0}, \lambda (n,k))}^{n^{\prime}}
			\overline{H_{m_{0}, \mu (n_{0}, \lambda(n,k))}^{n_{0}}} H_{m,k}^{n},\,\, k \in Y_d,
		\end{align}
		respectively.
		
		We call them {\it Latin twill} and {\it Hadamard twill} respectively.
		
		We may re-write Hadamard twill in another form as
		\begin{align}
			\lefteqn{H_{m, \mu (n_{0}, \lambda (n^{\prime}, k))}^{n} H_{m_{0},
					\mu (n_{0}, \lambda (n, k))}^{n_{0}} H_{m^{\prime},k}^{n^{\prime}}
			}\nonumber\\
			&= H_{m^{\prime}, \mu (n_{0}, \lambda (n,k))}^{n^{\prime}}
			H_{m_{0}, \mu (n_{0}, \lambda(n^{\prime},k))}^{n_{0}} H_{m,k}^{n}, k \in Y_d.
		\end{align}
		\item[(b)] For latin squares coming from group $G$ as in (iii) (e) above. Latin twill
		reduces to
		\begin{align}
			nn_{0}^{-1}n^{\prime} = n^{\prime} n_{0}^{-1} n.
		\end{align}
		For the inverse latin square arising as in (iii)(g) above, it is $n^{-1} n_{0}
		n^{\prime -1} = n^{\prime -1} n_{0} n^{-1},$ which on taking inverses, becomes
		$$n^{\prime} n_{0}^{-1} n = nn_{0}^{-1} n^{\prime}.$$
		Thus the two Latin twills are the same. Interestingly, the Latin twill remains
		the same for latin squares arising as in (f), (l), (m), (n) as well. We note an
		equivalent useful form of the Latin twill:
		\begin{align}
			(n^{-1}_0 n)(n_0^{-1}
			n^{\prime})=(n_0^{-1} n^{\prime})(n_0^{-1}n).
		\end{align}
		This Latin twill is satisfied
		automatically for abelian groups~$G.$
		\item[(c)] Hadamard twills can be written down which will be quite complicated for
		different cases.

		For the sake of a simple illustration, we consider the case in (v)(d) coming from $\mathbb{Z}_3.$ Then $\lambda \equiv
		\mu$ and $\lambda (n,k) = n-k =\mu (n,k)$ for $n,k=0,1,2.$ So Hadamard twill
		becomes: for $k \in \{0,1,2\},$
		\begin{align*}
			\lefteqn{H_{m, n_{0}-n^{\prime} + k}\,\, H_{m_{0}, n_{0} - n + k}\,\,
				H_{m^{\prime},k} } \\
			&= H_{m^{\prime}, n_{0} - n+ k} \,\,H_{m_{0}, n_{0}-n^{\prime}+k}\,\,
			H_{m,k},
		\end{align*}
		
		i.e., for $k \in \{0,1,2\},$
		\begin{align*}
			\lefteqn{ m(n_0-n^{\prime}+k) + m_0 (n_0-n+k)+ m^{\prime} k }\\
			&= m^{\prime} (n_0-n+k) + m_0 (n_0-n^{\prime}+k) + mk \quad (\mbox{mod}\,3)
		\end{align*}
		i.e., for $k \in \{0,1,2\},$
		\begin{align*}
			\lefteqn{m (n_0 - n^{\prime}) + m_0 (n_0-n) + (m+m_0+m^{\prime}) k }\\
			&= m^{\prime}(n_0-n) + m_0 (n_0 - n^{\prime})+ (m+m_0+m^{\prime}) k
		\end{align*}
		i.e.,
		$$(m-m_0) (n^{\prime}-n_0)-(m^{\prime}- m_0)(n-n_0)= 0 \quad (\mbox{mod}\,3).
		$$
		This gives us exactly $4$ $\mathbf{W}$-MASS's $\{(m_0, k) : k \neq n_0\},$
		$\{(j, n_0): j \neq m_0 \},$ $\{(m_0 + 1, n_0+1), (m_0+2, n_0+2)\}$ and
		$\{(m_0+1, n_0+2), (m_0+2, n_0+1) \}.$ They are all full-size and mutually
		disjoint.
		\end{enumerate}
	\end{itemize}
	
	\subsubsection{\it Examples of equivalent and non-equivalent unitary bases.}
	We will freely use 4.4.1 above.
	\begin{enumerate}
	\item[(i)] Computation details in  4.4.1 (vi)(c) above make it clear that
	$\mathbf{W}$-MASS's for tag at $(m_0, n_0)$ for the right subtraction latin
	square arising from $\mathbb{Z}_d$ and all Hadamard matrices $H^n$ same as in
	4.4.1(v)(c) are
	governed by the commuting rule
	$(m-m_0)(n^{\prime}-n_0)-(m^{\prime}-m_0)(n-n_0)=0~ (\mbox{mod}\,d).$ By
	Theorems \ref{thm2.7} and \ref{thm2.8} unitary basis for that and the one discussed in \S 2.3, \S 2.4 and Example \ref{e3.2.5} are equivalent.
	
	\item[(ii)] The descriptions of $\mathbf{W}$-MASS's for examples in 4.4.1(v)(e)  above
	and Example~\ref{e3.2.5}(b) (together with Remark \ref{r4.5.2} (i)(b) below) make it clear that
	the two unitary bases given in these two examples for $d=6$ are not equivalent.
	
	\item[(iii)] The context here is that of Example~\ref{e3.2.6}.\\
	\begin{enumerate}
	\item[(a)]~To within phases of $1$ and $-1$ and relabelling, all tags have
	the same underlying unitary system $\mathbf{W}.$
	\item[(b)]~To within PCUE, the fan system comes from 3.2.6(i).
	\item[(c)]~By Theorems~\ref{thm2.7} and \ref{thm2.8} (together with Remark~\ref{r4.5.2}(i) below), the unitary
	basis here is not equivalent to the one in the Example in 3.2.5(a).
	\end{enumerate}
	\item[(iv)] The question of phase equivalence in the examples
	above will not present significantly new points because it amounts to
	multiplying different rows of the Hadamard matrix by different numbers of
	modulus one.
	If the latin square has a right identity $k_0,$ then we can normalize this
	situation
	by keeping the $k_0$-column in each Hadamard matrix consisting of one's
	alone. In the particular case when $\lambda$ comes from a group, we may choose
	the
	identity to be the first element and thus insist on the first row and the first
	column of each Hadamard matrix to consist of one's alone.
	\end{enumerate}
	\subsection{Unitary error bases} For phase-equivalence the best set up is perhaps of nice unitary error bases
	defined by Knill \cite{EK}.
	\subsubsection{\it The Concepts}
	\begin{enumerate}
	\item[(i)] As in (\cite{EK}, \S 2) a \emph{nice unitary error basis} on a Hilbert space
	$\mathcal{H}$ of dimension $d$
	is defined as a set $\mathcal{E} = \{E_g\}_{g \in G}$ where $E_g$ is unitary
	on $\mathcal{H},$ $G$ is a group of order $d^2,$ $e$ its identity, ${\rm tr}E_g
	= d \delta_{g,e}$ and $E_gE_h = \omega_{g,h} E_{g h}.$ By renormalizing the
	operators of the error basis, it can be assumed that $\det E_g=1,$ in which case
	$\omega_{g,h}$ is a $d$-th root of unity. Error bases with this property are
	called {\it very nice.} Such error bases generate a finite group of unitary
	operators $\overline{\mathcal{E}}$ whose centre consists of scalar multiples of
	the identity. An {\it error group} is a finite group of unitary operators
	generated by a nice unitary error basis and certain multiples of the identity.
	The group $\mathbf{H}$ is an {\it abstract error group} if it is isomorphic to
	an error group.

	\item[(ii)] We quote Knill's Theorem without proof.
	\begin{theorem} (\cite{EK}, Theorem 2.1).\,\,  The finite group
		$\mathbf{H}$ is an abstract error group if and only if $\mathbf{H}$ has an
		irredcucible character supported on the centre and the kernel of the associated
		irredcucible representation is trivial.
	\end{theorem}
	\vskip.2em\noindent
	\item[(iii)] These concepts have been intensively and extensively studied by
	researchers and also very efficiently utilised by some of them for constructing
	interesting examples of error-detecting (correcting) quantum codes. For this
	purpose, the rich theory of group actions, Weyl operators, Weyl commutation
	relations, multipliers, cocycles, bicharacters, imprimitivity systems has been
	found to be of great importance by them, including K. R. Parthasarathy.
	For a good account we may refer to \cite{Pa} and
	references like \cite{dg}, \cite{vakrp} and \cite{vapkkrp} therein.
	\end{enumerate}
	\begin{remark}\label{r4.5.2}  The underlying projective representation in \S4.5.1 viz., $g \rightarrow E_g$
		leads to some very useful facts. 
		\begin{enumerate}
		\item[(i)]
		\begin{enumerate}
		\item[(a)] For $g \in G$, $\omega_{g,g^{-1}} = \omega_{g^{-1}, g}$ and
		$E^{\ast}_{g} =
		E^{-1}_{g} = \overline{{\omega}_{g, g^{-1}}}
		E_{g^{-1}}.$ 
		\item[(b)] For all tags $\mathbf{T},$ the underlying unitary system
		$\mathbf{W}$ is the same up to relabelling and phases.
		This permits us to consider
		the fan system the same as $\mathcal{V}_{\mathbf{W}}$ for any $\mathbf{W},$ so
		as
		to say. In particular, we may drop $\mathbf{W}$ from $\mathbf{W}$-MASS. In
		fact, it is enough to consider $\mathbf{W} = \{E_g : g \in G, g \neq e\}.$
		Further, figures 3,4,5 and 17 above display the respective fan systems as well.
		\item[(c)]
		$E_g, E_{g^{-1}}$ move
		together in any $\mathbf{W}$-MASS. We now proceed to strengthen this
		observation.
		\end{enumerate}
		\item[(ii)] Let $G$ be a group and $e$ its identity and $M$ a maximal
		commutative subset of $G.$ Then $M$ is a subgroup of $G.$ To see this
		well-known basic fact in group theory, we first
		note that $M$ can not be empty simply because for a in $G,$ $\phi \neq
		\{a\}$ which is
		commutative. Now let $g_1, g_2 \in M.$ Then for $h \in M,$ $(g_1 g_2^{-1})h =
		g_1 (g_2^{-1}h) = g_1 (h g_2^{-1}) = (g_1 h) g_2^{-1} = h(g_1g_2^{-1}).$ So by
		maximality of $M,$ we have $g_1 g_2^{-1} \in M.$ This gives that $M$ is a
		subgroup of $G.$ We may say that $M$ {\it is a maximal commutative subset of $G$
			if and only if it is a maximal abelian subgroup of $G.$}
		\item[(iii)] Let $\mathbf{H}$ be a nice error group arising from a very nice
		error basis
		as in (i) above.
		\vskip.2em
		We write $\omega_{g,h}$ by $\omega(g,h)$ and also $ \mathbf{U} =\{U_g : g \in
		G\}$ instead of $\mathcal{E}$ for notaional convenience. Let
		$\mathbf{T}_{\omega}$ be the subgroup of $S^1$ generated by the range of
		$\omega.$ Then $\mathbf{H} =
		\{(g,\alpha) : g \in G, \alpha \in \mathbf{T}_{\omega}\}$ and, for $(g,
		\alpha),
		(h, \beta) \in \mathbf{H},$ $(g, \alpha) (h, \beta)= (gh, \omega (g,h)\alpha
		\beta).$ Because $\mbox{tr}\,(U_g^{-1} U_h) = \delta_{g,h} d$ for $g,h$ in $G,$
		whenever $U_g = \lambda U_h$ for some $g,h$ in $G$ and scalar $\lambda,$ we
		must have $g=h$ and $\lambda = 1.$ So for $g, h \in G,$ $U_g U_h = U_h U_g$ if
		and only if $\omega (g,h) U_{gh} = \omega (h,g) U_{hg}$ if and only if $gh=hg$
		and $\omega(g,h)=\omega(h,g).$ So, this condition is further equivalent to
		$(g,1)(h,1) = (h,1)(g,1),$ which, in turn is equivalent to $(g, \alpha)(h,
		\beta)=(h, \beta)(g,\alpha)$ for $\alpha, \beta \in \mathbf{T}_{\omega}$ and
		that, in turn is equivalent to $(g,\alpha)(h, \beta)=(h,\beta)(g, \alpha)$ for
		some
		$\alpha, \beta \in \mathbf{T}_{\omega}.$ Thus, we have the following immediate
		consequences of (ii) above.
		\begin{enumerate}
		\item[(a)] $M \subset \mathbf{U}$ is an AUS if and only if $\mathbf{H}_M= \{(g,
		\alpha) : U_g \in M, \alpha \in \mathbf{T}_{\omega}\}$ is a commutative subset
		of $\mathbf{H}.$
		\item[(b)] $M$ is a $\mathbf{U}$-MASS if and only if $\mathbf{H}_M$ is a maximal
		abelian subgroup of $\mathbf{H}.$
		\item[(c)] Put $G_M = \{g \in G : U_g \in M\}$ and consider any function $\chi$ on
		$G_M$ to $\mathbf{T}_{\omega}.$ Set $T_{\chi} = \{(g, \chi(g)): g \in G_M \},$
		the $\chi$-transversal. We note that $G_M$ is the first projection of any such
		$T_{\chi}$ as also of $\mathbf{H}_M.$
		\item[(d)] Thus, the problem of finding MASS's in $\mathbf{U}$ is
		equivalent
		to that of finding maximal abelian subgroups of $\mathbf{H}$ with different
		first
		projections.
		\item[(e)] Further development of the theory of projective representations of finite
		groups
		studied thoroughly by I. Schur in early 1900s is very vast and deep. The survey
		article
		by Costache \cite{tlc} gives a readable account. We will
		not go into details or utilise or cite scholarly papers and monographs in this
		paper.
		\end{enumerate}
		\end{enumerate}
	\end{remark}
	\subsubsection{\it Equivalence question of nice error bases and Shift and Multiply Bases.}
	Klappenecker and Roetteler \cite{ar} studied the following
	question
	of Schlingemann and Werner: Is every nice error basis (phase-) equivalent to a
	basis of shift-and-multiply type? They answered it in the negative by concrete
	examples using the theory of Heisenberg groups, theory of characters and
	projective representations of finite groups.
	
	One can attempt alternate proofs
	using our results and details from the theory of finite groups.
	\subsection{Quantum Latin squares}
	Musto and Vicary \cite{MV1} introduced quantum Latin squares, combinatorial quantum objects
	which generalize classical Latin squares, and investigated their applications in quantum
	computer science.
	\subsubsection{\it Applications}
	In \cite{MV1}, their main results are on applications to unitary error bases (UEBs),
	basis structures in quantum information which lie at the heart of procedures such as
	teleportation, dense coding and error correction. Their new approach simultaneously
	generalizes the shift-and-multiply and Hadamard methods.
	They give  an explicit construction of a UEB using their techniques which cannot be obtained from any of these existing methods.
	\subsubsection{\it Constructing mutually unbiased bases from quantum Latin squares.}
	In \cite{MV2} Musto introduces orthogonal quantum Latin squares, which restrict to traditional
	orthogonal Latin squares, and investigates their application in quantum information
	science. He uses quantum Latin squares to build maximally entangled bases, and shows
	how mutually unbiased maximally entangled bases can be constructed in square dimension
	from orthogonal quantum Latin squares.
	
	\subsubsection{\it Biunitary constructions in quantum information.}
	Reutter and Vicary \cite{RV} present an infinite number of construction
	schemes involving unitray error bases,
	Hadamard matrices, quantum Latin squares and controlled families.
	Their results rely on biunitary connections, algebraic objects which
	play a central role in the theory of planar algebras. They have an attractive graphical
	calculus which allows simple correctness proofs for the constructions they present. They apply these techniques to construct a unitary error basis that cannot be built using any previously known method.
	
	These authors, Vijay Kodiyalam, Sruthymurali and V. S. Sunder are carrying this work further.

	\section{Conclusion}
	We started with a unitary basis $\mathbf{U}=\{U_x, x \in X\}$ on
	a Hilbert space $\mathcal{H}$ of finite dimension $d$ and an $x_0 \in X.$ We
	associated the tag $\mathbf{T}=(x_0, U_{x_{0}}, \mathbf{W})$ at
	$x_0,$ where $\mathbf{W} = \{W_x = U_{x_{0}}^{\ast} U_x, x \in X,
	x \neq x_0 \}$ is a so-called unitary system. We obtained a covering of
	$\mathbf{W}$ by maximal abelian subsets of $\mathbf{W}$ (called $\mathbf{W}$-MASS's).
	We obtained the set of $\mathbf{W}$-MASS's for different concrete
	$\mathbf{W}$'s displaying various patterns, like mutually disjoint, overlapping
	in different ways, and, therefore, called them fans. Varying $x_0,$ the whole
	collection was called a fan system of $\mathbf{U}.$ We showed that it is an
	invariant of $\mathbf{U}$ to within (phase) equivalence of unitary bases.\
	The concept of collective unitary equivalence was utilised for this purpose.
	Applications to distinguishing Maximally entangled states bases were indicated.
	
	The thrust  of the paper is on applications of fan representation
	to quantum tomography. For this basics were explained and techniques were developed to obtain
	optimal informationally complete pure  measurement sets for any density.\ Examples were given to
	illustrate the results. Details for the ideal size $d^2$ such measurement systems  for The  Schwinger basis and the
	Shift and multiply bases using latin squares and Hadamard matrices were provided. Quantum mechanical overlaps for
	them were displayed for some of them and compared in different set-ups , like bases arising
	from Weyl operators or Wigner distributions of finite state systems.
	
			\section*{Notation}
	\begin{tabular}{cl} \hline\hline\\
		& \\
		$\mathcal{H}$&  Hilbert space\\
		$\mathcal{B}(\mathcal{H})$ & Set of bounded operators on $\mathcal{H}$\\
		\bf{U}& Unitary basis \\
		\bf{W}& Unitary system\\
		{\bf W}-MASS & Maximal abelian subsystem of {\bf W}\\
		$\widetilde{\textbf{W}}$ & Unitization of $\textbf{W}$\\
		$\textbf{V}$& A $\textbf{W}$-MASS\\
		$\V_{ \textbf{W}}$ & A set of ${\bf \text{W}}$-MASS's that cover {\bf W}\\
		
		$\mathcal{E}_{{ \textbf{V}}}$ & A common orthonormal basis of $\textbf{V}$ \\
		$\mathcal{P}_{\textbf{V}}$& Set of one dimensional projections associated with $\mathcal{E}_{{ \textbf{V}}}$\\
		MUB& Mutually unbiased bases\\
		$\mathbb{Z}_d$ & Ring of integers modulo $d$\\
		$\mathbb{F}_d$ & Finite field of order $d$ for a prime $d$\\
		
		$W(a,x)$ & Weyl operators\\
		\textbf{Q}& A family of mutually orthogonal projections on a Hilbert space $\mathcal{K}$ adding up to
		$I_{\mathcal{K}}$\\
		$q_t$ & Rank of an element $\text{Q}_t$ of \textbf{Q}\\
		$\sigma(T)$ & Spectrum of a normal operator $T$ on a Hilbert space $\mathcal{K}$\\
		$\textbf{Q}^T$& Spectral projection set of $T$\\
		$\mathcal{M}_{\textbf{W}}$ & A minimal subset of $\mathcal{V}_{\textbf{W}}$ that covers $\textbf{W}$\\
		CUE & Collective unitary equivalence\\
		TUS& Tagged unitary system\\
		$H_{\textbf{W}}$ & Hadamard fan for $\mathbf{W}$ \\
		MES& Maximally entangled states\\
		POVM& Positive operator valued measure\\
		SIC-POVM& Symmetric informationally complete positive operator valued measure\\
		& \\
		\hline\hline
	\end{tabular}
\vskip5mm
	\section*{Acknowledgement}
	
	Ajit Iqbal Singh expresses her deep sense of gratitude to \linebreak K.
	R. Parthasarathy. She has learnt most of the basic concepts in this
	paper from him during his Seminar Series of Stat. Math. Unit at the Indian
	Statistical Institute, New Delhi, University of Delhi and elsewhere. She has
	gained immensely from insightful discussion sessions with him
	from time to time.
	
	She thanks V. S. Sunder and V. Kodiyalam for supporting her visit to The
	Institute of Mathematical Sciences (IMSc), Chennai to participate in their
	scholarly workshops on ``Functional Analysis of Quantum Information Theory'' in
	December, 2011 - January, 2012, ``Planar algebras'' in March-April, 2012 and
	Sunder Fest in April, 2012. This enabled her to learn more from experts at these
	events and initiate her interaction with Sibasish Ghosh, who invited her for further
	visits from time to time. She also thanks R.
	Balasubramanian, then
	Director, and V. Arvind, present Director, IMSc, for providing more such opportunities, kind hospitality,
	stimulating research atmosphere and encouragement
	all through.
	She would particularly like to mention discussion
	with V. Paulsen, R. Simon and Andreas Winter. Her visits to Hyderabad University
	and The Indian Institute of Science Education and Research, Bhopal (IISERB)
	facilitated interaction with S. Chaturvedi. She thanks him and Nikita Agarwal, Head, Department of Mathematics, IISERB for the kind hospitality during these visits.
	
	She thanks Indian National Science Academy for support under the INSA Senior
	Scientist and Honorary Scientist Programmes and Indian Statistical
	Institute, New Delhi for Visiting positions under these
	programmes
	together with excellent research atmosphere and facilities all the time.
	
	The authors thank Kenneth A. Ross for reading the paper and suggesting
	improvements.
	They also thank W.M. Kantor for useful comments and suggestions. They are grateful to the referees for their useful comments and
	suggestions which led to a massive  revision and development of a substantial new part with details in an accessible form.
	Finally, they thank Mr. Anil Kumar Shukla and M/s~Scientific Documentations for transforming the
	manuscript into its  \LaTeX \,\,form from its initial draft. They are also grateful to Mr Shamim ul Haque for his immense help in adapting the \LaTeX \,\, file to the Reviews in Mathematical Physics style.
	

	
\end{document}